\pdfoutput=1
\documentclass{article}
\usepackage{iclr2025_conference,times}

\usepackage{hyperref}
\usepackage{url}
\usepackage{enumitem}
\usepackage{booktabs}
\usepackage{graphicx}
\usepackage{amsmath}
\usepackage{xcolor}
\usepackage{adjustbox}
\usepackage{array}
\usepackage{multirow}
\usepackage{subcaption}

\title{Beyond Sequence: Impact of Geometric \\ Context for RNA Property Prediction}

\author{
Junjie Xu$^{1,2}$ \thanks{This work was done while the author was an intern at Johnson \& Johnson.} \  \thanks{Equal contribution as first authors} \ , 
Artem Moskalev$^1$ \footnotemark[2] \ , 
Tommaso Mansi$^1$, 
Mangal Prakash$^1$ \thanks{Equal contribution as last authors} \ , 
Rui Liao$^1$ \footnotemark[3]  \\
$^1$Johnson \& Johnson Innovative Medicine,  $^2$The Pennsylvania State University \\
\texttt{junjiexu@psu.edu} \\
\texttt{\{amoskal2, tmansi, mpraka12, rliao2\}@its.jnj.com} \\
}

\iclrfinalcopy 
\begin{document}

\maketitle

\begin{abstract}
Accurate prediction of RNA properties, such as stability and interactions, is crucial for advancing our understanding of biological processes and developing RNA-based therapeutics. RNA structures can be represented as 1D sequences, 2D topological graphs, or 3D all-atom models, each offering different insights into its function. Existing works predominantly focus on 1D sequence-based models, which overlook the geometric context provided by 2D and 3D geometries. This study presents the first systematic evaluation of incorporating explicit 2D and 3D geometric information into RNA property prediction, considering not only performance but also real-world challenges such as limited data availability, partial labeling, sequencing noise, and computational efficiency. To this end, we introduce a newly curated set of RNA datasets with enhanced 2D and 3D structural annotations, providing a resource for model evaluation on RNA data. Our findings reveal that models with explicit geometry encoding generally outperform sequence-based models, with an average prediction RMSE reduction of around 12\%\ across all various RNA tasks and excelling in low-data and partial labeling regimes, underscoring the value of explicitly incorporating geometric context. On the other hand, geometry-unaware sequence-based models are more robust under sequencing noise but often require around $2-5\times$ training data to match the performance of geometry-aware models. Our study offers further insights into the trade-offs between different RNA representations in practical applications and addresses a significant gap in evaluating deep learning models for RNA tasks.
\end{abstract}

\section{Introduction}
\label{sec:intro}

\begin{figure}
    \centering
    \begin{subfigure}[b]{0.75\linewidth}
        \centering
        \includegraphics[width=\linewidth]{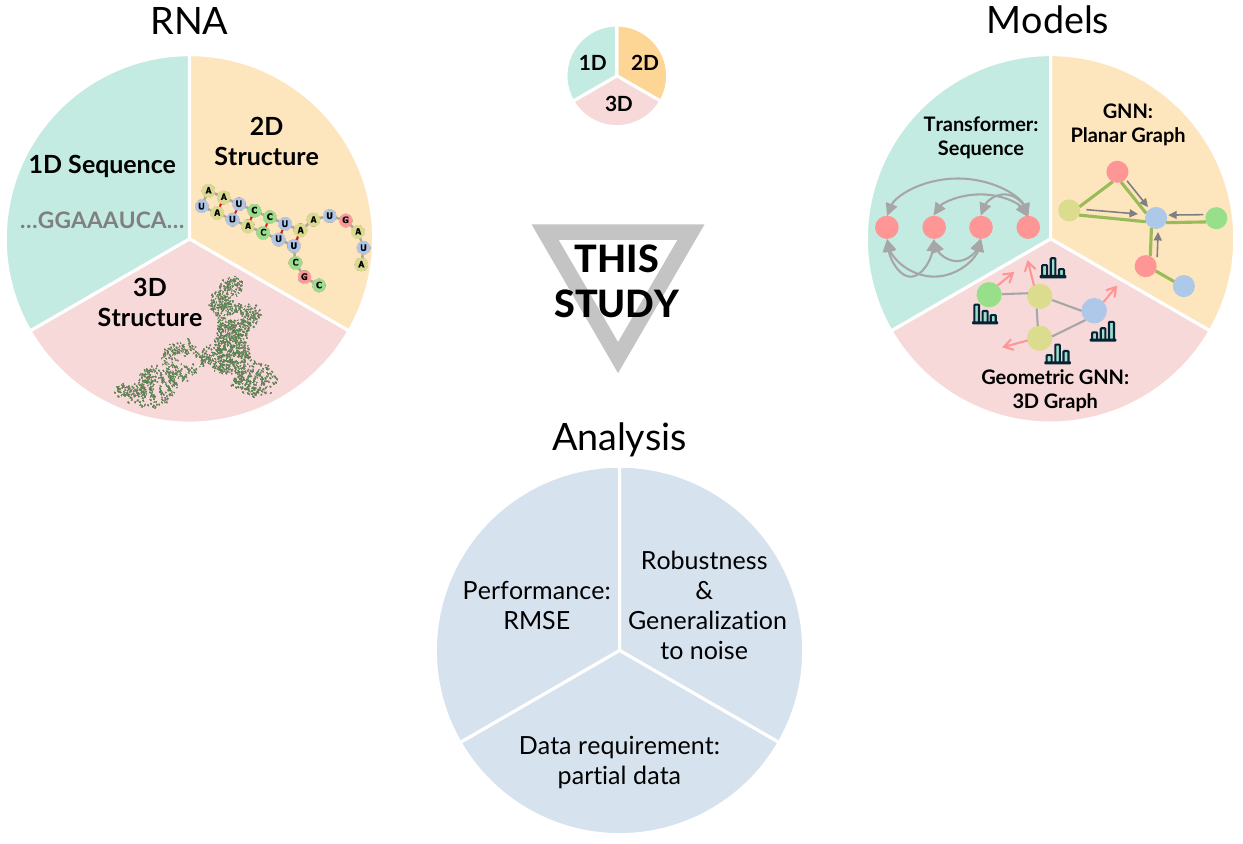}
    \end{subfigure}
    \hspace{0.4cm}
    \begin{subfigure}[b]{0.19427\linewidth}
        \centering
        \includegraphics[width=\linewidth]{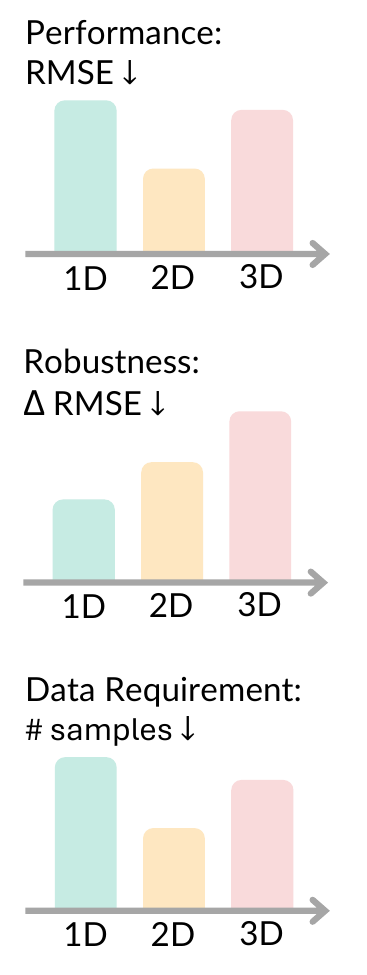}
    \end{subfigure}
    \caption{\textbf{Overview of the study.} (a) Left panel: RNA sequences represented in 1D, 2D, and 3D structures, processed by 1D sequence, 2D GNN, and 3D GNN models. Our analysis includes prediction error, robustness and generalization to sequencing noise, and performance under limited training data and partial labelings. (b) Right panel: Comparative performance of 1D, 2D, and 3D methods across experimental conditions. Histograms show RMSE performance, relative RMSE changes with increasing noise, and data requirements for optimal performance. Lower values indicate better performance in all metrics.}
    \vspace{-2em}
    \label{fig:teaser}
\end{figure}

RNA plays a central role in the machinery of life, serving as a crucial intermediary between nucleotide and amino acid worlds~\citep{sahin2014mrna}. Beyond its messenger role, RNA is involved in diverse biological processes, including gene regulation, catalytic activity, and structural support within ribosomes~\citep{sharp2009centrality, strobel2016rna}. This versatility makes RNA a key target for fundamental biological research and therapeutic interventions. As our understanding of RNA complexity grows, so does the need for advanced computational tools for its analysis.

Modeling RNA is challenging due to its intricate secondary and tertiary structures, dynamic conformational changes, and interactions with cellular components~\citep{han2015advanced}. Furthermore, RNA analysis is hindered by the practical challenges of RNA data acquisition which include sequencing errors~\citep{ozsolak2011rna}, batch effects~\citep{tran2020benchmark}, incomplete sequencing~\citep{alfonzo2021call}, partial labeling~\citep{wayment2022deep}, and high costs of obtaining large labeled datasets~\citep{byron2016translating}. Moreover, RNA molecule can be represented in different ways: as a 1D nucleotide sequence, a 2D graph of base pairings, or a 3D atomic structure. Each representation highlights different aspects of RNA, presenting both opportunities and challenges for model design and selection (Fig.~\ref{fig:teaser}).

In this work, we systematically study the performance of various machine learning models for RNA property prediction, extending beyond traditional sequence-based approaches~\citep{he2021nucleic, soylu2023bert2ome, prakash2024bridging, yazdani-jahromi2025helm} to include methods that process RNA with its 2D or 3D geometry. While 2D and 3D RNA representations offer potentially richer information, they also present unique challenges. In the absence of high-quality experimental data, accessing 2D or 3D RNA structures requires running structure prediction algorithms prone to noise and mistakes, especially in the presence of sequencing errors~\citep{schneider2023will, wang2023trrosettarna}. A few mutations in the nucleotide sequence owing to sequencing mistakes can hugely alter the 2D and 3D structure (Fig.~\ref{fig:rna_noise}), potentially undermining the benefit of using additional geometric context. Furthermore, real-world RNA datasets often suffer from partial labeling~\citep{wayment2022deep} and scarcity of training samples~\citep{wint2022kingdom}, which may affect geometry-aware methods differently than sequence-based approaches. These challenges raise a fundamental question: to what extent does explicit 2D and 3D geometry contribute to RNA property prediction, and under what circumstances might it offer advantages over geometry-free sequence models?


In this study, we seek to answer this question by making the following contributions: 
\begin{itemize}[left=0pt, itemsep=0pt, topsep=0pt]
\item We introduce a diverse collection of RNA datasets, including newly annotated 2D and 3D structures, covering various prediction tasks at nucleotide and sequence levels across multiple species.
\item We provide a unified testing environment to evaluate different types of machine learning models for RNA property prediction, including sequence models for 1D, graph neural networks for 2D, and equivariant geometric networks for 3D RNA representations.
\item We conduct a comprehensive analysis of how different models perform under various conditions, such as limited data and labels, different types of sequencing errors, and out-of-distribution scenarios. We highlight the trade-offs and contexts in which each modeling approach is most effective, guiding researchers in selecting suitable models for specific RNA analysis challenges.
\item We also introduce novel modifications to existing 3D geometric models based on biological prior, specifically optimizing them for handling large-scale point cloud RNA data, thus improving the efficiency and performance of 3D models significantly. 
\end{itemize}

Our study reveals several key insights: \textbf{(i)} 2D models generally outperform 1D models, with spectral GNNs reducing prediction error by about 12\%\ on average across all datasets, highlighting the importance of explicitly considering RNA structural information; \textbf{(ii)} 3D equivariant GNNs outperform 1D and some 2D methods in noise-free scenario but are sensitive to noise, exhibiting up to a 56\%\ decrease in prediction quality under high sequencing error rates; \textbf{(iii)} geometry-free sequence models remain the most robust to sequencing noise, showing only a 14-27\%\ increase in prediction error compared to noise-free conditions, however they require around $2-5\times$ more training data to match the performance of geometry-aware models.

\section{Datasets and Models}
\label{sec:dataset_model}

Here, we discuss datasets selected for our study with RNA-level prediction labels. These datasets are selected to vary from small to large-scale and to encompass both nucleotide-level tasks and sequence-level tasks. We perform an extensive evaluation across these datasets, leveraging three different model families (1D, 2D, 3D) spanning 9 representative models in total. 

\subsection{Datasets}
The datasets vary in size based on the number of sequences and sequence lengths: the small dataset Tc-Riboswitches~\citep{groher2018tuning}, the medium datasets Open Vaccine COVID-19~\citep{wayment2022deep} and Ribonanza-2k~\citep{he2024ribonanza}, and the large dataset Fungal~\citep{wint2022kingdom}. All datasets provide regression labels. Detailed statistics for these datasets are provided in Appendix~\ref{appen:dataset_stat}.

\begin{enumerate}[left=0pt, label=\arabic*.]
\item \textbf{Tc-Riboswitches}: 355 mRNA sequences (67-73 nucleotides) with sequence-level labels for tetracycline-dependent riboswitch switching behavior, important for optimizing gene regulation in synthetic biology and gene therapy.
\item \textbf{Open Vaccine COVID-19}: 4,082 RNA sequences (each of 107 nucleotides) with nucleotide-level degradation rate labels, crucial for predicting RNA stability in mRNA vaccine development.
\item \textbf{Ribonanza-2k}: 2,260 RNA sequences (each of 120 nucleotides) with nucleotide-level experimental reactivity labels, supporting RNA structure modeling and RNA-based drug design.
\item \textbf{Fungal}: 7,056 coding and tRNA sequences (150-3,000 nucleotides) from 450 fungal species, used for sequence-level protein expression prediction.
\end{enumerate}

\subsection{Data Preprocessing and Curation}
For the OpenVaccine COVID-19 dataset, we filter out sequences with a signal-to-noise ratio (SNR) below 1, as recommended by the dataset authors~\citep{wayment2022deep}, to ensure that only sequences with a significant signal relative to background noise are included, thereby enhancing the reliability of modeling. For the other datasets, we use the original sequences since no SNR annotations are available. 

Since all the RNA datasets come with sequences only, we employ EternaFold~\citep{wayment2022rna} and RhoFold~\citep{shen2022e2efold} to infer 2D and 3D molecular structures respectively. We selected EternaFold and RhoFold due to their state-of-the-art performances acknowledged in recent works~\citep{Wayment-Steele, wayment2022deep, he2024ribonanza} Additionally, RhoFold typically runs in seconds to a minute per sequence, unlike other 3D structure prediction tools which usually take hours, and hence not suitable for large datasets.

For 1D modeling, we use the original RNA sequences without structural augmentation which equates to processing a plain string of nucleotides. The 2D datasets represent each RNA sequence as a graph with nodes for nucleotides and edges for bonds between nucleotides. The node features are six-dimensional, incorporating one-hot nucleotide identity (`A', `C', `G', `U') alongside the sum and mean base-pairing probabilities (BPP), which are available from 2D structure prediction tools. In 3D, each RNA molecule is represented as a graph, with nodes corresponding to individual atoms. Node features represent one-hot atom identity.

\subsection{Models}
We select well-established model architectures recognized for their state-of-the-art performance for molecular property prediction tasks in various domains. \textbf{1D Model}: Transformer1D~\citep{honda2019smiles, he2021nucleic}; RNA-FM~\citep{chen2022interpretable}, SpliceBERT~\citep{chen2023self} \textbf{2D Models}: GCN~\citep{kipf2017semi, wieder2020compact}, GAT~\citep{velickovic2018graph, ye2022molecular}, ChebNet~\citep{defferrard2016convolutional, knyazev2018spectral}, Transformer1D2D~\citep{he2023rnadegformer}, Graph Transformer~\citep{shi2020masked, li2022kpgt}, and GraphGPS~\citep{rampasek2022GPS, zhu2023dual}; \textbf{3D Models}: SchNet~\citep{schutt2017schnet, han2022geometrically}, EGNN~\citep{satorras2021n}, FAENet~\citep{duval2023faenet}, DimeNet~\citep{gasteiger2020directional}, GVP~\citep{jing2020learning}, and FastEGNN~\citep{zhangimproving}. Detailed descriptions of all the models can be found in Appendix Sec.~\ref{appen:models_overview}.

\paragraph{Training and evaluation} All models were trained on a NVIDIA A100 GPU. To ensure hyperparameter parity for each baseline, hyperparameters were optimized using Optuna~\citep{akiba2019optuna}, restricting the search to models with fewer than 10 million parameters that fit within the GPU memory constraint of 80GB. All model hyperparameters, training, and evaluation details are reported in Appendix~\ref{appen:reproduction}. We ran all models for $5$ random data splits (train:val:test split of 70:15:15) and we report average performance with a standard deviation across splits. The mean column-wise root mean squared error (MCRMSE), introduced in~\cite{Wayment-Steele}, is used as the evaluation metric. It is defined as $\text{MCRMSE}(f, D) = \sqrt{\frac{1}{n} \sum_{i=1}^{n} \left( \hat{y}_i - y_i \right)^2 }$, where \( f \) represents a model, \( D \) is the dataset, and \( \hat{y}_i \) and \( y_i \) are the predicted and true values for data point \( i \).
\section{Task Definitions}
\label{sec:tasks}

Here, we introduce the downstream tasks for evaluating models for RNA property prediction. Each task is designed to quantify specific behaviors under various real-world experimental conditions.


\paragraph{Task 1: Impact of structural information on prediction performance} This task aims to evaluate how incorporating RNA structural information affects prediction quality. We compare the performance of models using 1D (sequence-only), 2D, and 3D RNA representations to determine if and to what extent geometric data improves property prediction.

\paragraph{Task 2: Model efficiency in limited training data settings} Acquiring high-quality comprehensive RNA datasets is often challenging and resource-intensive thus limiting the amount of labeled data for training~\citep{teufel2022reducing, byron2016translating}. This task aims to investigate how model performance depends on the amount of training data used, evaluating the sample efficiency of each family of models. In other words, given a dataset \( D = \{ X, Y \} \), let \( D_\alpha = \{ X_\alpha, Y_\alpha \} \) be a subset where the training set is reduced to a fraction \( \alpha \). We train models on different sets of $D_\alpha$ datasets with decreasing \( \alpha \).

\paragraph{Task 3: Performance with partial sequence labeling} Due to the high cost of measuring properties for every nucleotide in RNA sequence, real-world datasets often contain partial annotations~\citep{wayment2022deep} where labels are only available for the first small part of the sequence. This task is relevant for nucleotide-level datasets and it aims to investigate how well a model can generalize to a whole RNA sequence when labels are only available for a portion of it. 

\paragraph{Task 4: Robustness to sequencing noise} Acquiring RNA data requires sequencing. In practice sequencing procedure may introduce sequencing errors (random mutations of nucleotides) that vary depending on the sequencing technology and platform~\citep{ozsolak2011rna, fox2014accuracy}. These errors affect the raw sequence data, and propagate to structural noise in 2D and 3D. The goal of this task is to assess how well models can maintain reliable performance when trained and tested under the same distribution of realistic levels of sequencing noise observed in practice, ensuring robustness across a consistent noise environment. This reflects real-world cases where a specific sequencing method produces noisy data, but the noise characteristics are stable across training and deployment.

\paragraph{Task 5: Generalization to Out-of-Distribution (OOD) data} This task focuses on a different practical challenge: models trained on high-quality RNA sequences are often deployed in conditions where the data exhibits different noise characteristics due to batch effect~\citep{tran2020benchmark} or the use of different sequencing platforms~\citep{tom2017identifying}. Here, the objective is to evaluate how well models generalize to OOD datasets with different levels of sequencing noise, assessing the extent of performance degradation as noise levels increase. This task simulates the scenario where a model encounters noisier data than it was trained on, highlighting its ability to adapt to unexpected experimental conditions.

For detailed descriptions and motivations of the five task settings, please refer to Appendix \ref{appen:detailed_task_descriptions}.
\section{Experiments and Results}
\label{sec:exp}


\subsection{Impact of explicit geometry learning on model performance}
We begin by addressing Task 1, where we compare the performance of model families when trained and evaluated on the downstream RNA datasets. Additionally, we provide runtime and memory comparison in Appendix \ref{appen:memory_compute}.

\begin{table}[htbp]
\centering
\caption{\textbf{Comparison of 1D, 2D, and 3D models across datasets.} \textbf{Bold} indicates the best, \underline{underline} the second-best. `OOM' means out-of-memory. ChebNet excels by capturing global graph information. Overall, 2D models outperform 1D models, highlighting the value of structural information. Although 3D models face challenges with scalability and noisy predictions, our nucleotide pooling strategy, based on biological prior, enhances their performance on shorter sequences, allowing 3D encoding to occasionally surpass 1D models. See Sec.~\ref{limitations_3d} for details on nucleotide pooling strategy.}
\label{tab:model_comparison}
\resizebox{0.85\textwidth}{!}{%
\begin{tabular}{lcccc}
\toprule
\textbf{Model} & \textbf{COVID} & \textbf{Ribonanza} & \textbf{Tc-Ribo} & \textbf{Fungal} \\
\midrule
\multicolumn{5}{l}{\textit{1D model}} \\
\midrule
Transformer1D     & 0.361$\pm$0.017 & 0.705$\pm$0.015 & 0.705$\pm$0.079 & 1.417$\pm$0.005 \\ 
RNA-FM            & 0.591$\pm$0.081 & 0.990$\pm$0.144 & 0.693$\pm$0.001 & 1.420$\pm$0.028 \\ 
SpliceBERT        & 0.588$\pm$0.077 & 1.022$\pm$0.144 & 0.708$\pm$0.003 & $1.435\pm$0.059 \\ 
\midrule
\multicolumn{5}{l}{\textit{2D model}} \\
\midrule
Transformer1D2D   & 0.305$\pm$0.012 & \underline{0.514$\pm$0.004} & \underline{0.633$\pm$0.001} & OOM \\ 
GCN               & 0.359$\pm$0.009 & 0.595$\pm$0.006 & 0.701$\pm$0.004 & 1.192$\pm$0.077 \\ 
GAT               & \underline{0.315$\pm$0.006} & 0.534$\pm$0.006 & 0.685$\pm$0.024 & 1.112$\pm$0.035 \\ 
ChebNet           & \textbf{0.279$\pm$0.007} & \textbf{0.468$\pm$0.002} & \textbf{0.621$\pm$0.022} & \textbf{0.973$\pm$0.003} \\ 
Graph Transformer & 0.318$\pm$0.008 & 0.515$\pm$0.001 & 0.710$\pm$0.041 & 1.317$\pm$0.002 \\ 
GraphGPS          & 0.332$\pm$0.013 & 0.523$\pm$0.003 & 0.715$\pm$0.012 & \underline{1.025$\pm$0.081} \\ 
\midrule
\multicolumn{5}{l}{\textit{3D model (without pooling) }} \\
\midrule
EGNN        &0.480$\pm$0.025  &0.808$\pm$0.023  &0.725$\pm$0.002   & OOM     \\ 
SchNet      &0.499$\pm$0.003  &0.843$\pm$0.004  &0.696$\pm$0.008   & OOM     \\ 
FAENet      & 0.486$\pm$0.010 & 0.834$\pm$0.003 & 0.703$\pm$0.011 & OOM \\ 
DimeNet     & 0.497$\pm$0.012 & 0.855$\pm$0.006 & 0.712$\pm$0.004 & OOM \\ 
GVP         & 0.467$\pm$0.010 & 0.797$\pm$0.012 & 0.744$\pm$0.004 & OOM \\ 
FastEGNN    & 0.477$\pm$0.005 & 0.816$\pm$0.014 & 0.753$\pm$0.001 & OOM \\ 
\midrule
\multicolumn{5}{l}{\textit{3D model (with nucleotide pooling) }} \\
\midrule
EGNN (pooling)         & 0.364$\pm$0.003 & 0.619$\pm$0.007 & 0.663$\pm$0.010     & OOM    \\ 
SchNet (pooling)       & 0.390$\pm$0.006 & 0.685$\pm$0.006 & 0.655$\pm$0.038     & OOM    \\
FastEGNN (pooling)     & 0.444$\pm$0.003 & 0.753$\pm$0.015 & 0.710$\pm$0.011     & OOM \\ 
\bottomrule
\end{tabular}
}
\vspace{-1em}
\end{table}

\paragraph{2D models consistently outperform 1D model} Results in Table~\ref{tab:model_comparison} reveal that 2D methods consistently outperform the 1D sequence model across all datasets. Notably, the Transformer1D2D model, which simply augments the attention matrix with adjacency features alongside, achieves around 10\%\ lower prediction MCRMSE on average across datasets than its geometry-free counterpart. This suggests that explicitly incorporating structural information is crucial, as learning from sequence data alone proves to be insufficient. Further experiments, detailed in the Appendix~\ref{appen:comapre_1d_1d2d}, investigate the learned attention maps of both the Transformer1D2D and the Transformer1D model and their correlation with structural information and reveal that Transformer1D2D attention maps are much more closely aligned with the topological structure of nucleotide graph, reinforcing the conclusion that explicit encoding of structural information is essential for improved performance.

For the foundation models, RNA-FM and SpliceBERT, we observe that for 2 out of 4 datasets (Covid and Ribonanza-2k), RNA foundation models perform worse than the simple supervised transformer baseline whereas for the other two datasets (Tc-Riboswitches and Fungal), transformer and RNA foundation models achieve similar performance. This is consistent with recent works in multiple biology-related domains demonstrating specialized foundation models are yet to surpass simple supervised learning baselines~\citep{xu2024specialized, kedzierska2023assessing, yang2024convolutions}. Hence, we choose Transformer1D as the model of choice for subsequent experiments.

\paragraph{Spectral GNN outperforms spatial GNNs in 2D} ChebNet, a spectral method, outperforms spatial methods such as GCN, GAT, Graph Transformer, and GraphGPS, achieving a prediction MCRMSE 2.5\%\ lower than the next best 2D model across datasets. Spatial GNNs aggregate node features layer by layer, emphasizing local information within a fixed distance. While computationally efficient, these methods are limited by the 1-Weisfeiler-Lehman (WL) test, which constrains the expressive power of node-based updates~\citep{xu2018powerful}. In addition, spatial GNNs may suffer from a limited receptive field while spectral methods approximate global graph features, enabling a global receptive field since the first layer~\cite{wang2022powerful, bo2023survey, xu2024shape}. This allows ChebNet to effectively process global information which is important for RNA data due to potential long-range interaction between nucleotides.

\paragraph{Challenges of modeling geometric context in all-atom resolution}
\label{limitations_3d}
Contrary to our expectations, 3D models at all-atom resolution (EGNN w/o pooling, SchNet w/o pooling DimeNet (w/o pooling), FAENet (w/o pooling), GVP (w/o pooling) and FastEGNN (w/o pooling) in Table~\ref{tab:model_comparison}) show relatively high prediction MCRMSE across datasets, underperforming compared to 1D and 2D methods. We hypothesize this is due to two factors.

First, all-atom 3D models rely on a limited local neighborhood of adjacent atoms, limiting their receptive fields and preventing them from capturing long-range dependencies (see Appendix Table~\ref{appen:dataset_stat}), which can be crucial for determining RNA properties~\citep{shetty2013hepatitis, alshareedah2019interplay}. Expanding the local neighborhood for these methods becomes challenging due to the overwhelming scale of large molecular systems in all-atom resolution. Second, the performance of 3D models is often limited by the inherent inaccuracies in 3D structure prediction tools, which are generally less reliable compared to 2D structure prediction methods~\citep{ponce2019computational}.
While FastEGNN is designed for larger molecules, its reliance on the center of mass virtual node initialization may not align well with tasks like RNA property prediction, which inherently have a sequence prior. This may explain the limited performance of FastEGNN in this context and RNA-specific virtual node initialization and carefully designed pooling mechanisms may be needed to further adopt this architecture for large-molecules with a sequence prior.

To address the receptive field limitation of all-atom methods, we employ a biological prior by pooling atomic features into nucleotide-level representations after a few layers of all-atom operations. This strategy is aligned with RNA’s natural secondary structure, where atoms group into nucleotides, and nucleotides form the complete RNA molecule~\citep{deng2023rna}. This novel strategy allows us to maintain all-atom resolution in the initial layers while increasing the receptive field via pooling. By balancing the number of all-atom and nucleotide layers as hyperparameters, we can balance fine- and coarse-grained all-atom and nucleotide resolution. Compared to other 3D models, EGNN, FastEGNN, and SchNet perform slightly better, and hence we introduce the nucleotide pooled variants for these models. We call 3D models with pooling EGNN (nuc. pooling), FastEGNN (nuc. pooling), and SchNet (nuc. pooling). This strategy significantly enhances 3D model performance, reducing prediction MCRMSE by $\sim$10\%\ compared to the original EGNN, SchNet, and FastEGNN and outperforming 1D models on the Ribonanza-2k and Tc-Riboswitches datasets by 5\%\ on average. On the COVID dataset, EGNN (nuc. pooling) matches the Transformer1D model but still trails behind 2D models. All subsequent experiments report results using nucleotide-pooled versions of the 3D models.

Next, we investigated the second hypothesis regarding higher noise in 3D structures by quantifying variability in predicted structures across different 3D prediction tools in Appendix~\ref{appen:3d_structure_noise}. We observe substantial variability (between 11-45Å RMSD across structures given by different 3D structure prediction tools), suggesting considerable noise in 3D predictions, which likely contributes to the poorer performance of 3D models.


\subsection{Model efficiency under limited data and partial sequence labeling}
In this section, we combine the analysis of Tasks 2 and 3, assessing model performance in scenarios with limited training data or partial labels.

To analyze how the amount of training data influences model performance, we run experiments with varying portions of the full datasets (25\%,\ 50\%,\ 75\%,\ and 100\%)\ on the medium- and large-scale datasets: COVID, Ribonanza-2k, and Fungal (Appendix Fig.~\ref{fig:compare_partial}(a) for illustration). The small size of the Tc-Riboswitches dataset is excluded from the analysis, as training with lower ratios would have resulted in inadequate sample sizes for meaningful evaluation. Additionally, GPU memory constraints prevent the application of Transformer1D2D and 3D models on the Fungal dataset due to its large sequence length.

We also evaluate the impact of partial property labels for nucleotide-level tasks, a common occurrence owing to costly experimental measurements~\citep{Wayment-Steele} to identify which models are best suited to handle the challenges of incomplete labels in RNA property prediction. For this, we use the COVID and Ribonanza datasets as these datasets contain nucleotide-level labels. We train the models using all training data but with varying proportions of labeled nucleotides (20\%,\ 40\%,\ 60\%,\ 80\%,\ and 100\%)\ per sequence, thus simulating incomplete or sparse labeling, while testing on fully labeled test sets (Appendix Fig.~\ref{fig:compare_partial}(b) for illustration).

\begin{figure}[ht]
\begin{picture}(0,135)
  \put(0,110){\begin{minipage}[b]{\textwidth}
   \includegraphics[width=\linewidth]{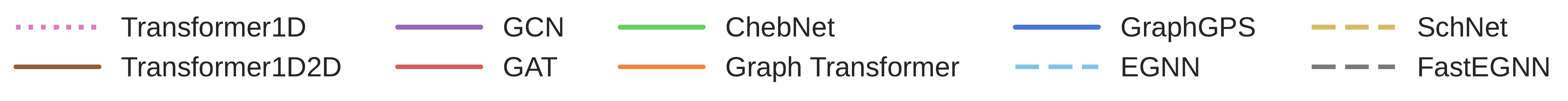}
    \end{minipage}}
    
  \put(0,0){\begin{minipage}[b]{.31\textwidth}
   \includegraphics[width=\linewidth]{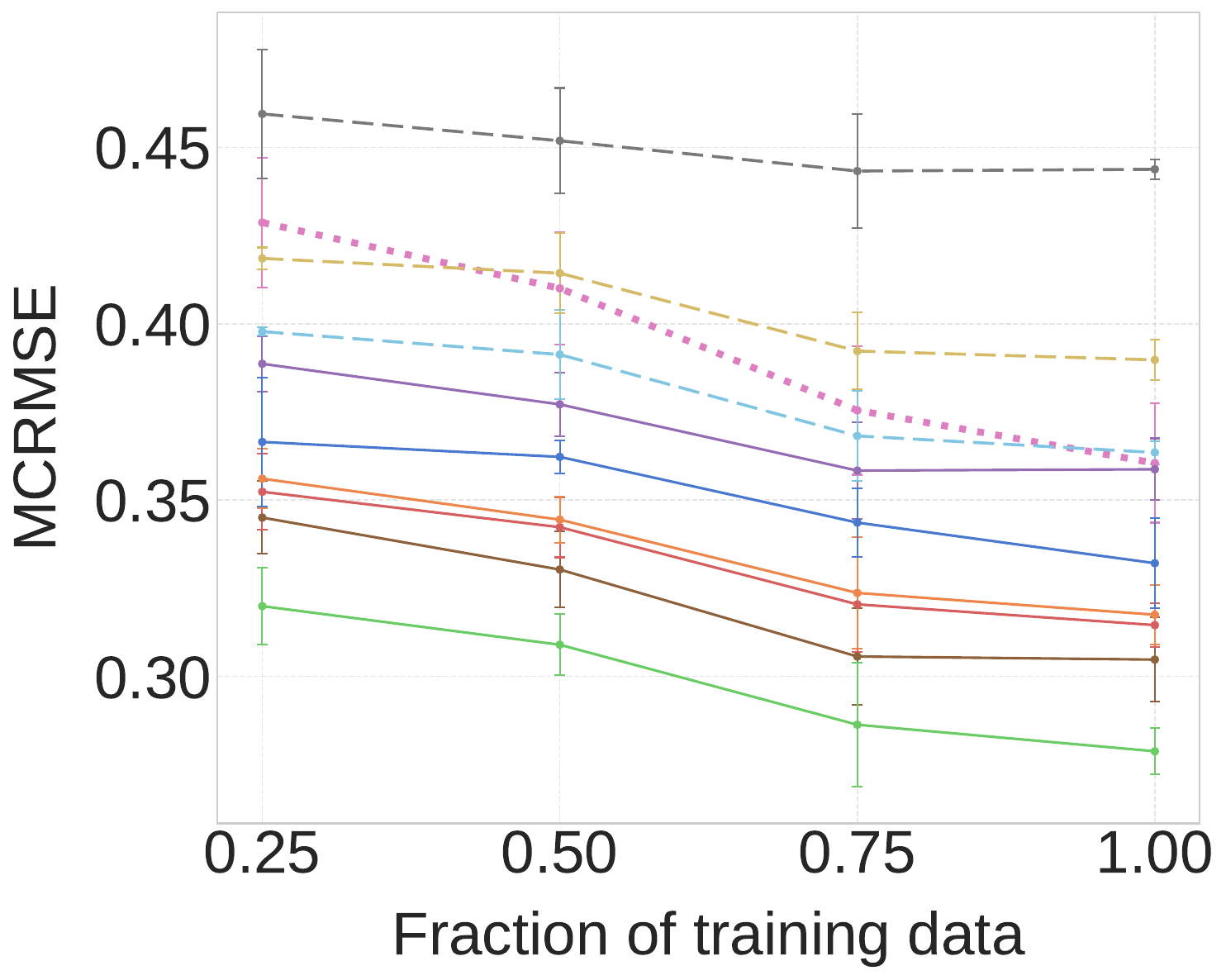}
   \vspace{-2em}
   \caption*{(a) COVID}
    \end{minipage}}
  \put(140,0){\begin{minipage}[b]{.31\textwidth}
   \includegraphics[width=\linewidth]{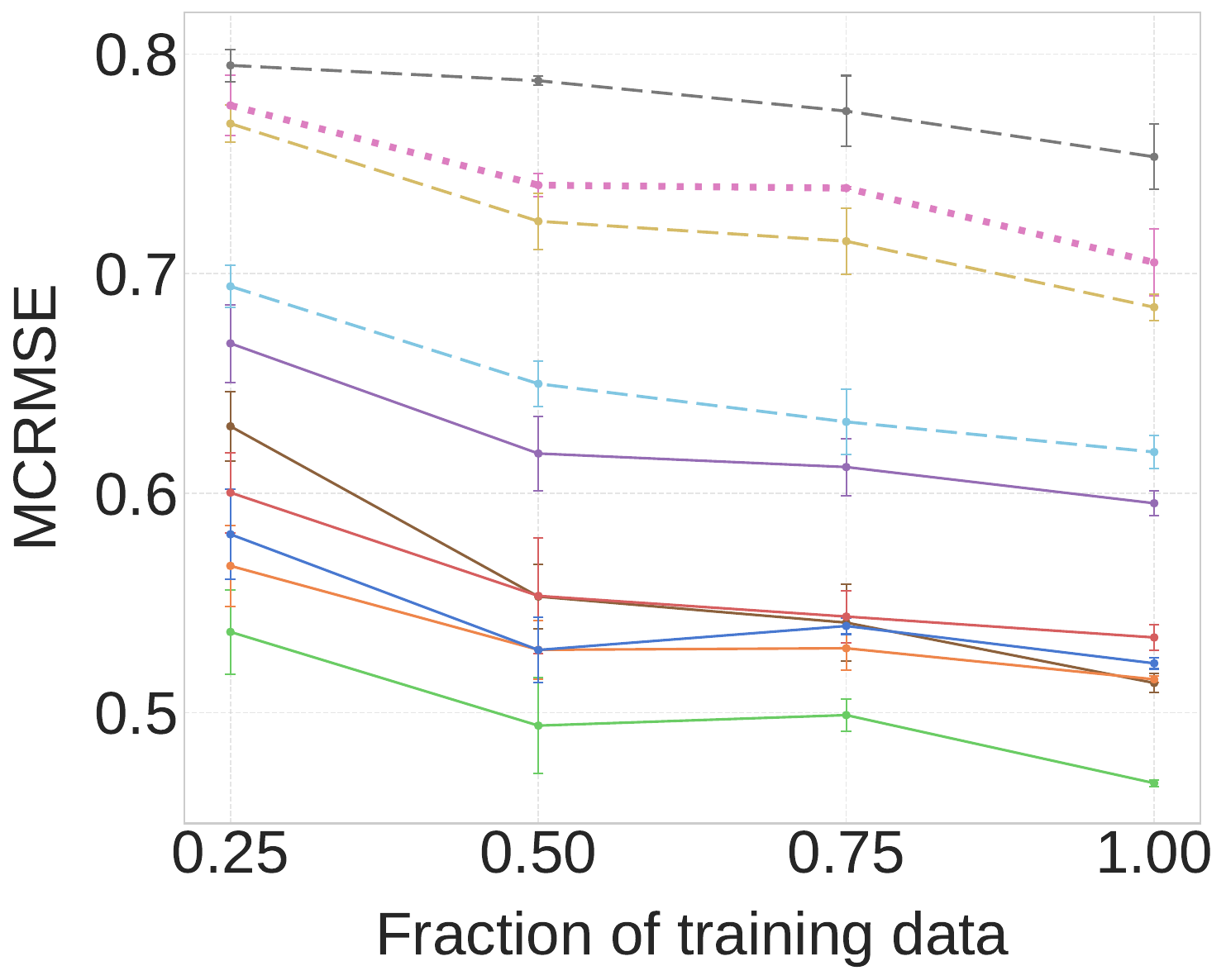}
   \vspace{-2em}
   \caption*{(b) Ribonanza}
    \end{minipage}}

  \put(280,0){\begin{minipage}[b]{.31\textwidth}
   \includegraphics[width=\linewidth]{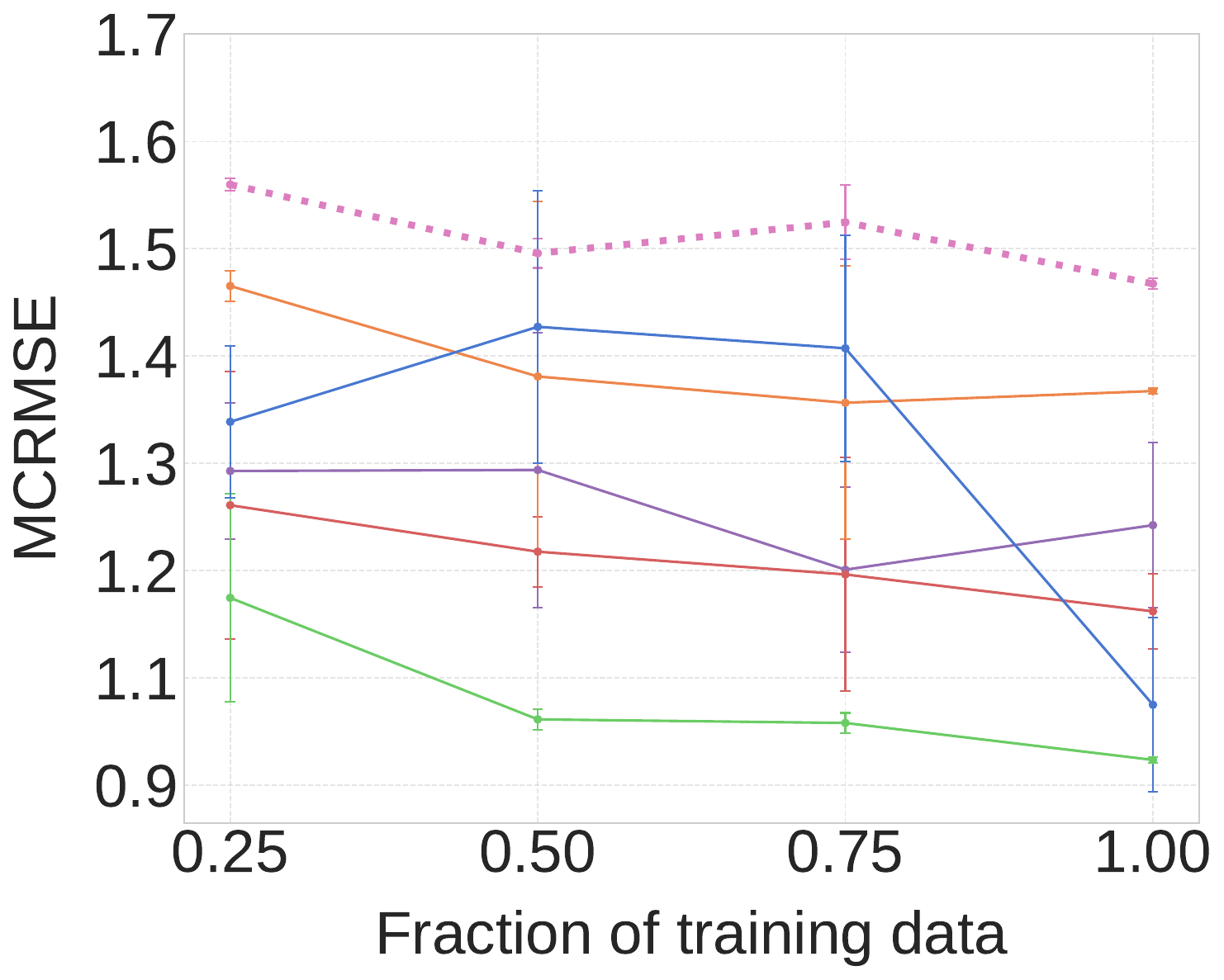}
   \vspace{-2em}
   \caption*{(c) Fungal}
    \end{minipage}}
  
\end{picture}
\vspace{-0.75em}
\caption{\textbf{Performance vs. fraction of training data across various datasets.} Model performance improves with increasing data, with lower MCRMSE across all models. 2D models consistently outperform 1D models, particularly in low-data regimes, underscoring the value of structural information for generalization. Dotted, solid, and dashed lines denote 1D, 2D, and 3D methods, respectively, which applies consistently throughout all figures in this paper.}
\label{fig:train_fraction}
\end{figure}

\vspace{0.1em}
\begin{figure}[ht]
\begin{picture}(0,110)
  \put(0,0){\begin{minipage}[b]{.33\textwidth}
   \includegraphics[width=\linewidth]{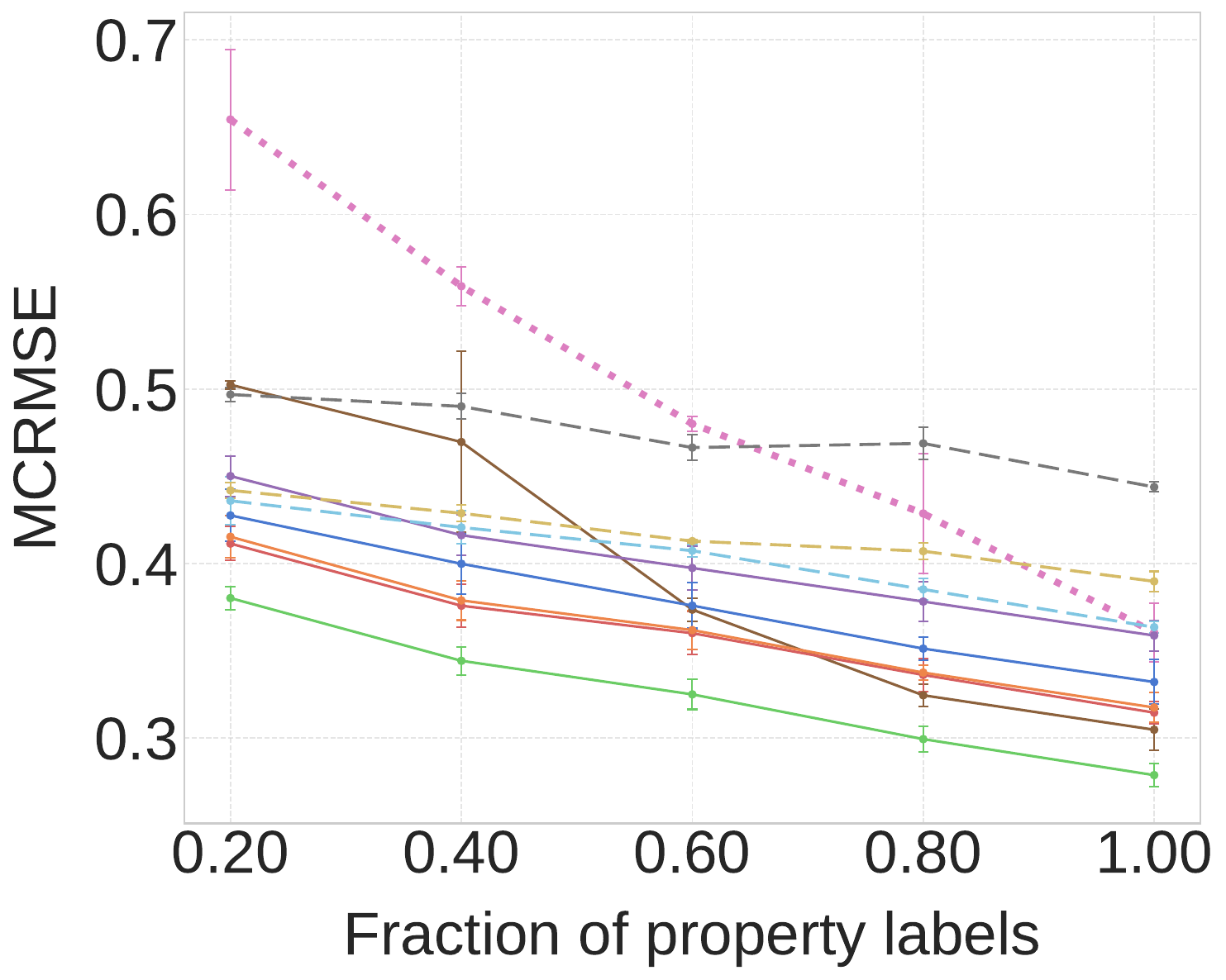}
   \vspace{-2em}
   \caption*{(a) COVID}
    \end{minipage}}
  \put(140,0){\begin{minipage}[b]{.33\textwidth}
   \includegraphics[width=\linewidth]{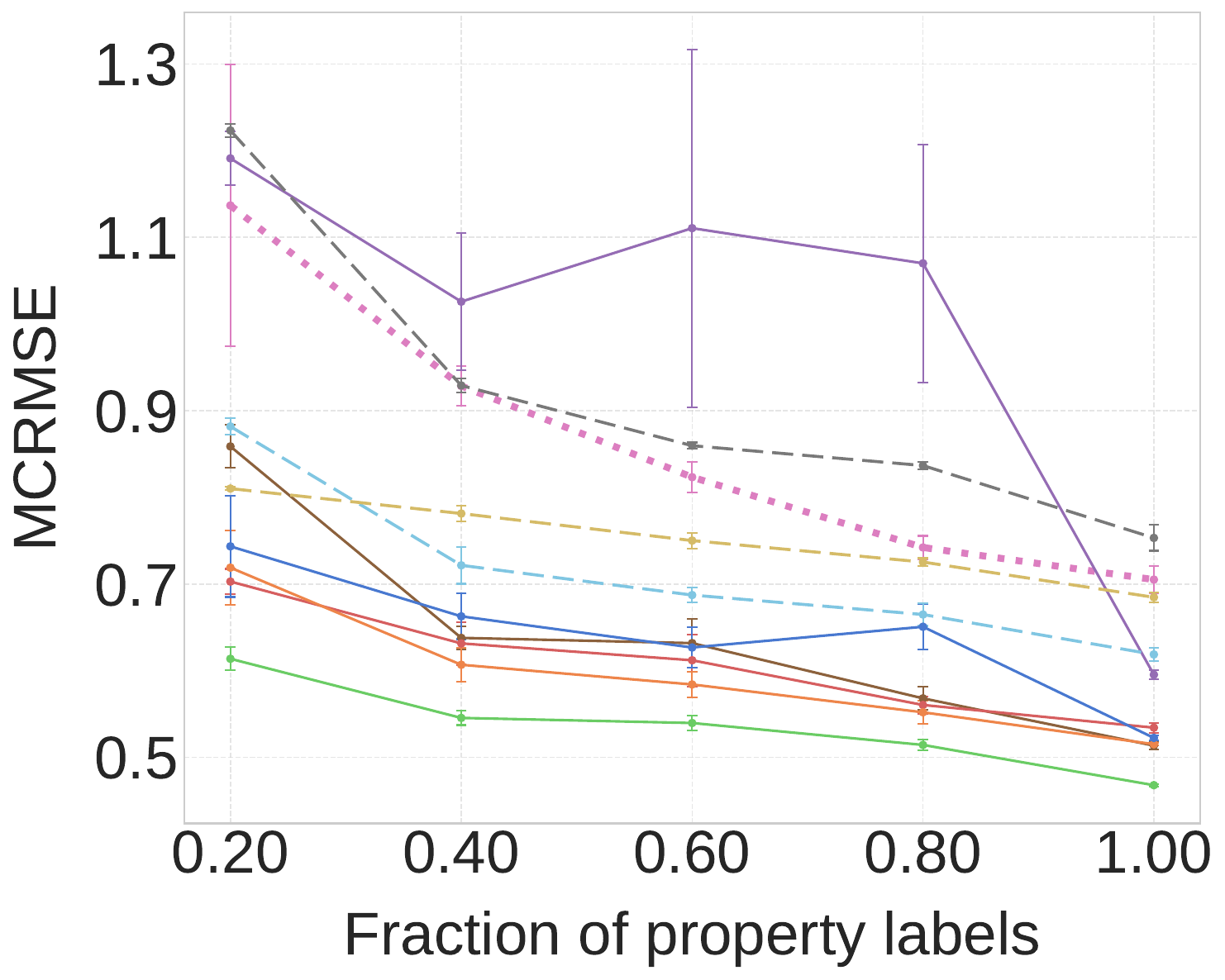}
   \vspace{-2em}
   \caption*{(b) Ribonanza}
    \end{minipage}}
  \put(278,17){\begin{minipage}[b]{0.23\textwidth}
   \includegraphics[width=\linewidth]{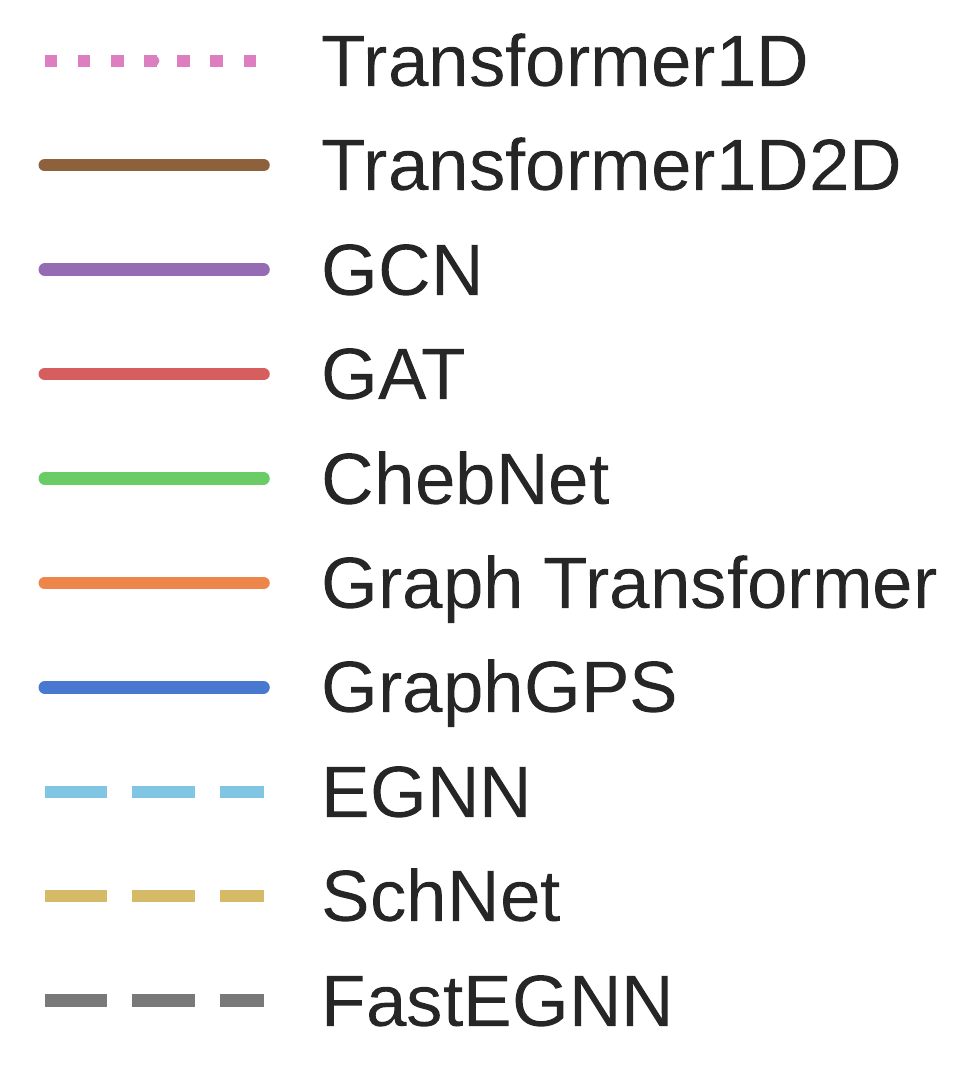}
    \end{minipage}}
\end{picture}
\vspace{-0.6em}
\caption{\textbf{Performance vs. partial property labels on COVID and Ribonanza datasets.} 2D models consistently outperform 1D models with sparse labeling, while Transformer1D and Transformer1D2D improve rapidly with denser supervision, emphasizing the need for more labels in transformer-based models.}
\vspace{-0.5em}
\label{fig:label-ratio-efficiency}
\end{figure}

\paragraph{More training data improves performance} Unsurprisingly, across all models and datasets, a clear trend emerges: increasing the amount of available data, whether through higher number of training data points or greater proportion of available labels leads to improved performance (Fig.~\ref{fig:train_fraction}, Fig.~\ref{fig:label-ratio-efficiency}, Appendix Tables in Sec.~\ref{app_sec:data_availability},~\ref{app_sec:partial_labeling}). However, the degree of performance improvement varies significantly between model types, as analyzed next. 

\paragraph{2D models excel in low data and partial label regimes} Evaluating model performance at different training and label ratios reveals a notable trend: 2D models consistently outperform 1D and 3D models under low data and incomplete labeling regimes. Between 20-50\% training data and label levels, 2D models such as ChebNet, Transformer1D2D, GraphGPS, and GAT significantly outperform Transformer1D, highlighting the role of additional structural information for model sample efficiency. Interestingly, Transformer1D and Transformer1D2D exhibit a faster rate of improvement when more labels are available (Fig.~\ref{fig:label-ratio-efficiency}), suggesting that transformer-based architectures benefit from denser supervision. Notably, Transformer1D requires $2-5\times$ more training data/labels to match the performance of the least effective 2D models, which achieve comparable results using only 20\%\ to 50\%\ of the training data needed by Transformer1D when trained on the full dataset.

\paragraph{3D models outperform 1D model in limited data regime despite structural noise} For the medium-scale datasets (COVID and Ribonanza), where 3D models can be evaluated, we observe that the 3D models generally outperform or are on par with the Transformer1D, even for lower data and labeling regimes. This suggests that despite the noise introduced by inaccuracies in 3D structure predictions, the explicit geometric encoding in 3D models still provides an advantage over 1D models. EGNN, in particular, is consistently better than or on par with Transformer1D across all training and label ratios. This further emphasizes that models incorporating explicit geometric encoding (whether 2D or 3D) are more data-efficient than those relying solely on sequence information. However, it is important to note that 3D models do not match the efficiency of 2D models in these scenarios, likely due to their susceptibility to noise as discussed in Section~\ref{limitations_3d}.

\begin{figure}[ht]
\centering
\renewcommand{\arraystretch}{0.1} 
\begin{tabular}{>{\centering\arraybackslash}m{0.2cm} >{\centering\arraybackslash}m{3cm} >{\centering\arraybackslash}m{3cm} >{\centering\arraybackslash}m{3cm} >{\centering\arraybackslash}m{3cm}}
    & \multicolumn{4}{c}{\textbf{Noise Ratio}}   \\
    &  \textbf{0} & \textbf{0.1} & \textbf{0.2} & \textbf{0.3} \\
    \textbf{1D }
    & \vspace{0pt}\includegraphics[width=0.21\textwidth]{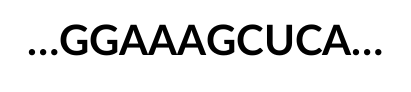} 
    & \vspace{0pt}\includegraphics[width=0.21\textwidth]{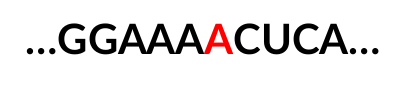} 
    & \vspace{0pt}\includegraphics[width=0.21\textwidth]{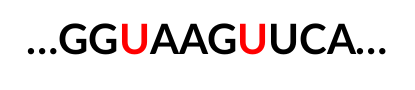} 
    & \vspace{0pt}\includegraphics[width=0.21\textwidth]{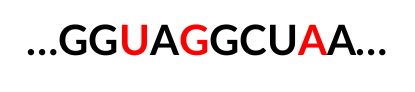} \\ 
    \textbf{2D}
    & \vspace{0pt}\includegraphics[width=0.15\textwidth]{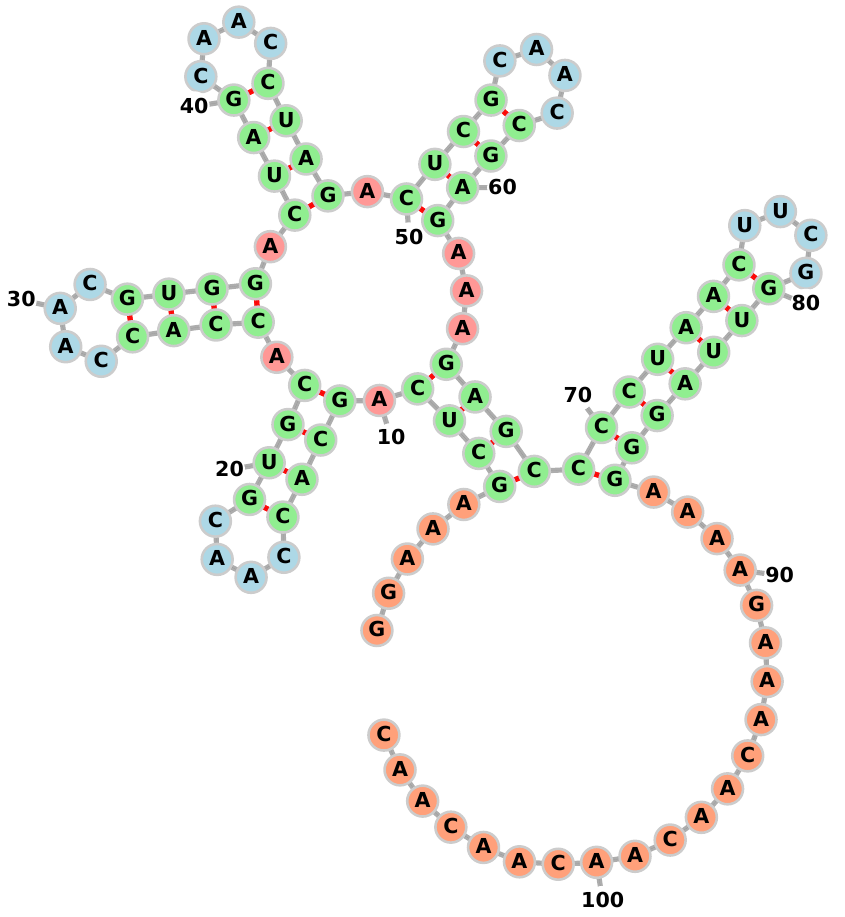} 
    & \vspace{0pt}\includegraphics[width=0.165\textwidth]{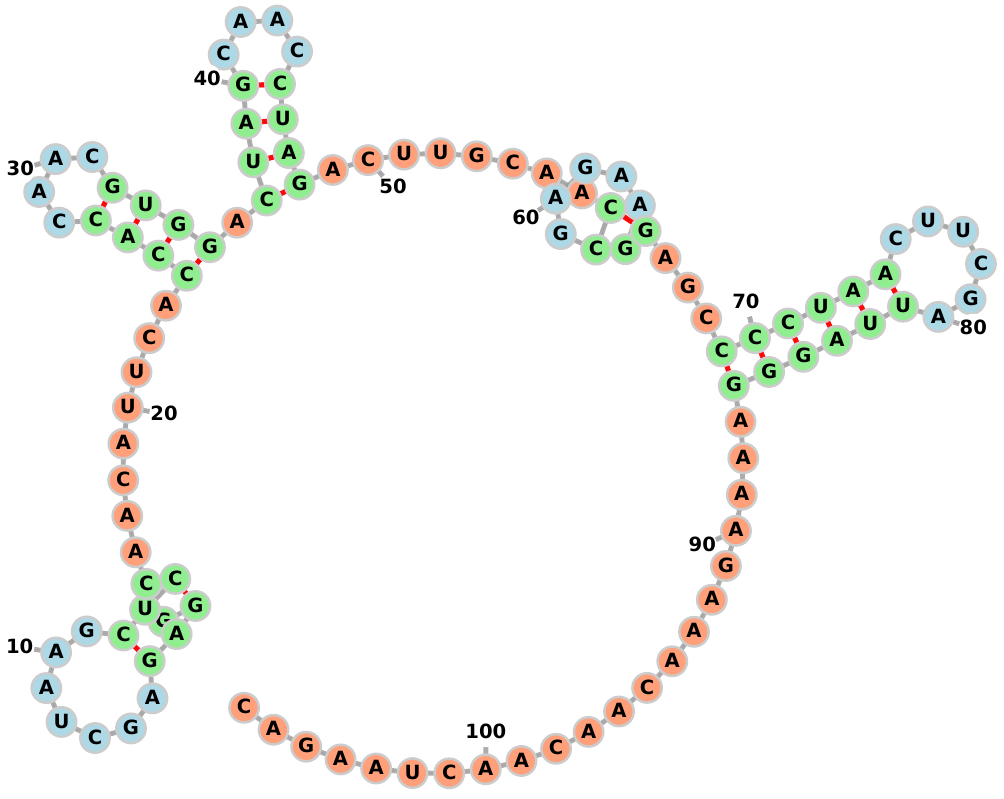} 
    & \vspace{0pt}\includegraphics[width=0.165\textwidth]{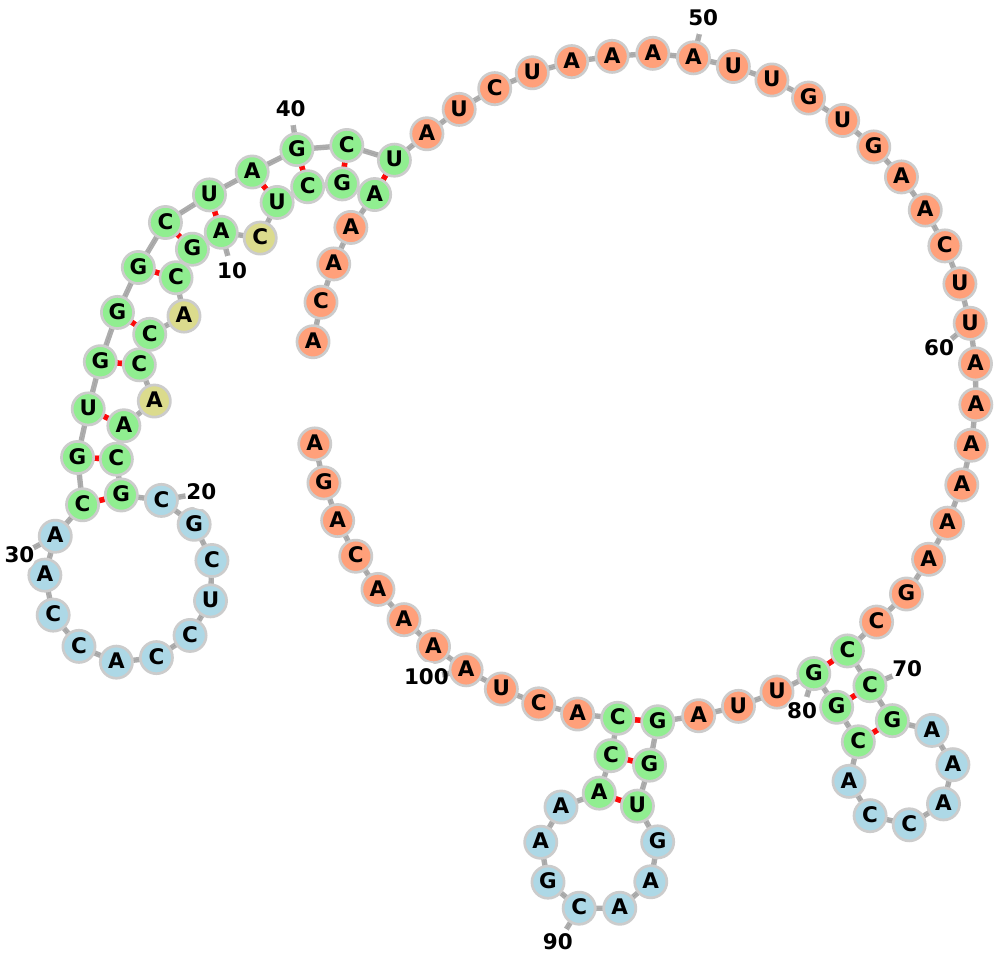} 
    & \vspace{0pt}\includegraphics[width=0.13\textwidth]{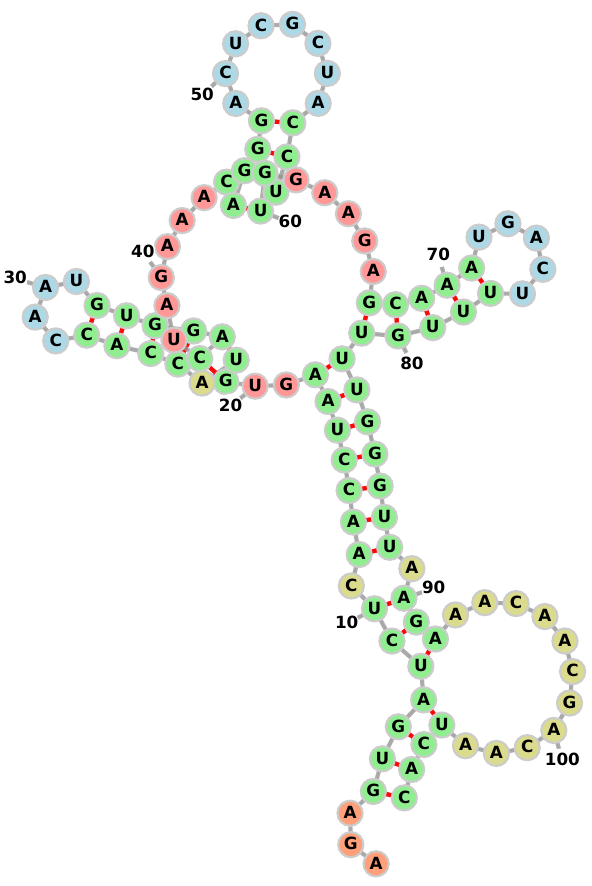} \\  [-1em]
    \textbf{3D}
    & \vspace{0pt}\includegraphics[width=0.155\textwidth]{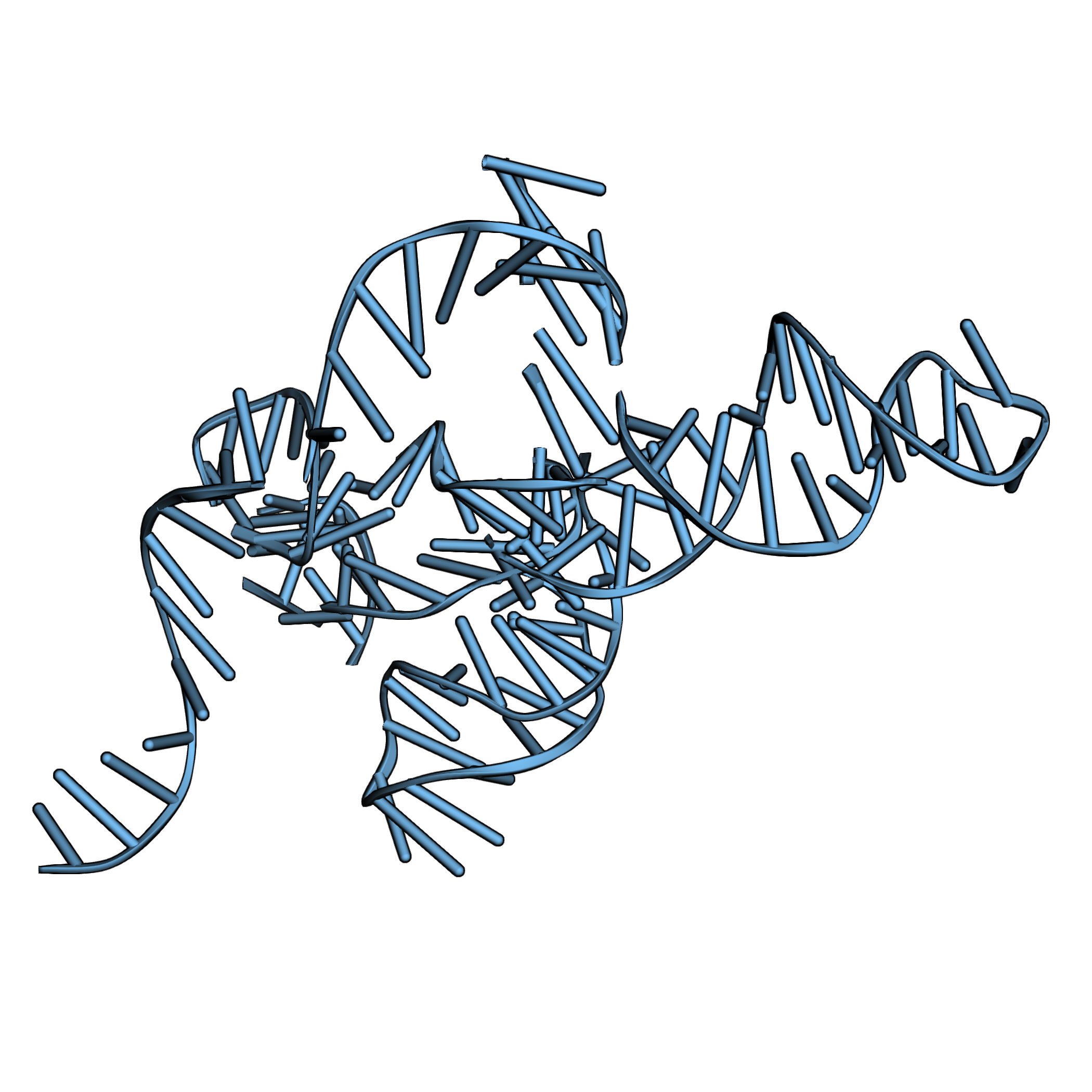} 
    & \vspace{0pt}\includegraphics[width=0.18\textwidth]{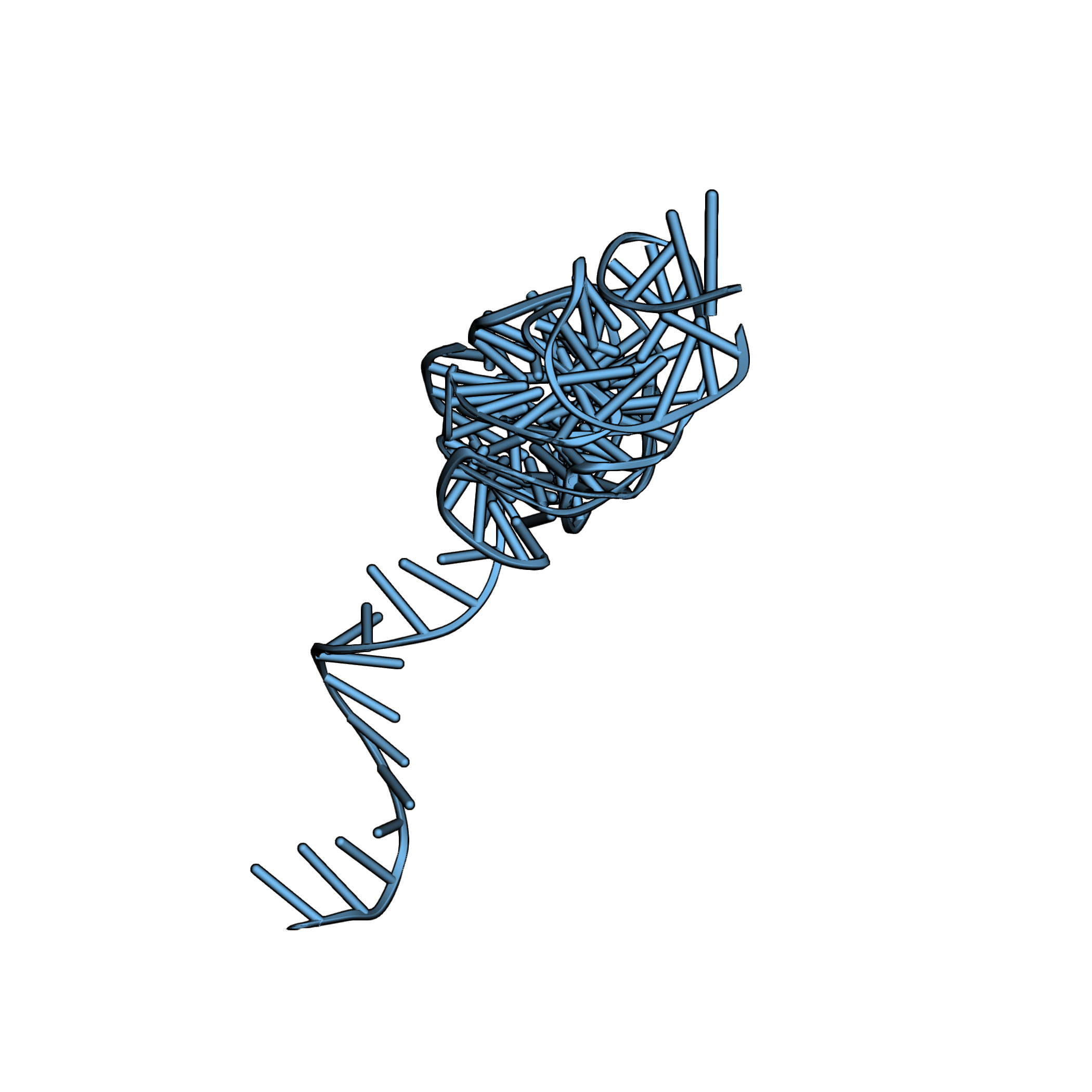} 
    & \vspace{0pt}\includegraphics[width=0.18\textwidth]{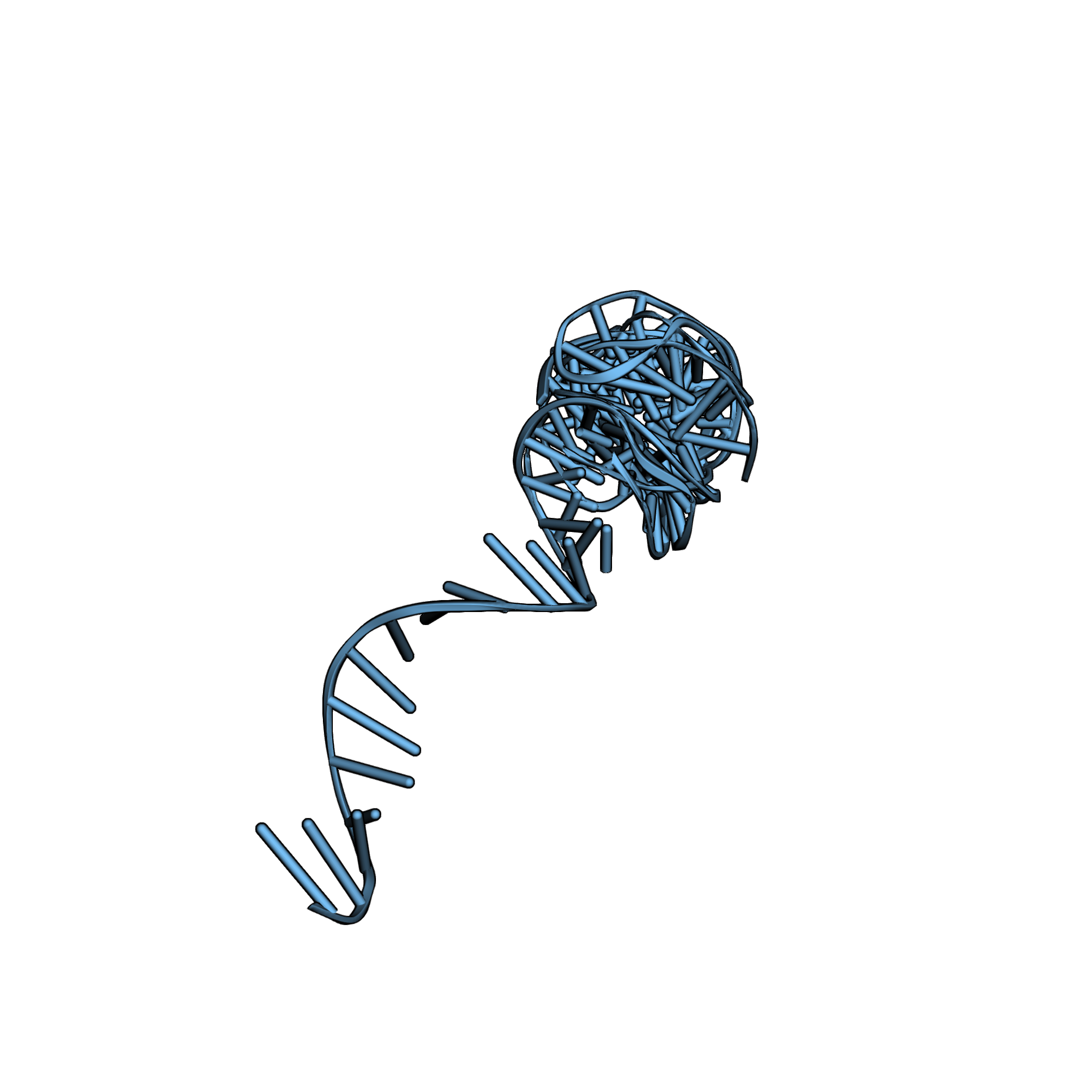} 
    & \vspace{0pt}\includegraphics[width=0.1\textwidth]{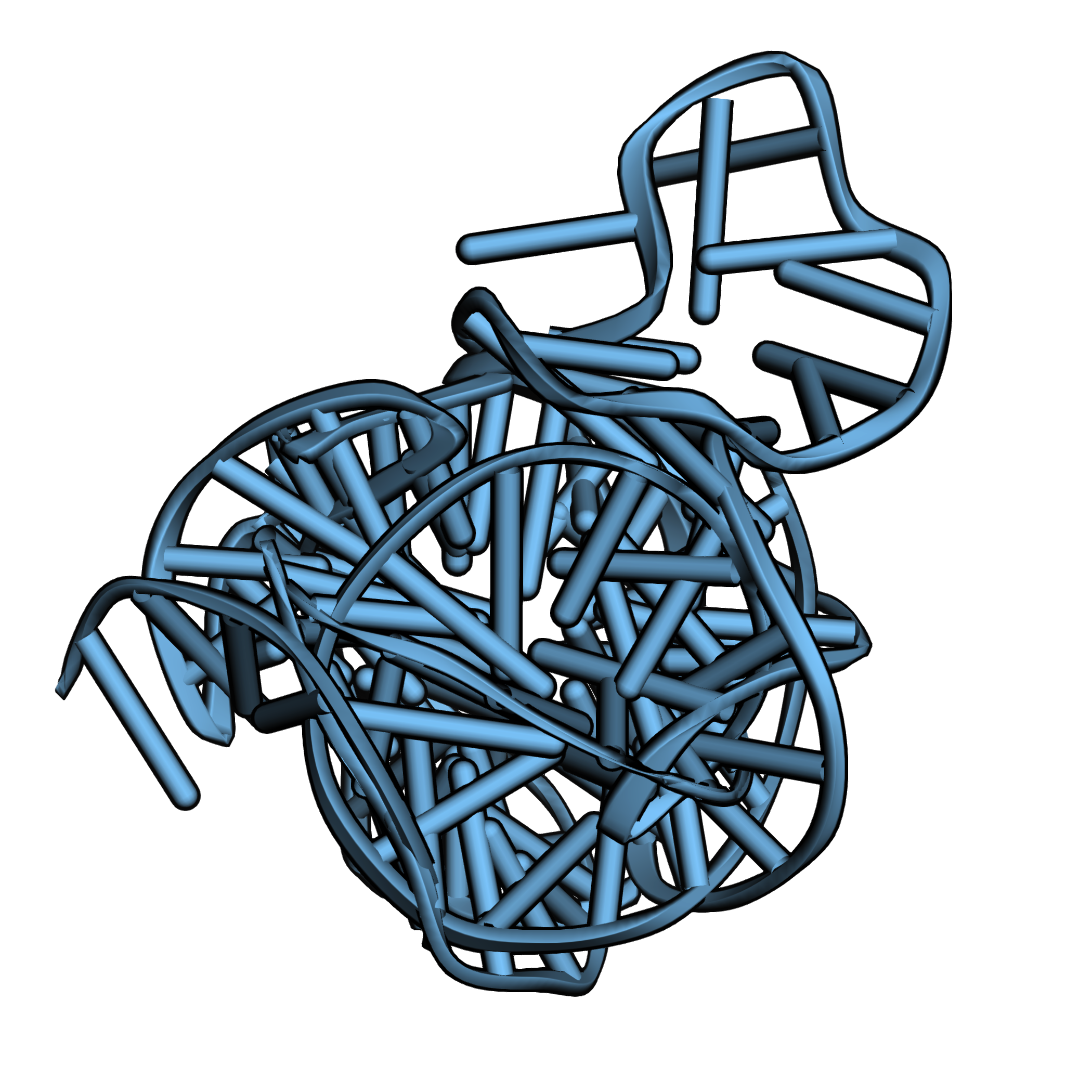}
\end{tabular}
\vspace{-0.5em}
\caption{\textbf{Visualization of 1D, 2D, and 3D structures under varying noise ratios (mutation errors during sequencing).} Each column represents a different noise ratio, showcasing the impact of noise on the structures across different dimensions.}
\vspace{-1.5em}
\label{fig:rna_noise}
\end{figure}

\subsection{Model Robustness and Generalization Under Sequencing Noise} 
Tasks 4 and 5 both deal with model performance with noise in the data, but focus on different aspects, robustness to noise, and ability to generalize across unseen noise distributions. As explained in Sec.~\ref{sec:tasks}, sequencing noise is common depending on the sequencing method and platform used~\citep{fox2014accuracy}, thereby introducing errors in sequences that propagate into 2D and 3D structures. Additionally one of the deployment scenarios involves models trained on high-quality clean data applied for datasets acquired under noisy conditions owing to different sequencing platforms or experimental batch effects~\citep{tom2017identifying}. 

To explore these practical aspects, we design two sets of experiments:

\begin{itemize}[left=0pt]
    \item \textit{Robustness:} We introduce sequencing noise into the training, validation, and test sequences to simulate realistic sequencing errors. Nucleotide mutations are applied with probabilities \{0.05, 0.1, 0.15, 0.2, 0.25, 0.3\}, mirroring typical sequencing error rates~\citep{pfeiffer2018systematic} which also propagates to the 2D and 3D structures (Fig.~\ref{fig:rna_noise}) via structure prediction tools. Importantly, these mutations are not random; the likelihood of a particular nucleotide mutating into another varies, as is well documented in sequencing studies~\cite {pfeiffer2018systematic}. Our noise model reflects these real-world mutation profiles. Crucially, while the input training, validation, and test sequences contain noise, the property labels remain clean. This again reflects practical scenarios where labels are experimentally determined independent of sequencing and thus unaffected by sequencing errors.
    
    \item \textit{Generalization:} Here, models are trained on clean, noise-free data corresponding to high-quality sequencing experiments but are tested on datasets with varying levels of noise simulated by sequencing mutation probabilities in \{0.05, 0.1, 0.15, 0.2, 0.25, 0.3\}. This setup reflects the real-world scenario where models trained on high-quality data will be deployed on OOD data that may come from different sequencers or have been affected by batch effects.
\end{itemize}

\begin{figure}[ht]
\begin{picture}(0,225)
  \put(0,115){\begin{minipage}[b]{.33\textwidth}
   \includegraphics[width=\linewidth]{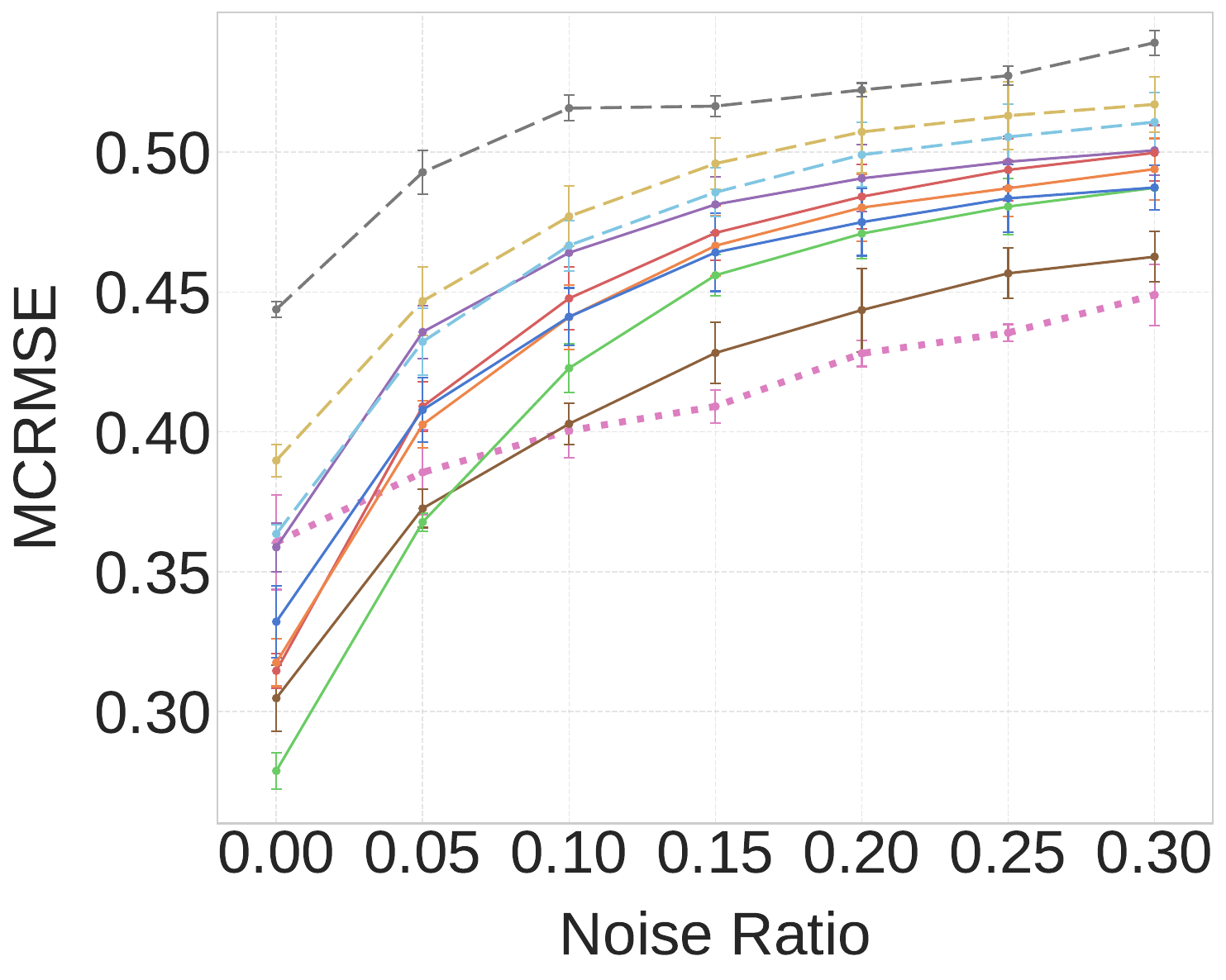}
   \vspace{-2em}
   \caption*{(a) COVID}
    \end{minipage}}
  \put(140,115){\begin{minipage}[b]{.33\textwidth}
   \includegraphics[width=\linewidth]{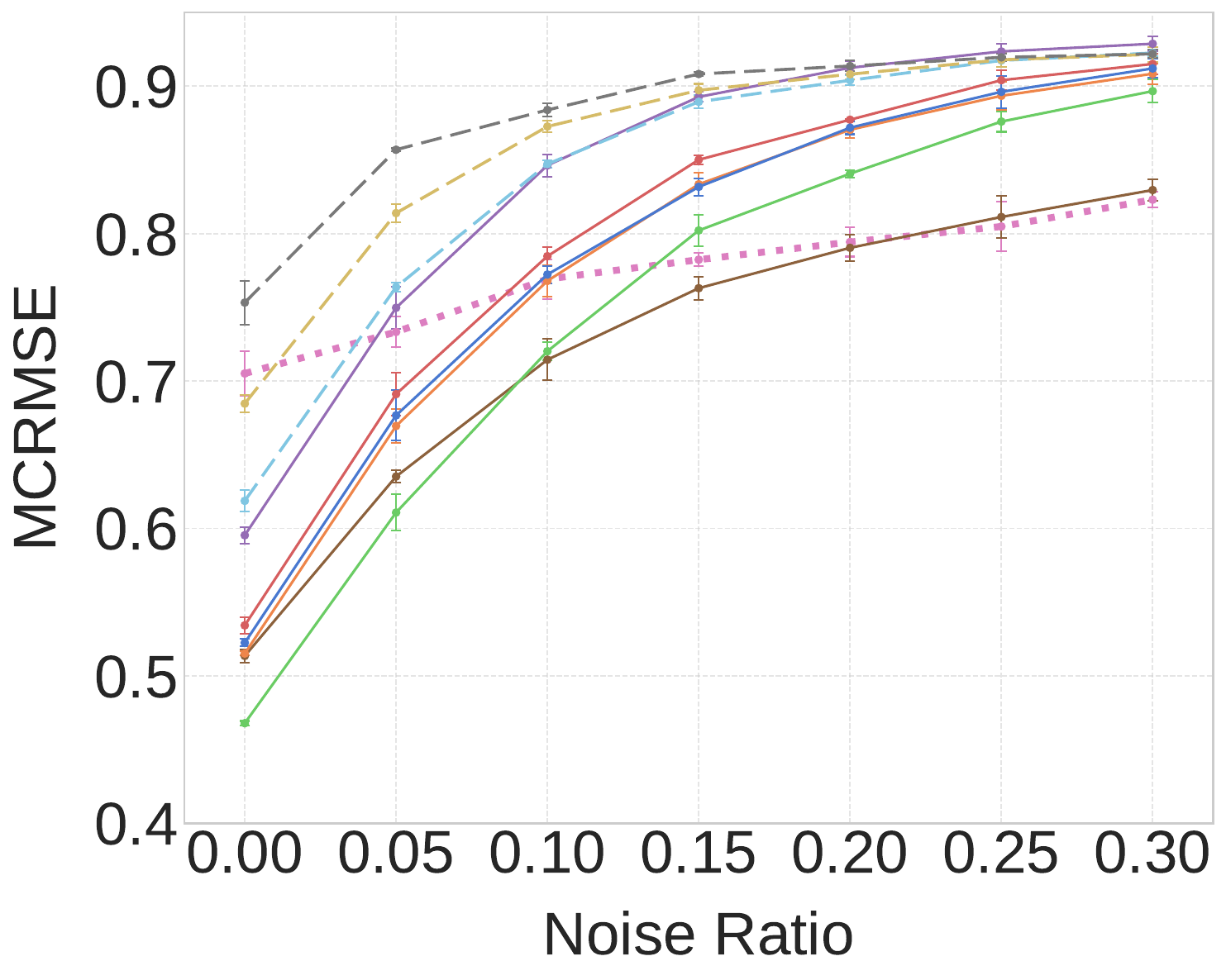}
   \vspace{-2em}
   \caption*{(b) Ribonanza}
    \end{minipage}}

  \put(275,45){\begin{minipage}[b]{.3\textwidth}
   \includegraphics[width=\linewidth]{figures/legend.pdf}
    \end{minipage}}

  \put(0, 0){\begin{minipage}[b]{.33\textwidth}
   \includegraphics[width=\linewidth]{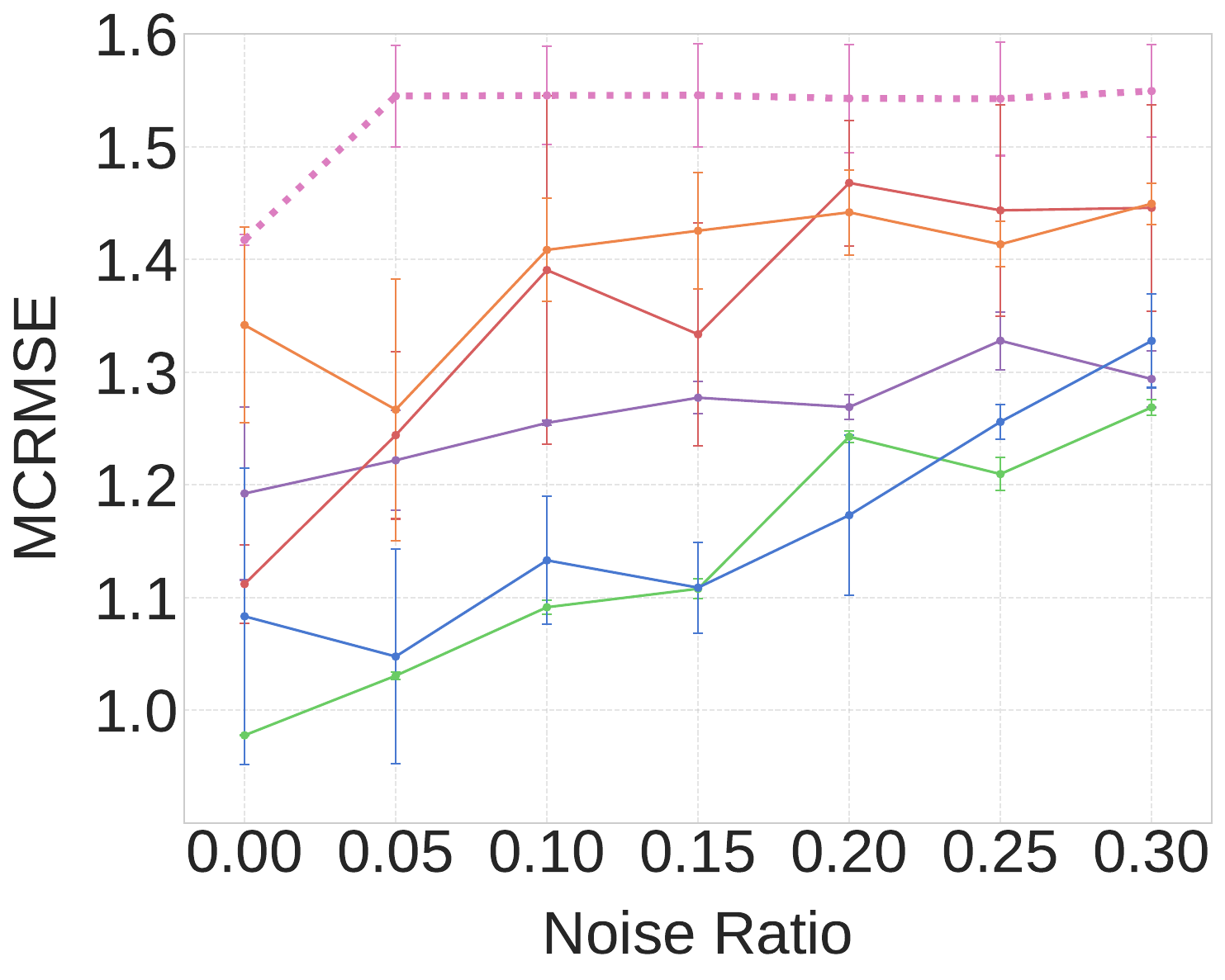}
   \vspace{-2em}
   \caption*{(c) Fungal}
    \end{minipage}}
  \put(140, 0){\begin{minipage}[b]{.33\textwidth}
   \includegraphics[width=\linewidth]{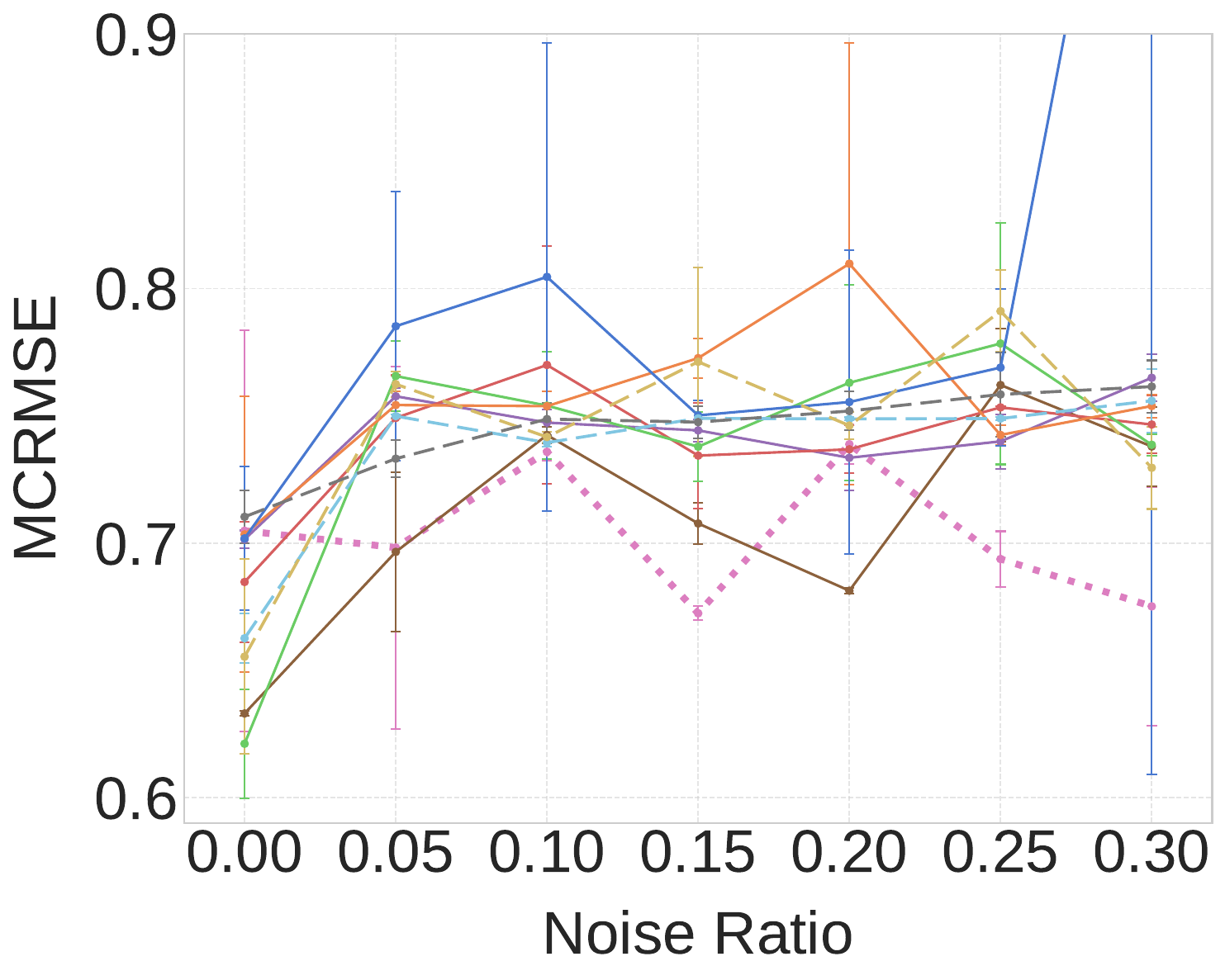}
   \vspace{-2em}
   \caption*{(d) Tc-Riboswitches}
    \end{minipage}}
\end{picture}
\vspace{-0.5em}
\caption{\textbf{Robustness experiments.} Transformer1D shows the least performance drop under increasing noise, maintaining the highest accuracy, with Transformer1D2D following closely. In contrast, 2D and 3D models, particularly ChebNet and 3D models, are more impacted by noise.}
\label{fig:robustness}
\end{figure}

\begin{figure}[ht]
\begin{picture}(0,225)
  \put(0,120){\begin{minipage}[b]{.33\textwidth}
   \includegraphics[width=\linewidth]{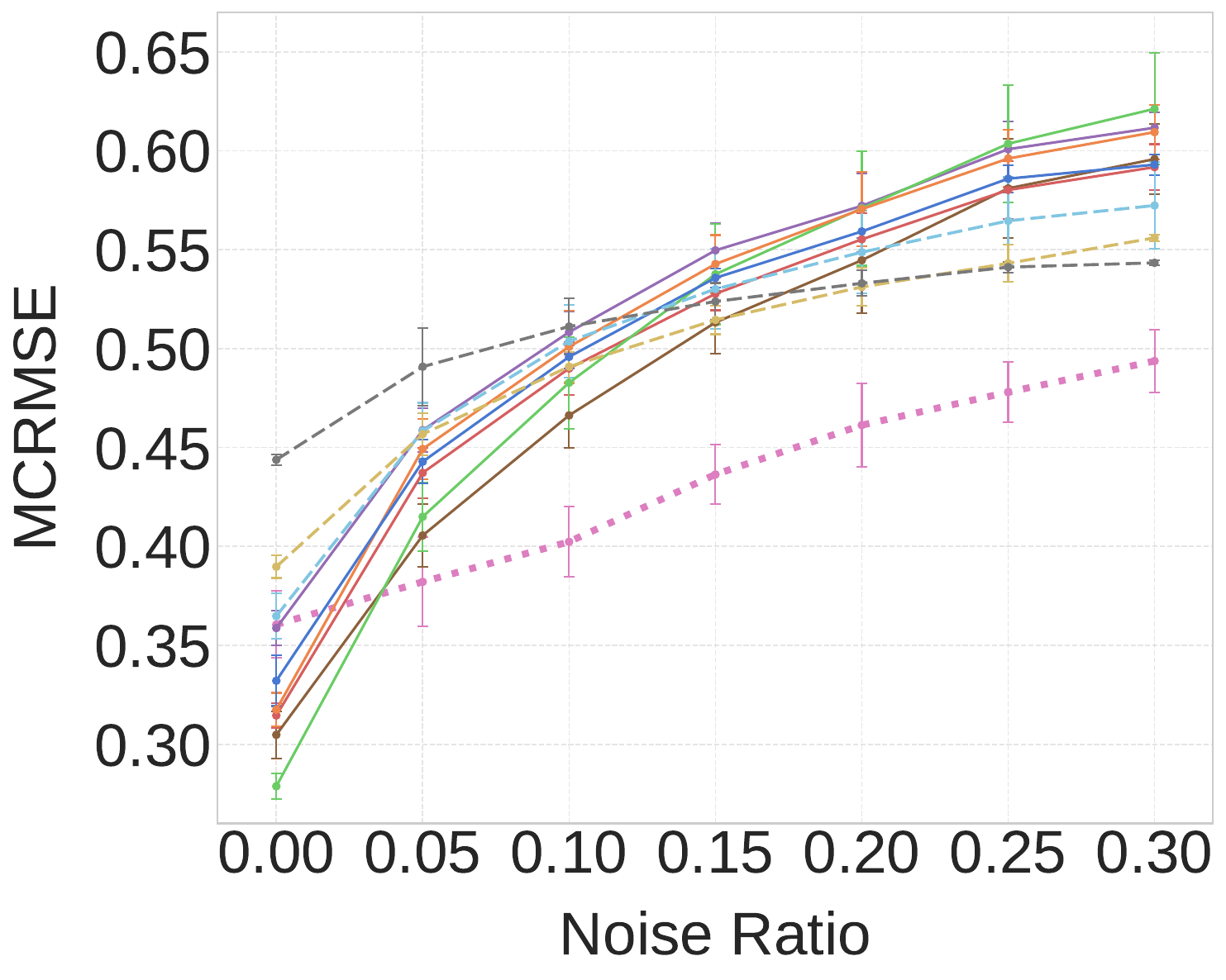}
   \vspace{-2em}
   \caption*{(a) COVID}
    \end{minipage}}
  \put(140,120){\begin{minipage}[b]{.33\textwidth}
   \includegraphics[width=\linewidth]{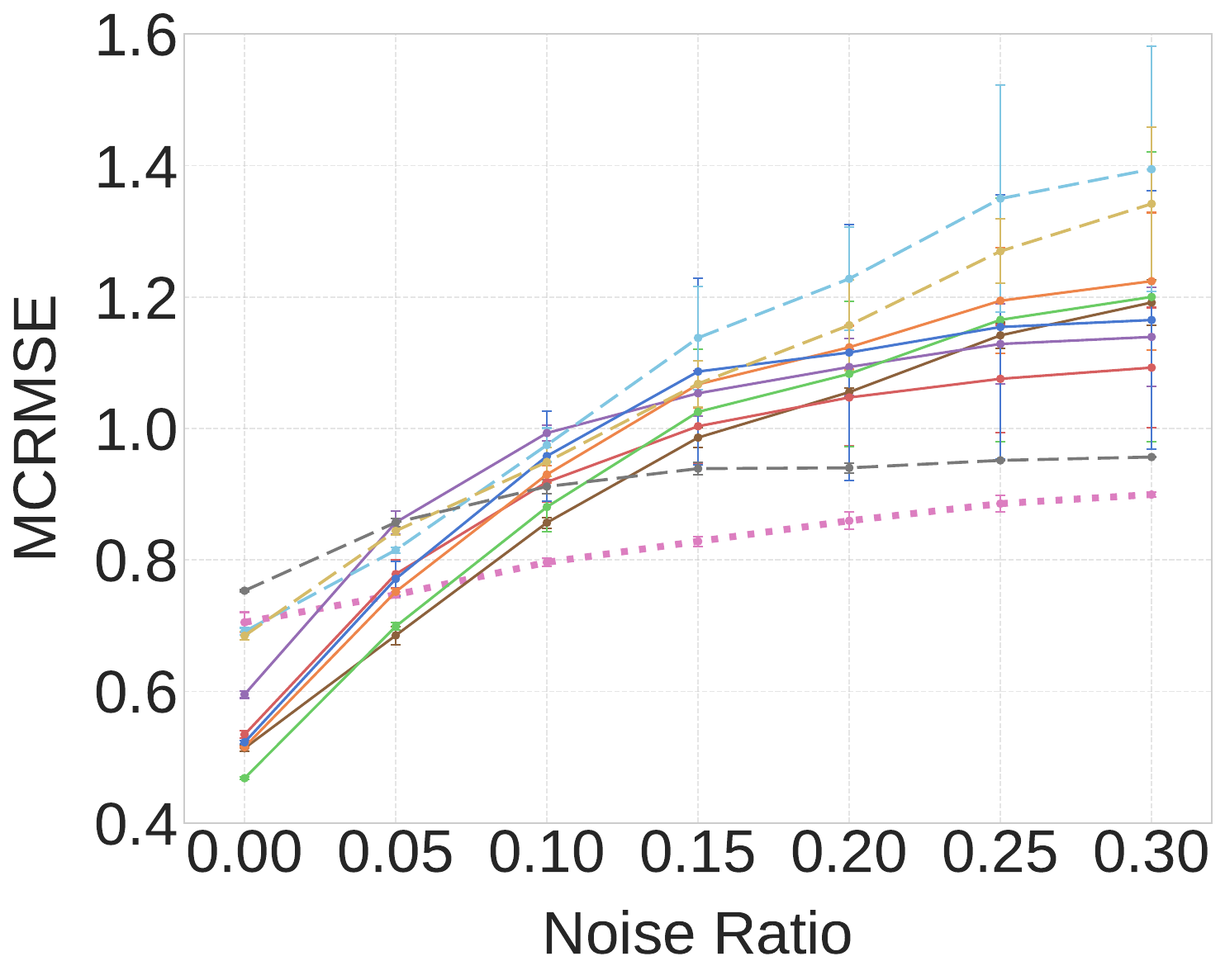}
   \vspace{-2em}
   \caption*{(b) Ribonanza}
    \end{minipage}}

  \put(275,45){\begin{minipage}[b]{.3\textwidth}
   \includegraphics[width=\linewidth]{figures/legend.pdf}
    \end{minipage}}

  \put(0, 0){\begin{minipage}[b]{.33\textwidth}
   \includegraphics[width=\linewidth]{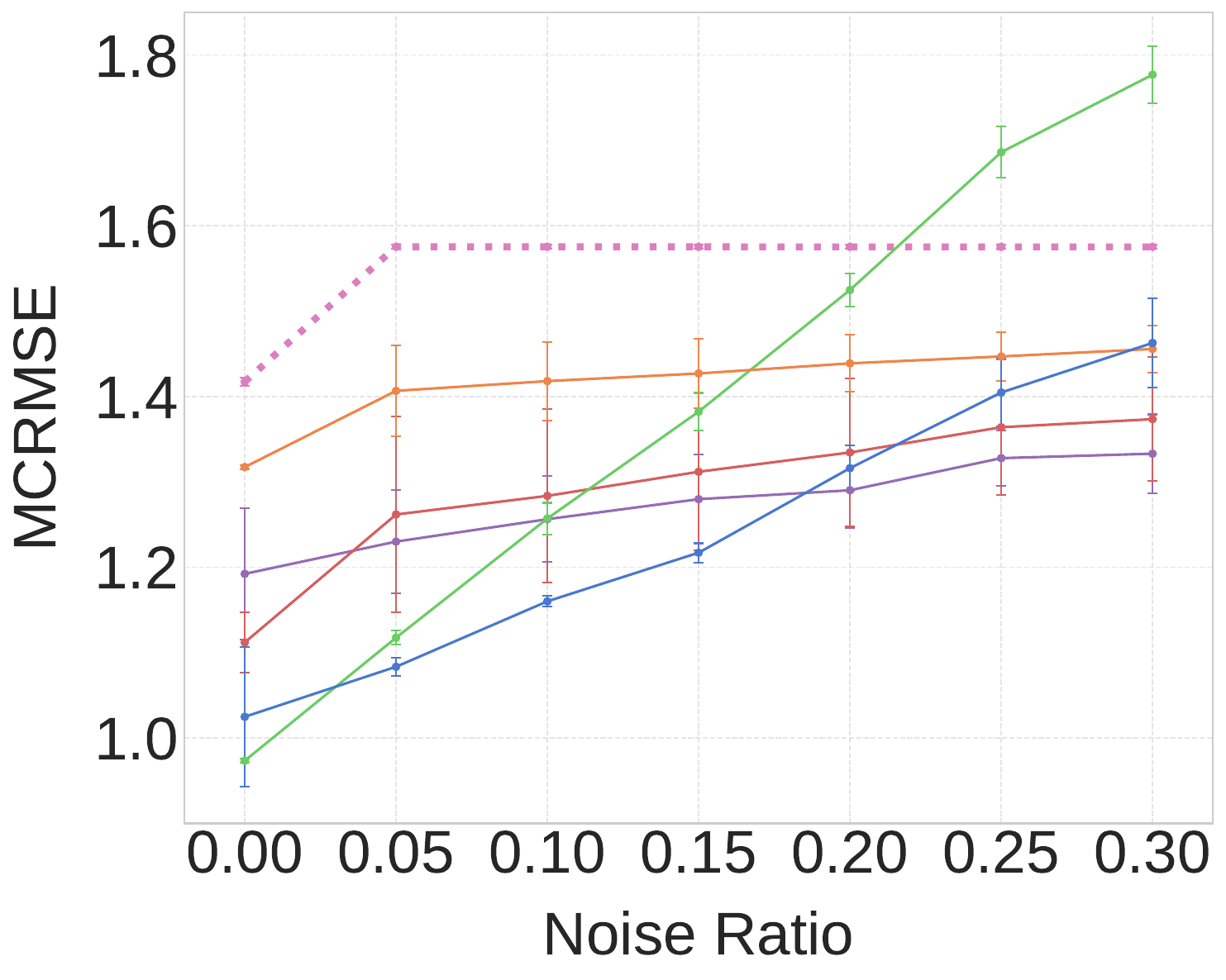}
   \vspace{-2em}
   \caption*{(c) Fungal}
    \end{minipage}}
  \put(140, 0){\begin{minipage}[b]{.33\textwidth}
   \includegraphics[width=\linewidth]{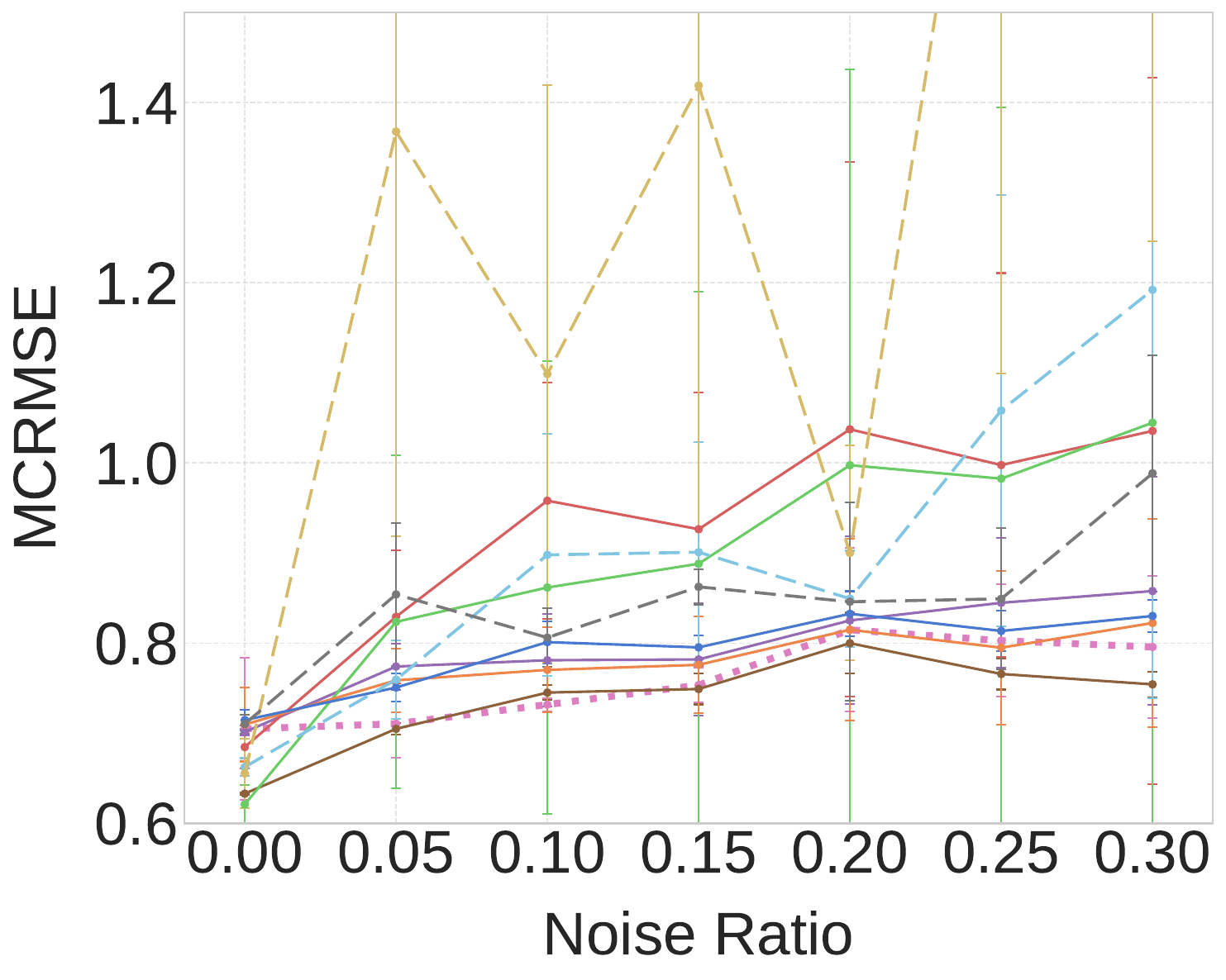}
   \vspace{-2em}
   \caption*{(d) Tc-Riboswitches}
    \end{minipage}}
\end{picture}
\vspace{-0.5em}
\caption{\textbf{Generalization experiments.} Transformer1D outperforms other models on noisy sequences, achieving the lowest RMSE at higher noise levels, particularly on COVID, Ribonanza, and Tc-Riboswitches. Transformer1D2D follows closely, showing that transformer-based models generalize better under noise than 2D and 3D models, especially in tasks with geometric representations.}
\vspace{-0.5em}
\label{fig:generalization}
\end{figure}

\paragraph{Transformer architectures demonstrate superior robustness and generalization under sequencing noise} 
Expectedly, across both tasks, increased noise levels generally leads to worse test MCRMSE for all models, indicating a decline in prediction performance (Fig.~\ref{fig:robustness}, Fig.~\ref{fig:generalization} and Appendix Tables in Sec.~\ref{app_sec:robustness},~\ref{app_sec:generalization_ood}). 

Among all models, Transformer1D demonstrates the highest robustness and generalization, exhibiting the least performance degradation as noise levels increase. Notably, in generalization experiments, Transformer1D achieves the best prediction MCRMSE on the COVID, Ribonanza, and Tc-Riboswitches datasets under higher noise levels (Fig.~\ref{fig:generalization} and Appendix~\ref{app_sec:generalization_ood}). The reliance of Transformer1D on sequence-only information without considering geometric context, while being a weakness in other scenarios, becomes a strength in case of noisy sequences as minor sequencing errors may severely alter the predicted RNA structures (Fig.~\ref{fig:rna_noise}). While its performance on the Fungal dataset is worse overall, Transformer1D maintains remarkably consistent performance as noise increases. Transformer1D2D ranks just behind Transformer1D outperforming other 2D and 3D models. This can be attributed to Transformer1D2D's ability to selectively focus on sequence information for noisy data rather than relying on structural data alone as the self-attention is only weakly conditioned on the graph topology.

More elaborated 2D and 3D models, which rigidly rely on structural information, are significantly more affected by noise, underperforming compared to plain sequence baseline in both robustness and generalization experiments for high noise levels. For low-to-moderate noise levels (5-10\% noise), 2D methods such as ChebNet still perform the best. Interestingly, ChebNet shows the worst generalization among 2D models, although it performs on par with other methods in robustness experiments. This suggests that while ChebNet struggles with OOD noise, its performance gets better when it is also trained on the same noise level used during testing, highlighting the need for retraining for different experimental data batches/noise levels in real-world applications. Across both experiments, 3D models have poor performance for more noisy conditions, particularly with the COVID and Ribonanza datasets, due to their dependence on 3D structures which are also sensitive to the propagation of sequencing errors (Fig.\ref{fig:rna_noise} bottom). In the smaller Tc-Riboswitches dataset, model performances vary more, likely due to limited data size, but transformer models still consistently demonstrate greater robustness to noise. 

Across both settings, as the noise ratio rises, the 1D model demonstrates the greatest resilience to noise, showing an average test MCRMSE increase of approximately 14\% and 27\%, respectively, relative to the train and test on clean data. In contrast, 2D models exhibit the highest sensitivity, with test MCRMSE increasing from 30\% to 82\%. Meanwhile, 3D models show intermediate performance, with MCRMSE increases of 29\% and 56\%. Our results reveal a higher vulnerability of 2D and 3D models to sequencing noise where geometric context becomes unreliable at a faster rate than a plain sequence of nucleotides.

\section{Conclusion}
We present the first comprehensive study on the benefits and challenges of the effect of geometric context for RNA property prediction models. With providing a curated set of RNA datasets with annotated 2D and 3D structures, we systematically evaluate the performance of 1D, 2D, and 3D models under various real-world conditions, such as limited data, partial labeling, sequencing errors and out-of-distribution generalization. Our results reveal that 2D models outperform 1D and 3D models, with spectral graph neural networks excelling even in low-data and partial labeling scenarios. For 3D models, we find that their potential benefits are hindered by the limited receptive field, computational complexity, and structural noise from RNA structure prediction tools. At the same time, 1D models demonstrate better robustness compared to 2D and 3D models in noisy and OOD conditions. This study highlights the value and limitations of using geometric context for RNA modeling. Future work could focus on ensembling 1D, 2D, and 3D models for complementary strengths, and on improving 2D and 3D models to better handle noise from structure prediction tools as elaborated in Appendix~\ref{appen:sequence_noise}. Another promising direction can be to investigate advanced 3D model architectures which incorporate high-degree steerable features as discussed in Appendix~\ref{appen:steerbale_features}.

\newpage
\section*{Ethics statement}
This work aims to advance the computational prediction of RNA properties through the development and evaluation of machine learning models utilizing diverse representations of RNA, including 1D sequences, 2D structures, and 3D geometries. In conducting this research, we are committed to upholding high ethical standards in all aspects of our study, ensuring that our work promotes scientific integrity, transparency, and responsible use of technology.

\begin{itemize}[left=0pt]
\item \textbf{Data Integrity and Fair Use:} All RNA datasets used in this study were acquired from publicly available and ethically sourced repositories. Where applicable, the original data sources have been duly cited, and all efforts have been made to ensure data privacy and compliance with relevant data-sharing agreements. Our curated datasets have been handled responsibly, with proper annotations to minimize errors and misinterpretations.

\item \textbf{Minimization of Bias:} We recognize that RNA data, particularly when it involves limited or incomplete datasets, can introduce biases in model predictions. To mitigate this, we employ a diverse set of RNA data and systematically analyze model performance under various conditions, including label scarcity and sequencing errors. By doing so, we aim to provide balanced insights into the potential advantages and limitations of different modeling approaches, fostering responsible model selection and deployment.

\item \textbf{Responsible Use of AI in Biology:} The machine learning techniques used in this study are intended to assist in scientific discovery and biological understanding, with potential applications in therapeutic interventions. However, we acknowledge the importance of caution in applying computational models in sensitive domains such as healthcare. While our models are designed to improve RNA property prediction, we emphasize that these predictions should not be used in isolation for clinical or therapeutic decision-making without further validation and consideration of ethical implications.  
\end{itemize}

\section*{Reproducibility Statement}
We are committed to the transparency and reproducibility of our findings. All methodologies, datasets, and benchmarking environments will be made publicly available to the research community, allowing others to reproduce, verify, and extend our work. This open approach aims to advance the field of RNA research and promote collaborative progress.
To ensure reproducibility, we also provide detailed descriptions of model architecture and training procedures in the main text. The appendix~\ref{appen:reproduction} contains full details on data preprocessing, hyperparameter searching and additional experimental information. All data and code will be provided to upon acceptance.

\bibliography{ref}

\begin{thebibliography}{89}
\providecommand{\natexlab}[1]{#1}
\providecommand{\url}[1]{\texttt{#1}}
\expandafter\ifx\csname urlstyle\endcsname\relax
  \providecommand{\doi}[1]{doi: #1}\else
  \providecommand{\doi}{doi: \begingroup \urlstyle{rm}\Url}\fi

\bibitem[Akiba et~al.(2019)Akiba, Sano, Yanase, Ohta, and Koyama]{akiba2019optuna}
Takuya Akiba, Shotaro Sano, Toshihiko Yanase, Takeru Ohta, and Masanori Koyama.
\newblock Optuna: A next-generation hyperparameter optimization framework.
\newblock In \emph{Proceedings of the 25th ACM SIGKDD international conference on knowledge discovery \& data mining}, pp.\  2623--2631, 2019.

\bibitem[Alfonzo et~al.(2021)Alfonzo, Brown, Byers, Cheung, Maraia, and Ross]{alfonzo2021call}
Juan~D Alfonzo, Jessica~A Brown, Peter~H Byers, Vivian~G Cheung, Richard~J Maraia, and Robert~L Ross.
\newblock A call for direct sequencing of full-length rnas to identify all modifications.
\newblock \emph{Nature genetics}, 53\penalty0 (8):\penalty0 1113--1116, 2021.

\bibitem[Alshareedah et~al.(2019)Alshareedah, Kaur, Ngo, Seppala, Kounatse, Wang, Moosa, and Banerjee]{alshareedah2019interplay}
Ibraheem Alshareedah, Taranpreet Kaur, Jason Ngo, Hannah Seppala, Liz-Audrey~Djomnang Kounatse, Wei Wang, Mahdi~Muhammad Moosa, and Priya~R Banerjee.
\newblock Interplay between short-range attraction and long-range repulsion controls reentrant liquid condensation of ribonucleoprotein--rna complexes.
\newblock \emph{Journal of the American Chemical Society}, 141\penalty0 (37):\penalty0 14593--14602, 2019.

\bibitem[Arora \& Sanguinetti(2022)Arora and Sanguinetti]{arora2022novo}
Viplove Arora and Guido Sanguinetti.
\newblock De novo prediction of rna--protein interactions with graph neural networks.
\newblock \emph{RNA}, 28\penalty0 (11):\penalty0 1469--1480, 2022.

\bibitem[Baek et~al.(2024)Baek, McHugh, Anishchenko, Jiang, Baker, and DiMaio]{baek2024accurate}
Minkyung Baek, Ryan McHugh, Ivan Anishchenko, Hanlun Jiang, David Baker, and Frank DiMaio.
\newblock Accurate prediction of protein--nucleic acid complexes using rosettafoldna.
\newblock \emph{Nature methods}, 21\penalty0 (1):\penalty0 117--121, 2024.

\bibitem[Batatia et~al.(2022)Batatia, Kovacs, Simm, Ortner, and Cs{\'a}nyi]{batatia2022mace}
Ilyes Batatia, David~P Kovacs, Gregor Simm, Christoph Ortner, and G{\'a}bor Cs{\'a}nyi.
\newblock Mace: Higher order equivariant message passing neural networks for fast and accurate force fields.
\newblock \emph{Advances in Neural Information Processing Systems}, 35:\penalty0 11423--11436, 2022.

\bibitem[Batzner et~al.(2022)Batzner, Musaelian, Sun, Geiger, Mailoa, Kornbluth, Molinari, Smidt, and Kozinsky]{batzner20223}
Simon Batzner, Albert Musaelian, Lixin Sun, Mario Geiger, Jonathan~P Mailoa, Mordechai Kornbluth, Nicola Molinari, Tess~E Smidt, and Boris Kozinsky.
\newblock E (3)-equivariant graph neural networks for data-efficient and accurate interatomic potentials.
\newblock \emph{Nature communications}, 13\penalty0 (1):\penalty0 2453, 2022.

\bibitem[Bo et~al.(2023)Bo, Wang, Liu, Fang, Li, and Shi]{bo2023survey}
Deyu Bo, Xiao Wang, Yang Liu, Yuan Fang, Yawen Li, and Chuan Shi.
\newblock A survey on spectral graph neural networks.
\newblock \emph{arXiv preprint arXiv:2302.05631}, 2023.

\bibitem[Boniecki et~al.(2016)Boniecki, Lach, Dawson, Tomala, Lukasz, Soltysinski, Rother, and Bujnicki]{boniecki2016simrna}
Michal~J Boniecki, Grzegorz Lach, Wayne~K Dawson, Konrad Tomala, Pawel Lukasz, Tomasz Soltysinski, Kristian~M Rother, and Janusz~M Bujnicki.
\newblock Simrna: a coarse-grained method for rna folding simulations and 3d structure prediction.
\newblock \emph{Nucleic acids research}, 44\penalty0 (7):\penalty0 e63--e63, 2016.

\bibitem[Byron et~al.(2016)Byron, Van Keuren-Jensen, Engelthaler, Carpten, and Craig]{byron2016translating}
Sara~A Byron, Kendall~R Van Keuren-Jensen, David~M Engelthaler, John~D Carpten, and David~W Craig.
\newblock Translating rna sequencing into clinical diagnostics: opportunities and challenges.
\newblock \emph{Nature Reviews Genetics}, 17\penalty0 (5):\penalty0 257--271, 2016.

\bibitem[Cen et~al.(2024)Cen, Li, Lin, Ren, Wang, and Huang]{cen2024high}
Jiacheng Cen, Anyi Li, Ning Lin, Yuxiang Ren, Zihe Wang, and Wenbing Huang.
\newblock Are high-degree representations really unnecessary in equivariant graph neural networks?
\newblock \emph{arXiv preprint arXiv:2410.11443}, 2024.

\bibitem[Chen et~al.(2022)Chen, Hu, Sun, Tan, Wang, Yu, Zong, Hong, Xiao, Shen, et~al.]{chen2022interpretable}
Jiayang Chen, Zhihang Hu, Siqi Sun, Qingxiong Tan, Yixuan Wang, Qinze Yu, Licheng Zong, Liang Hong, Jin Xiao, Tao Shen, et~al.
\newblock Interpretable rna foundation model from unannotated data for highly accurate rna structure and function predictions.
\newblock \emph{arXiv preprint arXiv:2204.00300}, 2022.

\bibitem[Chen et~al.(2023)Chen, Zhou, Ding, Wang, Ren, and Yang]{chen2023self}
Ken Chen, Yue Zhou, Maolin Ding, Yu~Wang, Zhixiang Ren, and Yuedong Yang.
\newblock Self-supervised learning on millions of pre-mrna sequences improves sequence-based rna splicing prediction.
\newblock \emph{bioRxiv}, pp.\  2023--01, 2023.

\bibitem[Chen et~al.(2020)Chen, Li, Umarov, Gao, and Song]{chen2020rna}
Xinshi Chen, Yu~Li, Ramzan Umarov, Xin Gao, and Le~Song.
\newblock Rna secondary structure prediction by learning unrolled algorithms.
\newblock \emph{arXiv preprint arXiv:2002.05810}, 2020.

\bibitem[Das \& Baker(2007)Das and Baker]{das2007automated}
Rhiju Das and David Baker.
\newblock Automated de novo prediction of native-like rna tertiary structures.
\newblock \emph{Proceedings of the National Academy of Sciences}, 104\penalty0 (37):\penalty0 14664--14669, 2007.

\bibitem[Defferrard et~al.(2016)Defferrard, Bresson, and Vandergheynst]{defferrard2016convolutional}
Micha{\"e}l Defferrard, Xavier Bresson, and Pierre Vandergheynst.
\newblock Convolutional neural networks on graphs with fast localized spectral filtering.
\newblock In \emph{Advances in Neural Information Processing Systems}, pp.\  3844--3852, 2016.

\bibitem[Deng et~al.(2023)Deng, Fang, Huang, Li, Xu, Ye, Zhang, Zhang, and Zhang]{deng2023rna}
Jie Deng, Xianyang Fang, Lin Huang, Shanshan Li, Lilei Xu, Keqiong Ye, Jinsong Zhang, Kaiming Zhang, and Qiangfeng~Cliff Zhang.
\newblock Rna structure determination: From 2d to 3d.
\newblock \emph{Fundamental Research}, 3\penalty0 (5):\penalty0 727--737, 2023.

\bibitem[Duval et~al.(2023)Duval, Schmidt, Hern{\'a}ndez-Garc{\i}a, Miret, Malliaros, Bengio, and Rolnick]{duval2023faenet}
Alexandre~Agm Duval, Victor Schmidt, Alex Hern{\'a}ndez-Garc{\i}a, Santiago Miret, Fragkiskos~D Malliaros, Yoshua Bengio, and David Rolnick.
\newblock Faenet: Frame averaging equivariant gnn for materials modeling.
\newblock In \emph{International Conference on Machine Learning}, pp.\  9013--9033. PMLR, 2023.

\bibitem[Fey \& Lenssen(2019)Fey and Lenssen]{Fey/Lenssen/2019}
Matthias Fey and Jan~E. Lenssen.
\newblock Fast graph representation learning with {PyTorch Geometric}.
\newblock In \emph{ICLR Workshop on Representation Learning on Graphs and Manifolds}, 2019.

\bibitem[Fox et~al.(2014)Fox, Reid-Bayliss, Emond, and Loeb]{fox2014accuracy}
Edward~J Fox, Kate~S Reid-Bayliss, Mary~J Emond, and Lawrence~A Loeb.
\newblock Accuracy of next generation sequencing platforms.
\newblock \emph{Next generation, sequencing \& applications}, 1, 2014.

\bibitem[Frank et~al.(2024)Frank, Unke, M{\"u}ller, and Chmiela]{frank2024euclidean}
J~Thorben Frank, Oliver~T Unke, Klaus-Robert M{\"u}ller, and Stefan Chmiela.
\newblock A euclidean transformer for fast and stable machine learned force fields.
\newblock \emph{Nature Communications}, 15\penalty0 (1):\penalty0 6539, 2024.

\bibitem[Fu et~al.(2022)Fu, Cao, Wu, Peng, Nie, and Xie]{fu2022ufold}
Laiyi Fu, Yingxin Cao, Jie Wu, Qinke Peng, Qing Nie, and Xiaohui Xie.
\newblock Ufold: fast and accurate rna secondary structure prediction with deep learning.
\newblock \emph{Nucleic acids research}, 50\penalty0 (3):\penalty0 e14--e14, 2022.

\bibitem[Gasteiger et~al.(2020)Gasteiger, Gro{\ss}, and G{\"u}nnemann]{gasteiger2020directional}
Johannes Gasteiger, Janek Gro{\ss}, and Stephan G{\"u}nnemann.
\newblock Directional message passing for molecular graphs.
\newblock \emph{arXiv preprint arXiv:2003.03123}, 2020.

\bibitem[Groher et~al.(2018)Groher, Jager, Schneider, Groher, Hamacher, and Suess]{groher2018tuning}
Ann-Christin Groher, Sven Jager, Christopher Schneider, Florian Groher, Kay Hamacher, and Beatrix Suess.
\newblock Tuning the performance of synthetic riboswitches using machine learning.
\newblock \emph{ACS synthetic biology}, 8\penalty0 (1):\penalty0 34--44, 2018.

\bibitem[Han et~al.(2022)Han, Rong, Xu, and Huang]{han2022geometrically}
Jiaqi Han, Yu~Rong, Tingyang Xu, and Wenbing Huang.
\newblock Geometrically equivariant graph neural networks: A survey.
\newblock \emph{arXiv preprint arXiv:2202.07230}, 2022.

\bibitem[Han et~al.(2015)Han, Gao, Muegge, Zhang, and Zhou]{han2015advanced}
Yixing Han, Shouguo Gao, Kathrin Muegge, Wei Zhang, and Bing Zhou.
\newblock Advanced applications of rna sequencing and challenges.
\newblock \emph{Bioinformatics and biology insights}, 9:\penalty0 BBI--S28991, 2015.

\bibitem[He et~al.(2021)He, Gao, Sabnis, and Sun]{he2021nucleic}
Shujun He, Baizhen Gao, Rushant Sabnis, and Qing Sun.
\newblock Nucleic transformer: Deep learning on nucleic acids with self-attention and convolutions.
\newblock \emph{bioRxiv}, pp.\  2021--01, 2021.

\bibitem[He et~al.(2023)He, Gao, Sabnis, and Sun]{he2023rnadegformer}
Shujun He, Baizhen Gao, Rushant Sabnis, and Qing Sun.
\newblock Rnadegformer: accurate prediction of mrna degradation at nucleotide resolution with deep learning.
\newblock \emph{Briefings in Bioinformatics}, 24\penalty0 (1):\penalty0 bbac581, 2023.

\bibitem[He et~al.(2024)He, Huang, Townley, Kretsch, Karagianes, Cox, Blair, Penzar, Vyaltsev, Aristova, et~al.]{he2024ribonanza}
Shujun He, Rui Huang, Jill Townley, Rachael~C Kretsch, Thomas~G Karagianes, David~BT Cox, Hamish Blair, Dmitry Penzar, Valeriy Vyaltsev, Elizaveta Aristova, et~al.
\newblock Ribonanza: deep learning of rna structure through dual crowdsourcing.
\newblock \emph{bioRxiv}, 2024.

\bibitem[Hofacker et~al.(1994)Hofacker, Fontana, Stadler, Bonhoeffer, Tacker, Schuster, et~al.]{hofacker1994fast}
Ivo~L Hofacker, Walter Fontana, Peter~F Stadler, L~Sebastian Bonhoeffer, Manfred Tacker, Peter Schuster, et~al.
\newblock Fast folding and comparison of rna secondary structures.
\newblock \emph{Monatshefte fur chemie}, 125:\penalty0 167--167, 1994.

\bibitem[Honda et~al.(2019)Honda, Shi, and Ueda]{honda2019smiles}
Shion Honda, Shoi Shi, and Hiroki~R Ueda.
\newblock Smiles transformer: Pre-trained molecular fingerprint for low data drug discovery.
\newblock \emph{arXiv preprint arXiv:1911.04738}, 2019.

\bibitem[Jing et~al.(2020)Jing, Eismann, Suriana, Townshend, and Dror]{jing2020learning}
Bowen Jing, Stephan Eismann, Patricia Suriana, Raphael~JL Townshend, and Ron Dror.
\newblock Learning from protein structure with geometric vector perceptrons.
\newblock \emph{arXiv preprint arXiv:2009.01411}, 2020.

\bibitem[Joshi et~al.(2023)Joshi, Bodnar, Mathis, Cohen, and Liò]{joshi2023expressive}
Chaitanya~K. Joshi, Cristian Bodnar, Simon~V. Mathis, Taco Cohen, and Pietro Liò.
\newblock On the expressive power of geometric graph neural networks.
\newblock In \emph{International Conference on Machine Learning}, 2023.

\bibitem[Kabsch(1976)]{kabsch1976solution}
Wolfgang Kabsch.
\newblock A solution for the best rotation to relate two sets of vectors.
\newblock \emph{Acta Crystallographica Section A: Crystal Physics, Diffraction, Theoretical and General Crystallography}, 32\penalty0 (5):\penalty0 922--923, 1976.

\bibitem[Kedzierska et~al.(2023)Kedzierska, Crawford, Amini, and Lu]{kedzierska2023assessing}
Kasia~Z Kedzierska, Lorin Crawford, Ava~P Amini, and Alex~X Lu.
\newblock Assessing the limits of zero-shot foundation models in single-cell biology.
\newblock \emph{bioRxiv}, pp.\  2023--10, 2023.

\bibitem[Kipf \& Welling(2017)Kipf and Welling]{kipf2017semi}
Thomas~N Kipf and Max Welling.
\newblock Semi-supervised classification with graph convolutional networks.
\newblock In \emph{5th International Conference on Learning Representations, ICLR 2017, Toulon, France, April 24-26, 2017, Conference Track Proceedings}, 2017.

\bibitem[Knyazev et~al.(2018)Knyazev, Lin, Amer, and Taylor]{knyazev2018spectral}
Boris Knyazev, Xiao Lin, Mohamed~R Amer, and Graham~W Taylor.
\newblock Spectral multigraph networks for discovering and fusing relationships in molecules.
\newblock \emph{arXiv preprint arXiv:1811.09595}, 2018.

\bibitem[Krishnan et~al.(2024)Krishnan, Roy, and Gromiha]{krishnan2024reliable}
Sowmya~R Krishnan, Arijit Roy, and M~Michael Gromiha.
\newblock Reliable method for predicting the binding affinity of rna-small molecule interactions using machine learning.
\newblock \emph{Briefings in Bioinformatics}, 25\penalty0 (2):\penalty0 bbae002, 2024.

\bibitem[Kulkarni et~al.(2023)Kulkarni, Thangappan, Deb, and Wu]{kulkarni2023comparative}
Mandar Kulkarni, Jayaraman Thangappan, Indrajit Deb, and Sangwook Wu.
\newblock Comparative analysis of rna secondary structure accuracy on predicted rna 3d models.
\newblock \emph{Plos one}, 18\penalty0 (9):\penalty0 e0290907, 2023.

\bibitem[Li et~al.(2022)Li, Zhao, and Zeng]{li2022kpgt}
Han Li, Dan Zhao, and Jianyang Zeng.
\newblock Kpgt: knowledge-guided pre-training of graph transformer for molecular property prediction.
\newblock In \emph{Proceedings of the 28th ACM SIGKDD Conference on Knowledge Discovery and Data Mining}, pp.\  857--867, 2022.

\bibitem[Moskalev et~al.(2024)Moskalev, Prakash, Liao, and Mansi]{moskalev2024sehyena}
Artem Moskalev, Mangal Prakash, Rui Liao, and Tommaso Mansi.
\newblock {SE}(3)-hyena operator for scalable equivariant learning.
\newblock In \emph{ICML 2024 Workshop on Geometry-grounded Representation Learning and Generative Modeling}, 2024.
\newblock URL \url{https://openreview.net/forum?id=rUXsKeN6dG}.

\bibitem[Nithin et~al.(2024)Nithin, Kmiecik, B{\l}aszczyk, Nowicka, and Tuszy{\'n}ska]{nithin2024comparative}
Chandran Nithin, Sebastian Kmiecik, Roman B{\l}aszczyk, Julita Nowicka, and Irina Tuszy{\'n}ska.
\newblock Comparative analysis of rna 3d structure prediction methods: towards enhanced modeling of rna--ligand interactions.
\newblock \emph{Nucleic Acids Research}, 52\penalty0 (13):\penalty0 7465--7486, 2024.

\bibitem[Ozsolak \& Milos(2011)Ozsolak and Milos]{ozsolak2011rna}
Fatih Ozsolak and Patrice~M Milos.
\newblock Rna sequencing: advances, challenges and opportunities.
\newblock \emph{Nature reviews genetics}, 12\penalty0 (2):\penalty0 87--98, 2011.

\bibitem[Pearce et~al.(2022)Pearce, Omenn, and Zhang]{pearce2022novo}
Robin Pearce, Gilbert~S Omenn, and Yang Zhang.
\newblock De novo rna tertiary structure prediction at atomic resolution using geometric potentials from deep learning.
\newblock \emph{BioRxiv}, pp.\  2022--05, 2022.

\bibitem[Peni{\'c} et~al.(2024)Peni{\'c}, Vla{\v{s}}i{\'c}, Huber, Wan, and {\v{S}}iki{\'c}]{penic2024rinalmo}
Rafael~Josip Peni{\'c}, Tin Vla{\v{s}}i{\'c}, Roland~G Huber, Yue Wan, and Mile {\v{S}}iki{\'c}.
\newblock Rinalmo: General-purpose rna language models can generalize well on structure prediction tasks.
\newblock \emph{arXiv preprint arXiv:2403.00043}, 2024.

\bibitem[Pfeiffer et~al.(2018)Pfeiffer, Gr{\"o}ber, Blank, H{\"a}ndler, Beyer, Schultze, and Mayer]{pfeiffer2018systematic}
Franziska Pfeiffer, Carsten Gr{\"o}ber, Michael Blank, Kristian H{\"a}ndler, Marc Beyer, Joachim~L Schultze, and G{\"u}nter Mayer.
\newblock Systematic evaluation of error rates and causes in short samples in next-generation sequencing.
\newblock \emph{Scientific reports}, 8\penalty0 (1):\penalty0 10950, 2018.

\bibitem[Ponce-Salvatierra et~al.(2019)Ponce-Salvatierra, Astha, Merdas, Nithin, Ghosh, Mukherjee, and Bujnicki]{ponce2019computational}
Almudena Ponce-Salvatierra, Astha, Katarzyna Merdas, Chandran Nithin, Pritha Ghosh, Sunandan Mukherjee, and Janusz~M Bujnicki.
\newblock Computational modeling of rna 3d structure based on experimental data.
\newblock \emph{Bioscience reports}, 39\penalty0 (2):\penalty0 BSR20180430, 2019.

\bibitem[Prakash et~al.(2024)Prakash, Moskalev, Jr., Combs, Mansi, Scheer, and Liao]{prakash2024bridging}
Mangal Prakash, Artem Moskalev, Peter~DiMaggio Jr., Steven Combs, Tommaso Mansi, Justin Scheer, and Rui Liao.
\newblock Bridging biomolecular modalities for knowledge transfer in bio-language models.
\newblock In \emph{Neurips 2024 Workshop Foundation Models for Science: Progress, Opportunities, and Challenges}, 2024.
\newblock URL \url{https://openreview.net/forum?id=dicOSQVPLm}.

\bibitem[Ramp\'{a}\v{s}ek et~al.(2022)Ramp\'{a}\v{s}ek, Galkin, Dwivedi, Luu, Wolf, and Beaini]{rampasek2022GPS}
Ladislav Ramp\'{a}\v{s}ek, Mikhail Galkin, Vijay~Prakash Dwivedi, Anh~Tuan Luu, Guy Wolf, and Dominique Beaini.
\newblock {Recipe for a General, Powerful, Scalable Graph Transformer}.
\newblock \emph{Advances in Neural Information Processing Systems}, 35, 2022.

\bibitem[Sahin et~al.(2014)Sahin, Karik{\'o}, and T{\"u}reci]{sahin2014mrna}
Ugur Sahin, Katalin Karik{\'o}, and {\"O}zlem T{\"u}reci.
\newblock mrna-based therapeutics—developing a new class of drugs.
\newblock \emph{Nature reviews Drug discovery}, 13\penalty0 (10):\penalty0 759--780, 2014.

\bibitem[Satorras et~al.(2021)Satorras, Hoogeboom, and Welling]{satorras2021n}
V{\i}ctor~Garcia Satorras, Emiel Hoogeboom, and Max Welling.
\newblock E (n) equivariant graph neural networks.
\newblock In \emph{International conference on machine learning}, pp.\  9323--9332. PMLR, 2021.

\bibitem[Schlick \& Pyle(2017)Schlick and Pyle]{schlick2017opportunities}
Tamar Schlick and Anna~Marie Pyle.
\newblock Opportunities and challenges in rna structural modeling and design.
\newblock \emph{Biophysical journal}, 113\penalty0 (2):\penalty0 225--234, 2017.

\bibitem[Schneider et~al.(2023)Schneider, Sweeney, Bateman, Cerny, Zok, and Szachniuk]{schneider2023will}
Bohdan Schneider, Blake~Alexander Sweeney, Alex Bateman, Jiri Cerny, Tomasz Zok, and Marta Szachniuk.
\newblock When will rna get its alphafold moment?
\newblock \emph{Nucleic Acids Research}, 51\penalty0 (18):\penalty0 9522--9532, 2023.

\bibitem[Sch{\"u}tt et~al.(2017)Sch{\"u}tt, Kindermans, Sauceda~Felix, Chmiela, Tkatchenko, and M{\"u}ller]{schutt2017schnet}
Kristof Sch{\"u}tt, Pieter-Jan Kindermans, Huziel~Enoc Sauceda~Felix, Stefan Chmiela, Alexandre Tkatchenko, and Klaus-Robert M{\"u}ller.
\newblock Schnet: A continuous-filter convolutional neural network for modeling quantum interactions.
\newblock \emph{Advances in neural information processing systems}, 30, 2017.

\bibitem[Sharma et~al.(2008)Sharma, Ding, and Dokholyan]{sharma2008ifoldrna}
Shantanu Sharma, Feng Ding, and Nikolay~V Dokholyan.
\newblock ifoldrna: three-dimensional rna structure prediction and folding.
\newblock \emph{Bioinformatics}, 24\penalty0 (17):\penalty0 1951--1952, 2008.

\bibitem[Sharp(2009)]{sharp2009centrality}
Phillip~A Sharp.
\newblock The centrality of rna.
\newblock \emph{Cell}, 136\penalty0 (4):\penalty0 577--580, 2009.

\bibitem[Shen et~al.(2022)Shen, Hu, Peng, Chen, Xiong, Hong, Zheng, Wang, King, Wang, et~al.]{shen2022e2efold}
Tao Shen, Zhihang Hu, Zhangzhi Peng, Jiayang Chen, Peng Xiong, Liang Hong, Liangzhen Zheng, Yixuan Wang, Irwin King, Sheng Wang, et~al.
\newblock E2efold-3d: End-to-end deep learning method for accurate de novo rna 3d structure prediction.
\newblock \emph{arXiv preprint arXiv:2207.01586}, 2022.

\bibitem[Shetty et~al.(2013)Shetty, Stefanovic, and Mihailescu]{shetty2013hepatitis}
Sumangala Shetty, Snezana Stefanovic, and Mihaela~Rita Mihailescu.
\newblock Hepatitis c virus rna: molecular switches mediated by long-range rna--rna interactions?
\newblock \emph{Nucleic acids research}, 41\penalty0 (4):\penalty0 2526--2540, 2013.

\bibitem[Shi et~al.(2020)Shi, Huang, Feng, Zhong, Wang, and Sun]{shi2020masked}
Yunsheng Shi, Zhengjie Huang, Shikun Feng, Hui Zhong, Wenjin Wang, and Yu~Sun.
\newblock Masked label prediction: Unified message passing model for semi-supervised classification.
\newblock \emph{arXiv preprint arXiv:2009.03509}, 2020.

\bibitem[Singh et~al.(2021)Singh, Paliwal, Zhang, Singh, Litfin, and Zhou]{singh2021improved}
Jaswinder Singh, Kuldip Paliwal, Tongchuan Zhang, Jaspreet Singh, Thomas Litfin, and Yaoqi Zhou.
\newblock Improved rna secondary structure and tertiary base-pairing prediction using evolutionary profile, mutational coupling and two-dimensional transfer learning.
\newblock \emph{Bioinformatics}, 37\penalty0 (17):\penalty0 2589--2600, 2021.

\bibitem[Soylu \& Sefer(2023)Soylu and Sefer]{soylu2023bert2ome}
Necla~Nisa Soylu and Emre Sefer.
\newblock Bert2ome: Prediction of 2-o-methylation modifications from rna sequence by transformer architecture based on bert.
\newblock \emph{IEEE/ACM Transactions on Computational Biology and Bioinformatics}, 20\penalty0 (3):\penalty0 2177--2189, 2023.

\bibitem[Strobel et~al.(2016)Strobel, Watters, Loughrey, and Lucks]{strobel2016rna}
Eric~J Strobel, Kyle~E Watters, David Loughrey, and Julius~B Lucks.
\newblock Rna systems biology: uniting functional discoveries and structural tools to understand global roles of rnas.
\newblock \emph{Current opinion in biotechnology}, 39:\penalty0 182--191, 2016.

\bibitem[Teufel \& Sobetzko(2022)Teufel and Sobetzko]{teufel2022reducing}
Marc Teufel and Patrick Sobetzko.
\newblock Reducing costs for dna and rna sequencing by sample pooling using a metagenomic approach.
\newblock \emph{BMC genomics}, 23\penalty0 (1):\penalty0 613, 2022.

\bibitem[Thomas et~al.(2018)Thomas, Smidt, Kearnes, Yang, Li, Kohlhoff, and Riley]{thomas2018tensor}
Nathaniel Thomas, Tess Smidt, Steven Kearnes, Lusann Yang, Li~Li, Kai Kohlhoff, and Patrick Riley.
\newblock Tensor field networks: Rotation-and translation-equivariant neural networks for 3d point clouds.
\newblock \emph{arXiv preprint arXiv:1802.08219}, 2018.

\bibitem[Tom et~al.(2017)Tom, Reeder, Forrest, Graham, Hunkapiller, Behrens, and Bhangale]{tom2017identifying}
Jennifer~A Tom, Jens Reeder, William~F Forrest, Robert~R Graham, Julie Hunkapiller, Timothy~W Behrens, and Tushar~R Bhangale.
\newblock Identifying and mitigating batch effects in whole genome sequencing data.
\newblock \emph{BMC bioinformatics}, 18:\penalty0 1--12, 2017.

\bibitem[Tran et~al.(2020)Tran, Ang, Chevrier, Zhang, Lee, Goh, and Chen]{tran2020benchmark}
Hoa Thi~Nhu Tran, Kok~Siong Ang, Marion Chevrier, Xiaomeng Zhang, Nicole Yee~Shin Lee, Michelle Goh, and Jinmiao Chen.
\newblock A benchmark of batch-effect correction methods for single-cell rna sequencing data.
\newblock \emph{Genome biology}, 21:\penalty0 1--32, 2020.

\bibitem[Vaswani et~al.(2017)Vaswani, Shazeer, Parmar, Uszkoreit, Jones, Gomez, Kaiser, and Polosukhin]{vaswani2017attention}
Ashish Vaswani, Noam Shazeer, Niki Parmar, Jakob Uszkoreit, Llion Jones, Aidan~N Gomez, {\L}ukasz Kaiser, and Illia Polosukhin.
\newblock Attention is all you need.
\newblock In \emph{Advances in Neural Information Processing Systems}, pp.\  5998--6008, 2017.

\bibitem[Veli{\v{c}}kovi{\'c} et~al.(2018)Veli{\v{c}}kovi{\'c}, Cucurull, Casanova, Romero, Li{\`o}, and Bengio]{velickovic2018graph}
Petar Veli{\v{c}}kovi{\'c}, Guillem Cucurull, Arantxa Casanova, Adriana Romero, Pietro Li{\`o}, and Yoshua Bengio.
\newblock Graph attention networks.
\newblock In \emph{6th International Conference on Learning Representations, ICLR 2018, Vancouver, BC, Canada, April 30--May 3, 2018, Conference Track Proceedings}, 2018.

\bibitem[Wang et~al.(2022)Wang, Wang, Zhang, Bi, and Zhu]{wang2022brief}
Hong Wang, Shuyu Wang, Yong Zhang, Shoudong Bi, and Xiaolei Zhu.
\newblock A brief review of machine learning methods for rna methylation sites prediction.
\newblock \emph{Methods}, 203:\penalty0 399--421, 2022.

\bibitem[Wang et~al.(2023)Wang, Feng, Han, Wang, Ye, Du, Wei, Zhang, Peng, and Yang]{wang2023trrosettarna}
Wenkai Wang, Chenjie Feng, Renmin Han, Ziyi Wang, Lisha Ye, Zongyang Du, Hong Wei, Fa~Zhang, Zhenling Peng, and Jianyi Yang.
\newblock trrosettarna: automated prediction of rna 3d structure with transformer network.
\newblock \emph{Nature Communications}, 14\penalty0 (1):\penalty0 7266, 2023.

\bibitem[Wang \& Zhang(2022)Wang and Zhang]{wang2022powerful}
Xiyuan Wang and Muhan Zhang.
\newblock How powerful are spectral graph neural networks.
\newblock In \emph{International conference on machine learning}, pp.\  23341--23362. PMLR, 2022.

\bibitem[Wayment-Steele et~al.(2020)Wayment-Steele, Kladwang, and Das]{Wayment-Steele}
Hannah~K. Wayment-Steele, Wipapat Kladwang, and Rhiju Das.
\newblock Rna secondary structure packages ranked and improved by high-throughput experiments.
\newblock \emph{bioRxiv}, 2020.
\newblock \doi{10.1101/2020.05.29.124511}.
\newblock URL \url{https://www.biorxiv.org/content/early/2020/05/31/2020.05.29.124511}.

\bibitem[Wayment-Steele et~al.(2022{\natexlab{a}})Wayment-Steele, Kladwang, Strom, Lee, Treuille, Becka, Participants, and Das]{wayment2022rna}
Hannah~K Wayment-Steele, Wipapat Kladwang, Alexandra~I Strom, Jeehyung Lee, Adrien Treuille, Alex Becka, Eterna Participants, and Rhiju Das.
\newblock Rna secondary structure packages evaluated and improved by high-throughput experiments.
\newblock \emph{Nature methods}, 19\penalty0 (10):\penalty0 1234--1242, 2022{\natexlab{a}}.

\bibitem[Wayment-Steele et~al.(2022{\natexlab{b}})Wayment-Steele, Kladwang, Watkins, Kim, Tunguz, Reade, Demkin, Romano, Wellington-Oguri, Nicol, et~al.]{wayment2022deep}
Hannah~K Wayment-Steele, Wipapat Kladwang, Andrew~M Watkins, Do~Soon Kim, Bojan Tunguz, Walter Reade, Maggie Demkin, Jonathan Romano, Roger Wellington-Oguri, John~J Nicol, et~al.
\newblock Deep learning models for predicting rna degradation via dual crowdsourcing.
\newblock \emph{Nature Machine Intelligence}, 4\penalty0 (12):\penalty0 1174--1184, 2022{\natexlab{b}}.

\bibitem[Wieder et~al.(2020)Wieder, Kohlbacher, Kuenemann, Garon, Ducrot, Seidel, and Langer]{wieder2020compact}
Oliver Wieder, Stefan Kohlbacher, M{\'e}laine Kuenemann, Arthur Garon, Pierre Ducrot, Thomas Seidel, and Thierry Langer.
\newblock A compact review of molecular property prediction with graph neural networks.
\newblock \emph{Drug Discovery Today: Technologies}, 37:\penalty0 1--12, 2020.

\bibitem[Wint et~al.(2022)Wint, Salamov, and Grigoriev]{wint2022kingdom}
Rhondene Wint, Asaf Salamov, and Igor~V Grigoriev.
\newblock Kingdom-wide analysis of fungal protein-coding and trna genes reveals conserved patterns of adaptive evolution.
\newblock \emph{Molecular biology and evolution}, 39\penalty0 (2):\penalty0 msab372, 2022.

\bibitem[Xu et~al.(2024{\natexlab{a}})Xu, Dai, Luo, Zhang, and Wang]{xu2024shape}
Junjie Xu, Enyan Dai, Dongsheng Luo, Xiang Zhang, and Suhang Wang.
\newblock Shape-aware graph spectral learning.
\newblock In \emph{Proceedings of the 33rd ACM International Conference on Information and Knowledge Management}, pp.\  2692--2701, 2024{\natexlab{a}}.

\bibitem[Xu et~al.(2018)Xu, Hu, Leskovec, and Jegelka]{xu2018powerful}
Keyulu Xu, Weihua Hu, Jure Leskovec, and Stefanie Jegelka.
\newblock How powerful are graph neural networks?
\newblock In \emph{International Conference on Learning Representations}, 2018.

\bibitem[Xu et~al.(2014)Xu, Zhao, and Chen]{xu2014vfold}
Xiaojun Xu, Peinan Zhao, and Shi-Jie Chen.
\newblock Vfold: a web server for rna structure and folding thermodynamics prediction.
\newblock \emph{PloS one}, 9\penalty0 (9):\penalty0 e107504, 2014.

\bibitem[Xu et~al.(2024{\natexlab{b}})Xu, Gupta, Cheng, Shen, Shen, Talwalkar, and Khodak]{xu2024specialized}
Zongzhe Xu, Ritvik Gupta, Wenduo Cheng, Alexander Shen, Junhong Shen, Ameet Talwalkar, and Mikhail Khodak.
\newblock Specialized foundation models struggle to beat supervised baselines.
\newblock \emph{arXiv preprint arXiv:2411.02796}, 2024{\natexlab{b}}.

\bibitem[Yan et~al.(2020)Yan, Hamilton, and Blanchette]{yan2020graph}
Zichao Yan, William~L Hamilton, and Mathieu Blanchette.
\newblock Graph neural representational learning of rna secondary structures for predicting rna-protein interactions.
\newblock \emph{Bioinformatics}, 36\penalty0 (Supplement\_1):\penalty0 i276--i284, 2020.

\bibitem[Yang et~al.(2024)Yang, Fusi, and Lu]{yang2024convolutions}
Kevin~K Yang, Nicolo Fusi, and Alex~X Lu.
\newblock Convolutions are competitive with transformers for protein sequence pretraining.
\newblock \emph{Cell Systems}, 15\penalty0 (3):\penalty0 286--294, 2024.

\bibitem[Yang et~al.(2023)Yang, Li, Pang, Cao, Li, and Zhang]{yang2023deciphering}
Yuning Yang, Gen Li, Kuan Pang, Wuxinhao Cao, Xiangtao Li, and Zhaolei Zhang.
\newblock Deciphering 3'utr mediated gene regulation using interpretable deep representation learning.
\newblock \emph{bioRxiv}, pp.\  2023--09, 2023.

\bibitem[Yazdani-Jahromi et~al.(2025)Yazdani-Jahromi, Prakash, Mansi, Moskalev, and Liao]{yazdani-jahromi2025helm}
Mehdi Yazdani-Jahromi, Mangal Prakash, Tommaso Mansi, Artem Moskalev, and Rui Liao.
\newblock {HELM}: Hierarchical encoding for m{RNA} language modeling.
\newblock In \emph{The Thirteenth International Conference on Learning Representations}, 2025.
\newblock URL \url{https://openreview.net/forum?id=MMHqnUOnl0}.

\bibitem[Ye et~al.(2022)Ye, Guan, Luo, Fang, Lai, and Wang]{ye2022molecular}
Xian-bin Ye, Quanlong Guan, Weiqi Luo, Liangda Fang, Zhao-Rong Lai, and Jun Wang.
\newblock Molecular substructure graph attention network for molecular property identification in drug discovery.
\newblock \emph{Pattern Recognition}, 128:\penalty0 108659, 2022.

\bibitem[Zhang et~al.(2020)Zhang, Wang, and Xiao]{zhang20203drna}
Yi~Zhang, Jun Wang, and Yi~Xiao.
\newblock 3drna: Building rna 3d structure with improved template library.
\newblock \emph{Computational and structural biotechnology journal}, 18:\penalty0 2416--2423, 2020.

\bibitem[Zhang et~al.(2024)Zhang, Cen, Han, Zhang, Zhou, and Huang]{zhangimproving}
Yuelin Zhang, Jiacheng Cen, Jiaqi Han, Zhiqiang Zhang, Jun Zhou, and Wenbing Huang.
\newblock Improving equivariant graph neural networks on large geometric graphs via virtual nodes learning.
\newblock In \emph{Forty-first International Conference on Machine Learning}, 2024.

\bibitem[Zheng et~al.(2022)Zheng, Qian, He, Kang, and Deng]{zheng2022graph}
Jingjing Zheng, Yurong Qian, Jie He, Zerui Kang, and Lei Deng.
\newblock Graph neural network with self-supervised learning for noncoding rna--drug resistance association prediction.
\newblock \emph{Journal of Chemical Information and Modeling}, 62\penalty0 (15):\penalty0 3676--3684, 2022.

\bibitem[Zhu et~al.(2023)Zhu, Xia, Wu, Xie, Zhou, Qin, Li, and Liu]{zhu2023dual}
Jinhua Zhu, Yingce Xia, Lijun Wu, Shufang Xie, Wengang Zhou, Tao Qin, Houqiang Li, and Tie-Yan Liu.
\newblock Dual-view molecular pre-training.
\newblock In \emph{Proceedings of the 29th ACM SIGKDD Conference on Knowledge Discovery and Data Mining}, pp.\  3615--3627, 2023.

\end{thebibliography}
\bibliographystyle{iclr2025_conference}

\newpage
\appendix

\section*{Appendix Overview}
\begin{itemize}[leftmargin=*]
\item \textbf{Appendix~\ref{sec:related_work}}: Related Work.
\item \textbf{Appendix~\ref{appen:models_overview}}: Models Overview.
\item \textbf{Appendix~\ref{appen:memory_compute}}: Memory and Computational Constraints.
\item \textbf{Appendix~\ref{appen:detailed_task_descriptions}}: Detailed task descriptions.
\item \textbf{Appendix~\ref{appen:discussion}}: Discussion on Advanced Model Architectures.
\item \textbf{Appendix~\ref{appen:extra_exp}}: Additional Experimental Information.
\item \textbf{Appendix~\ref{appen:3d_structure_noise}}: Analysis of Noise in 3D Structures.
\item \textbf{Appendix~\ref{appen:reproduction}}: Reproduction.
\item \textbf{Appendix~\ref{appen:detailed_data}}: Additional Results.
\end{itemize}

\section{Related Work}
\label{sec:related_work}

\paragraph{RNA property prediction} RNA-specific models remain scarce, likely due to limited specialized datasets. Recent advancements in sequence modeling have shown promise, particularly with foundation models like RNA-FM~\citep{chen2022interpretable}, UTRBERT~\citep{yang2023deciphering}, and RINALMO~\citep{penic2024rinalmo}, which use transformer-based architectures pre-trained on large RNA sequence corpora to predict various RNA functions and structures. While RNNs and CNNs have been applied to tasks like RNA methylation and protein binding~\citep{wang2022brief}, they struggle with long-range dependencies~\citep{moskalev2024sehyena}. Hybrid models like RNAdegformer~\citep{he2023rnadegformer}, combining convolutional layers with self-attention, improve predictions by capturing both local and global dependencies. Although some efforts integrate 2D structures with transformers, explicit 2D and 3D geometric modeling for RNA remains underexplored, with graph-based models mainly focusing on RNA-protein and RNA-drug interaction tasks rather than property prediction~\citep{krishnan2024reliable,yan2020graph,arora2022novo,zheng2022graph}.

\paragraph{RNA structure prediction} RNA 2D structure prediction has progressed from dynamic programming methods like Vienna RNAfold~\citep{hofacker1994fast} to deep learning-based tools like SPOT-RNA2~\citep{singh2021improved} and UFold~\citep{fu2022ufold}, which enhance accuracy by using neural networks and evolutionary data. Models such as E2Efold~\citep{chen2020rna} and RNA-FM~\citep{chen2022interpretable} employ transformer architectures to achieve state-of-the-art results in secondary structure prediction.

RNA 3D structure prediction has progressed through ab initio, template-based, and deep learning approaches. Ab initio methods (e.g., iFoldRNA~\citep{sharma2008ifoldrna}, SimRNA~\citep{boniecki2016simrna}) balance detail and efficiency but struggle with non-canonical interactions. Template-based models (e.g., FARNA/FARFAR~\citep{das2007automated}, 3dRNA~\citep{zhang20203drna}) depend on existing structures but are limited by available data. Deep learning models like DeepFoldRNA~\citep{pearce2022novo}, RhoFold~\citep{shen2022e2efold}, RoseTTAFoldNA~\citep{baek2024accurate}, and trRosettaRNA~\citep{wang2023trrosettarna} show promise in predicting 3D structures from sequence data but face challenges with novel RNA families due to RNA’s conformational flexibility~\citep{kulkarni2023comparative}.

Despite these advances, there is a gap in applying 2D and 3D modeling techniques to RNA property prediction. Most works focus on 1D representations and overlook the potential of geometric information from 2D and 3D structures. This study is the first to systematically explore the benefits and limitations of incorporating explicit structural data in deep learning-based RNA property prediction.

\section{Models overview}
\label{appen:models_overview}
\begin{enumerate}[left=0pt, label=\arabic*.]
\item \textbf{Transformer1D}: The Transformer1D model is a standard Transformer architecture adopted for RNA sequence processing. It includes an embedding layer to convert input tokens into dense vectors, positional encoding (PE) to retain sequence order and a multi-layer Transformer encoder to capture long-range dependencies within the sequence. 

\item \textbf{Transformer1D2D}: An adaptation of Transformer1D that integrates sequence and 2D graph structure information. The model encodes each nucleotide and incorporates BPP features, combining standard Transformer with positional encoding and a convolutional layer on the graph adjacency matrix. This convolutional output is added to the attention matrix, enabling the model to capture both sequential and structural dependencies.

\item \textbf{Graph Convolutional Network (GCN)}: A basic model in graph learning that aggregates and processes node features from local neighborhoods to capture both node characteristics and graph structure, making it effective for tasks like node classification.

\item \textbf{Graph Attention Network (GAT)}: Enhances graph convolutions by assigning different importance to neighboring nodes through local attention mechanisms, allowing the model to focus on more relevant nodes during feature aggregation.

\item \textbf{ChebNet}: A spectral GNN that utilizes Chebyshev polynomials to approximate the graph Laplacian, enabling graph convolutions with global structural context. This approach allows ChebNet to approximate global graph features with lower computational complexity.

\item \textbf{Graph Transformer}: This model uses Laplacian positional encoding to integrate structural information from the graph’s Laplacian into node features, which are then processed by Transformer layers. This enables aligning the sequential nature of transformer layers with graph topology.

\item \textbf{GraphGPS}: A hybrid model combining GNNs with transformers to capture both local and global graph information. It uses GNNs for local feature aggregation and transformers for long-range dependencies, making it effective for complex graph tasks requiring both local and global context.

\item \textbf{SchNet}: An SE(3)-invariant network designed for molecular property prediction on geometric graphs of atomic systems. It operates by modeling interactions through continuous-filter convolutional layers. Since the continuous filter in Schnet is conditioned on distance features, it maintains invariance to rotations and translations of atom coordinates.

\item \textbf{E(n)-Equivariant Graph Neural Network (EGNN)}: An equivariant network for geometric graphs with rotations, translations, and reflections symmetry. The EGNN operates as a non-linear message passing between scalar-invariant and vector equivariant quantities. 

\item \textbf{FAENet:} FAENet, or Frame Averaging Equivariant Network, is a lightweight and scalable Graph Neural Network designed for materials modeling. It introduces Stochastic Frame Averaging to achieve E(3)-equivariance through data transformations rather than architectural constraints, enabling superior accuracy and scalability on tasks like molecular and materials property prediction.

\item \textbf{GVPGNN}: Geometric Vector Perceptron GNN is a graph-based model designed to learn from 3D molecular structures by integrating geometric and relational information. It uses Geometric Vector Perceptrons to encode scalar and vector features, enabling effective modeling of both protein structure and interactions.

\item \textbf{DimeNet}: Directional Message Passing Neural Network leverages directional message passing to encode both distance and angular information between atoms. It utilizes spherical harmonics and Bessel functions for rotation equivariant message representations, enabling precise modeling of molecular interactions.

\item \textbf{FastEGNN}: FastEGNN introduces virtual nodes to traditional equivariant GNNs, enabling global message passing while maintaining scalability. These virtual nodes are designed to approximate and enhance real node interactions, ensuring E(3)-equivariance for efficient modeling of large geometric graphs.

\end{enumerate}

\section{Memory and Computational Constraints}
\label{appen:memory_compute}
\begin{figure}[ht]
    \centering
    \begin{subfigure}[b]{0.45\linewidth}
        \centering
        \includegraphics[width=\linewidth]{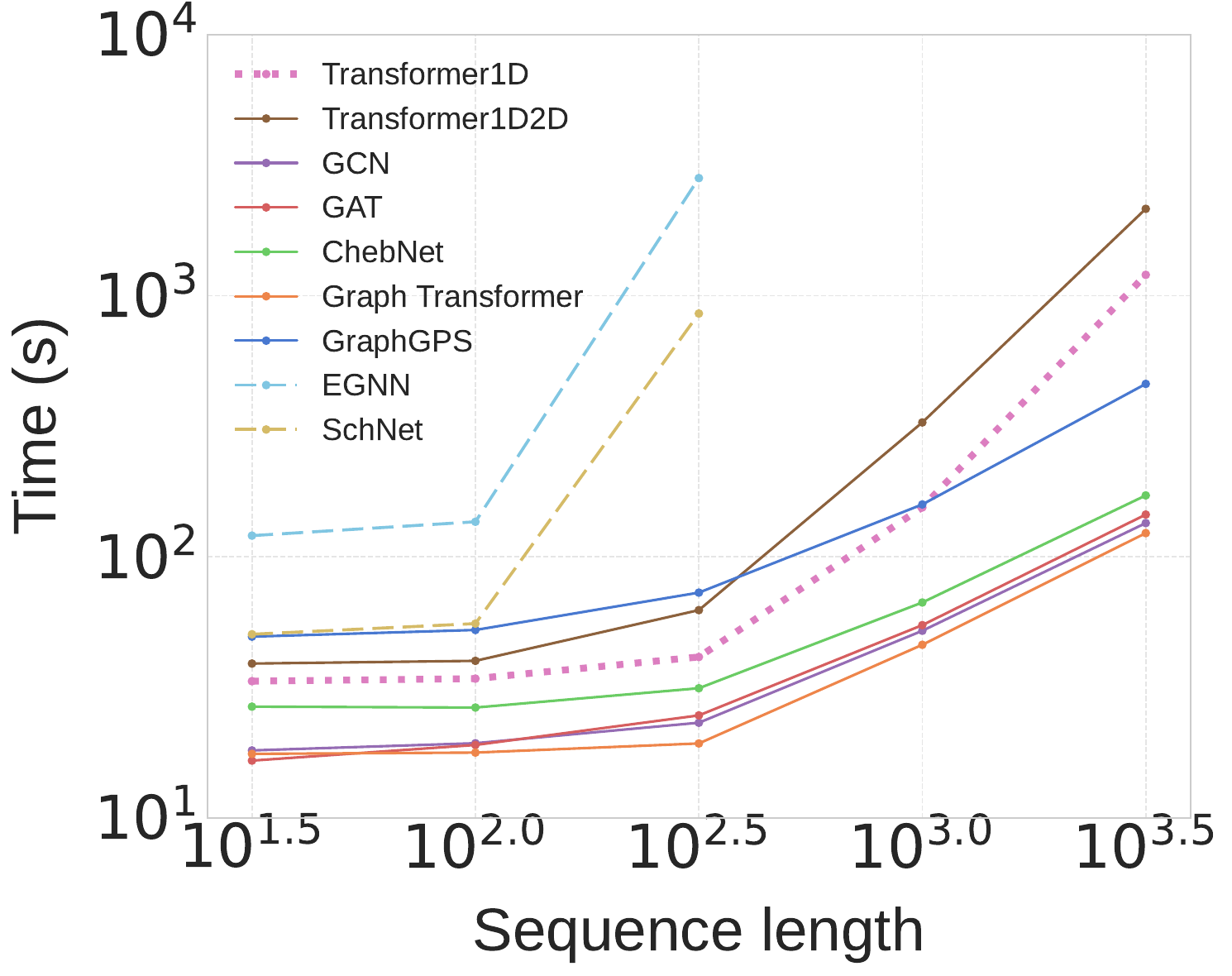}
        \label{fig:runningtime}
    \end{subfigure}
    \hspace{0.02\linewidth}
    \begin{subfigure}[b]{0.45\linewidth}
        \centering
        \includegraphics[width=\linewidth]{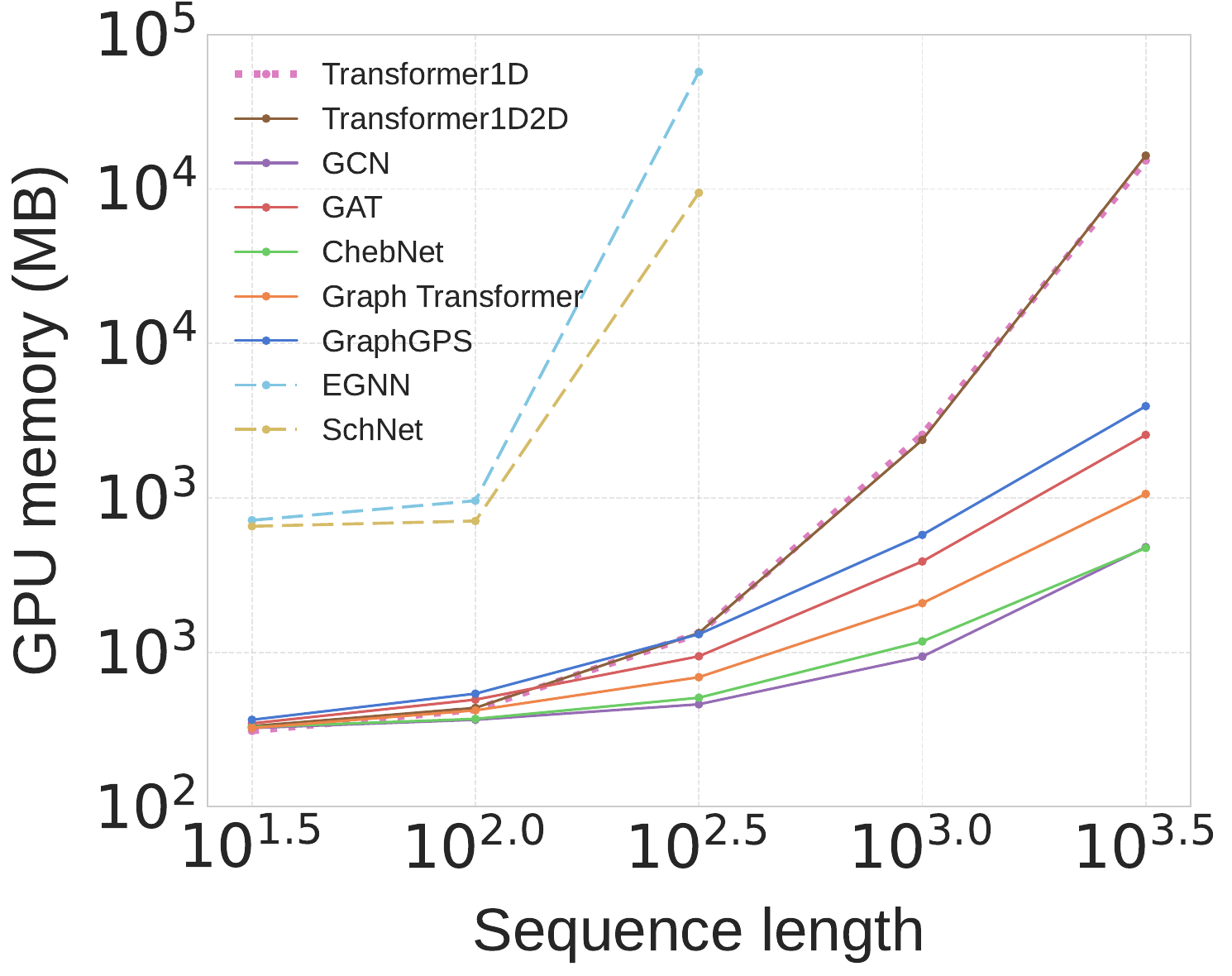}
        \label{fig:memory}
    \end{subfigure}
    \vspace{-1.5em}
    \caption{\textbf{Running time and memory usage comparison across models.} (1) Running time vs. sequence length (left): Transformer1D and Transformer1D2D scale poorly with sequence length due to expensive attention mechanisms and simultaneous processing of sequences and adjacency matrices, while simpler 2D models like GCN, GAT, and ChebNet are more efficient. (2) Memory usage vs. sequence length (right): Transformer1D2D and 3D models face out-of-memory issues with longer sequences, especially in the largest Fungal dataset, whereas 2D models use memory more efficiently, balancing computational demands and structural encoding.}
    \label{fig:runtime_memory_comparison}
\end{figure}

In this section, we compare the models based on run times and GPU memory. Both Transformer1D2D and 3D models (even with nucleotide pooling) encounter out-of-memory (OOM) issues when processing longer sequences, such as those in the Fungal dataset (Table~\ref{tab:model_comparison}). This highlights the need for optimization to handle longer sequences. Figure~\ref{fig:runtime_memory_comparison} shows that Transformer1D scales poorly in both runtime and memory due to its expensive attention mechanism, and Transformer1D2D faces additional challenges by processing the sequence and adjacency matrix simultaneously. In contrast, simpler 2D models like GCN, GAT, and ChebNet are more efficient. 3D models also scale poorly with sequence length due to the increasing number of atoms. Overall, 2D models provide a good balance between computational demands and performance for encoding structural information.

\section{Detailed Task Descriptions}
\label{appen:detailed_task_descriptions}

To comprehensively evaluate RNA property prediction models, we define five tasks that simulate real-world experimental challenges. These tasks are designed to assess the performance, efficiency, and robustness of models under various conditions encountered in RNA research and application. Below, we provide detailed explanations for each task, their scientific motivation, and their relevance to practical RNA property prediction.

\subsection{Task 1: Impact of Structural Information on Prediction Performance}

\textbf{Motivation:}  
RNA molecules fold into complex secondary (2D) and tertiary (3D) structures that dictate their functional properties. While traditional ML approaches often rely solely on the nucleotide sequence (1D representation), it is well-known in chemistry and biology that structural information influences RNA properties~\citep{schlick2017opportunities}. However, the extent to which geometric representations can improve performance of ML models remains unclear.

\textbf{Task Definition:} 
We evaluate multiple models across three classes (models described in main text):  
\begin{itemize}[left=0pt]
\item \textbf{1D models:} Utilize only the linear sequence of nucleotides.  
\item \textbf{2D models:} Incorporate RNA secondary structure, such as base-pairing interactions.  
\item \textbf{3D models:} Leverage full tertiary structural information, including spatial arrangements of nucleotides.  
\end{itemize}

By comparing their performance across diverse datasets, this task quantifies the benefits of using geometric information. Specifically, we aim to determine if models with explicit 2D or 3D inputs outperform sequence-only models, thereby highlighting the added value of structural representations in RNA property prediction.

\textbf{Significance:}
This analysis addresses the gap in existing literature, where the impact of structural information on RNA property prediction remains underexplored. The findings inform whether additional computational costs associated with structural modeling are justified by performance gains.

\subsection{Task 2: Model Efficiency in Limited Training Data Settings}

\textbf{Motivation:} 
High-quality RNA datasets with experimentally measured properties are scarce due to the technical difficulty and cost of experimental data acquisition~\citep{byron2016translating, teufel2022reducing}. In such scenarios, models must be sample-efficient, learning effectively from small datasets.

\textbf{Task Definition:}
Let \( D = \{X, Y\} \) represent a dataset, where \( X \) is the RNA input and \( Y \) is the property label. For this task, we define subsets \( D_\alpha \) by sampling a fraction \( \alpha \) of the original dataset. We evaluate models trained on varying \( \alpha \) values (e.g., \( \alpha = 25\%, 50\%, 75\% \)) to assess how performance scales with reduced training data.

\textbf{Significance:} 
This task simulates real-world scenarios where large labeled datasets are unavailable. It evaluates the ability of different models to generalize from limited data, providing insights into their efficiency and practical applicability.

\subsection{Task 3: Performance with Partial Sequence Labeling} 

\textbf{Motivation:} 
In nucleotide-level RNA datasets, labels for molecular properties (e.g., binding affinities or reactivity) are often available only for a subset of nucleotides in a sequence~\citep{wayment2022deep}. This is due to the high cost of annotating every nucleotide experimentally.

\textbf{Task Definition:} 
We consider datasets where only the first \( \ell \)-nucleotide positions in each sequence are labeled, representing partial sequence labeling. Models are trained using these incomplete labels and evaluated on their ability to predict properties across the full sequence. Performance is assessed using datasets with varying \( \ell \)-label fractions.

\textbf{Significance:}
This task reflects a critical real-world challenge where complete annotations are unavailable. It assesses a model’s ability to generalize from partial labels to unseen portions of RNA sequences, a valuable capability for applications with sparse experimental data.

\subsection{Task 4: Robustness to Sequencing Noise} 

\textbf{Motivation:}
RNA sequencing technologies often introduce noise in the form of random nucleotide errors (e.g., insertions, deletions, or substitutions)~\citep{ozsolak2011rna, fox2014accuracy}. This noise propagates through derived RNA structures (2D and 3D), potentially degrading model performance. Understanding how well models handle noise is crucial for their reliable deployment.

\textbf{Task Definition:} 
To simulate realistic sequencing errors, we introduce controlled noise levels into RNA sequences during both training and testing. Models are evaluated for their ability to maintain performance under consistent noise conditions, representing a scenario where noise characteristics are stable across experimental phases.

\textbf{Significance:}  
This task assesses a model’s robustness in practical settings where sequencing noise is unavoidable. It provides insights into the resilience of RNA models to variations in input quality.

\subsection{Task 5: Generalization to Out-of-Distribution (OOD) Data}

\textbf{Motivation:}
RNA models are often trained on high-quality experimental datasets but deployed in conditions where data characteristics differ significantly due to variations in sequencing platforms or protocols~\citep{tran2020benchmark, tom2017identifying}. This mismatch can lead to performance degradation.

\textbf{Task Definition:}
We evaluate models trained on clean RNA data and tested on datasets with higher levels of noise (representing OOD conditions). Performance metrics are analyzed as a function of increasing noise, quantifying the models’ ability to generalize to unseen distributions.

\textbf{Significance:}
This task mirrors real-world deployment scenarios where models encounter noisy or biased data. It highlights the limitations of models trained on idealized datasets and informs strategies for improving generalization under OOD conditions.

\section{Discussion on Advanced Model Architectures}
\label{appen:discussion}

\subsection{Enhancing 3D RNA Prediction with High-Degree Steerable Features}
\label{appen:steerbale_features}
Apart from the 3D models considered in this work, models using high-degree steerable features such as TFN~\citep{thomas2018tensor}, MACE~\citep{batatia2022mace} and NequIP~\citep{batzner20223} represent an important aspect of equivariant models, with potential to enhance model expressiveness by incorporating higher-order information. While memory constraints have limited their practical implementation in our current work, techniques such as scalarization of high-degree steerable features, as demonstrated in works such as HEGNN~\citep{frank2024euclidean} and SO3krates~\citep{cen2024high}, could address this challenge. For instance, HEGNN can improve the expressivity of equivariant GNNs by mitigating expressivity degeneration on symmetric graphs and leveraging higher-order representations. Although 3D RNA graphs are too complex to be considered symmetric, the ability to handle high-degree steerable features in large RNA graphs, with hundreds of nucleotides and thousands of atoms, remains valuable and has the potential to improve performance. Future work can explore such methods to improve 3D RNA prediction.

\subsection{Mitigating Sequencing Noise}
\label{appen:sequence_noise}
Notably, most studies in this field do not explicitly address sequencing noise common for RNA in model architecture design, prompting a need to explore effective strategies for mitigating its impact. An effective way to handle noise can be through ensemble methods. For instance, combining 1D, 2D, and 3D models by independently learning representations and integrating them via attention mechanisms that dynamically weigh each modality based on task relevance and noise can leverage the robustness of 1D methods while benefiting from the strengths of 2D and 3D approaches and be a good direction for future research.

\section{Additional Experimental Information}
\label{appen:extra_exp}

\subsection{Dataset Statistics}
\label{appen:dataset_stat}

Here, we present the statistics for each dataset used in the paper in Table~\ref{tab:data_stat}. The datasets are categorized as small, medium, or large based on the number of sequences and sequence length. ``Target" refers to the task the dataset is designed to predict, and ``\# Avg. Atoms" indicates the average number of atoms used in 3D models.

\begin{table}[h]
\centering
\caption{The statistics of Tc-Riboswitches, Ribonanza, COVID, and Fungal datasets.}
\label{tab:data_stat}
\begin{tabular}{ccccc}
\toprule
& \textbf{Tc-Riboswitches} & \textbf{Ribonanza} & \textbf{COVID} & \textbf{Fungal} \\ 
\midrule
Dataset Size                                                             & Small                    & Medium                & Medium            & Large           \\
Task Level                                                               & RNA-level                & Nucleotide-level      & Nucleotide-level  & RNA-level       \\
Target                                                                   & Switching Factor         & Degradation           & Degradation       & Expression      \\
\# Sequences                                                             & 355                      & 2260                  & 4082              & 7089            \\
Sequence Length                                                          & 66 - 75                  & 177                   & 107 - 130         & 150 - 3063      \\
\# Labels                                                                & 1                        & 2                     & 3                 & 1               \\
\begin{tabular}[c]{@{}c@{}}\# Avg. Atoms \\ (for 3D models)\end{tabular} & 1531                     & 3791                  & 2598              & N/A              \\
\bottomrule
\end{tabular}
\end{table}

\subsection{Comparison of partial training data and partial sequence labeling}
\label{appen:compare_partial}
\begin{figure}[ht]
    \centering
    \begin{subfigure}[b]{0.545\linewidth}
        \centering
        \includegraphics[width=\linewidth]{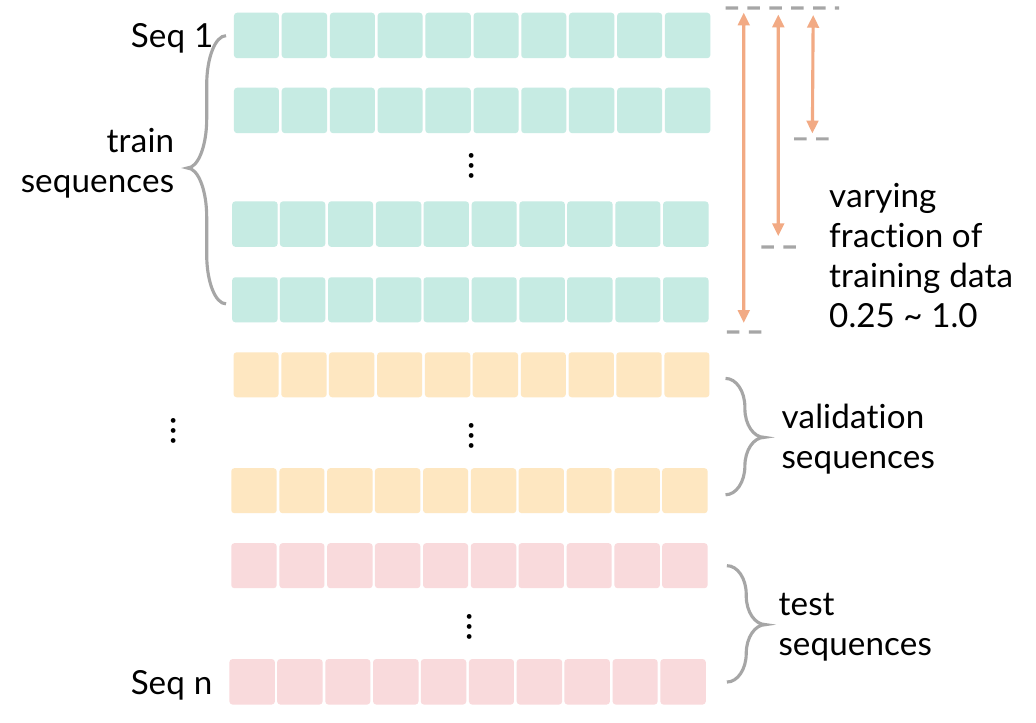}
        \caption{Varying fraction of training data.}
    \end{subfigure}
    \begin{subfigure}[b]{0.445\linewidth}
        \centering
        \includegraphics[width=\linewidth]{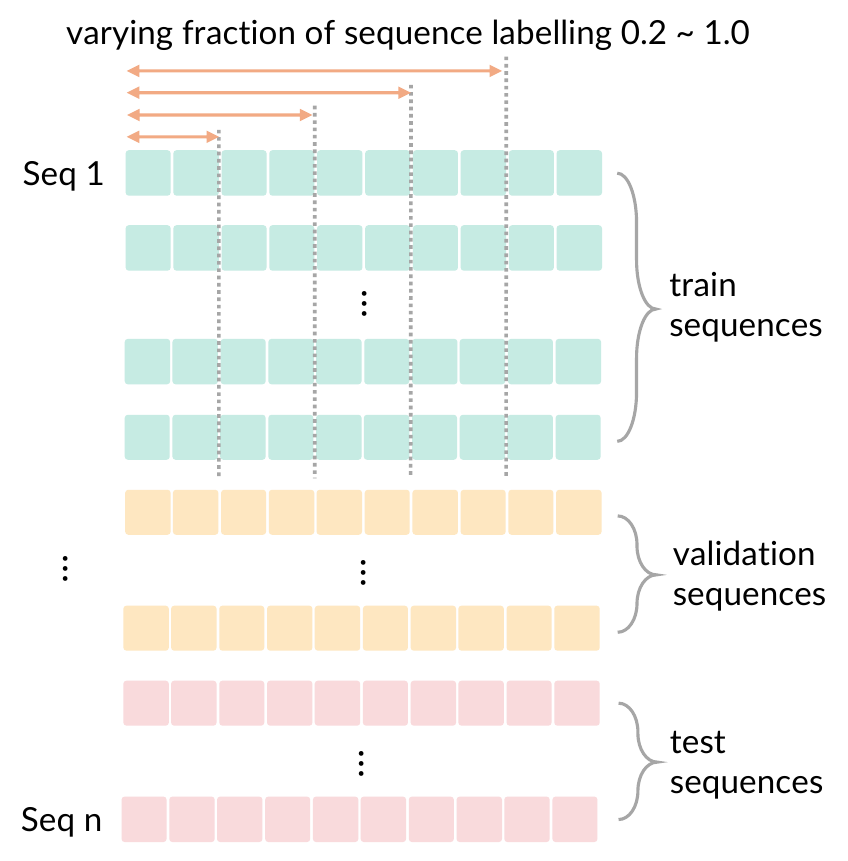}
        \caption{Varying fraction of sequence labeling.}
    \end{subfigure}
    \caption{\textbf{Comparison of partial training data and partial sequence labeling.} The orange arrows indicate the varying components. (a) utilizes full nucleotide labels but trains on varying fractions of RNA sequences (0.25, 0.5, 0.75, 1.0). (b) uses all training sequences but with varying fractions of nucleotide labels (0.2, 0.4, 0.6, 0.8, 1.0). Therefore, (b) is only for nucleotide-level tasks.}
    \label{fig:compare_partial}
\end{figure}

To clarify the differences and provide a more detailed explanation, we illustrate two experiments: partial training data and partial sequence labeling (Figure~\ref{fig:compare_partial}).

\subsection{Details of noisy experiments: robustness and generalization}
\begin{figure}[ht]
    \centering
    \includegraphics[width=0.75\linewidth]{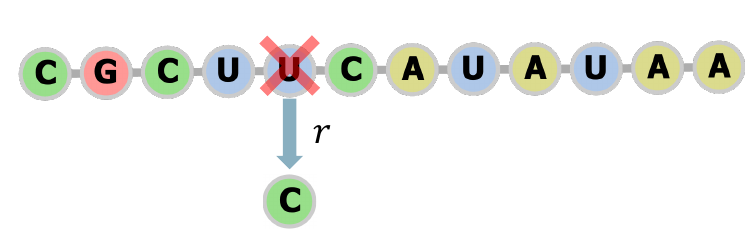}
    \caption{Mutation the nucleotide at each position independently with the probability $r$.}
    \vspace{1em}
    \label{fig:mutation}
\end{figure}

To create the noisy datasets, we vary the noise ratio \( r \) across the values \{0.05, 0.1, 0.15, 0.2, 0.25, 0.3\}. For each given noise ratio, we independently mutate the nucleotide at each position in a sequence with probability \( r \), as illustrated in Figure~\ref{fig:mutation}. The resulting mutated sequence is then passed to the 2D and 3D prediction tools to generate the corresponding structures. Figure~\ref{fig:rna_noise} gives a comprehensive illustration of getting noisy 1D, 2D, and 3D structures.

For the robustness experiments, all training, validation, and testing are conducted on the noisy datasets. In contrast, for the generalization experiments, the model is trained and validated on clean datasets, and its performance is tested on noisy datasets with varying noise ratios.

\subsection{Analysis of Transformer1D and Transformer1D2D}
\label{appen:comapre_1d_1d2d}

\begin{figure}[ht]
\centering
\begin{tabular}{cc}
    \begin{subfigure}[b]{0.4\textwidth}
        \centering
        \includegraphics[width=\textwidth]{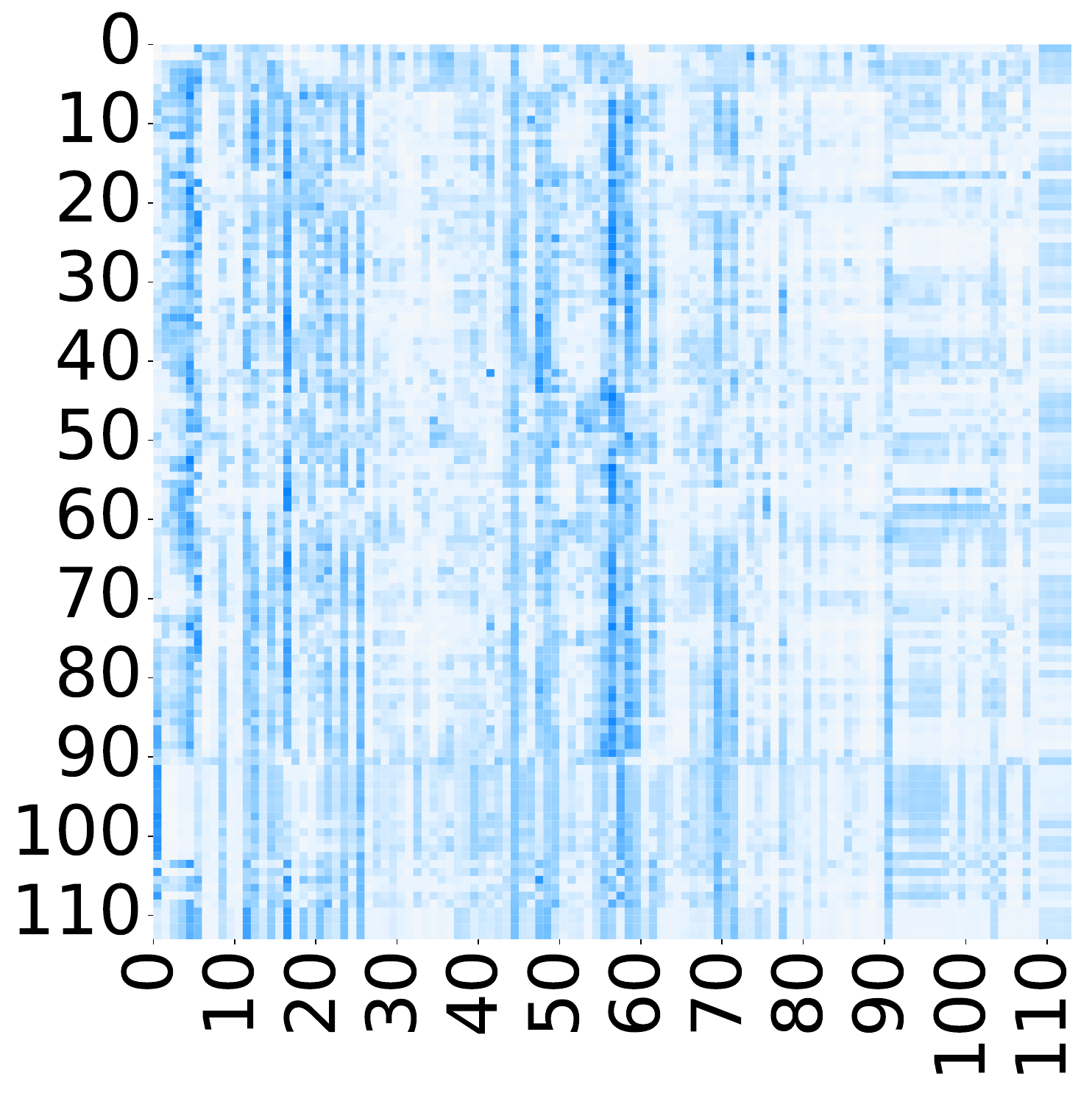}
        \caption{Transformer1D attention matrix.}
        \label{fig:attention_main_tsfm1d}
    \end{subfigure} & 
    \hspace{30pt}
    \begin{subfigure}[b]{0.48285\textwidth}
        \centering
        \includegraphics[width=\textwidth]{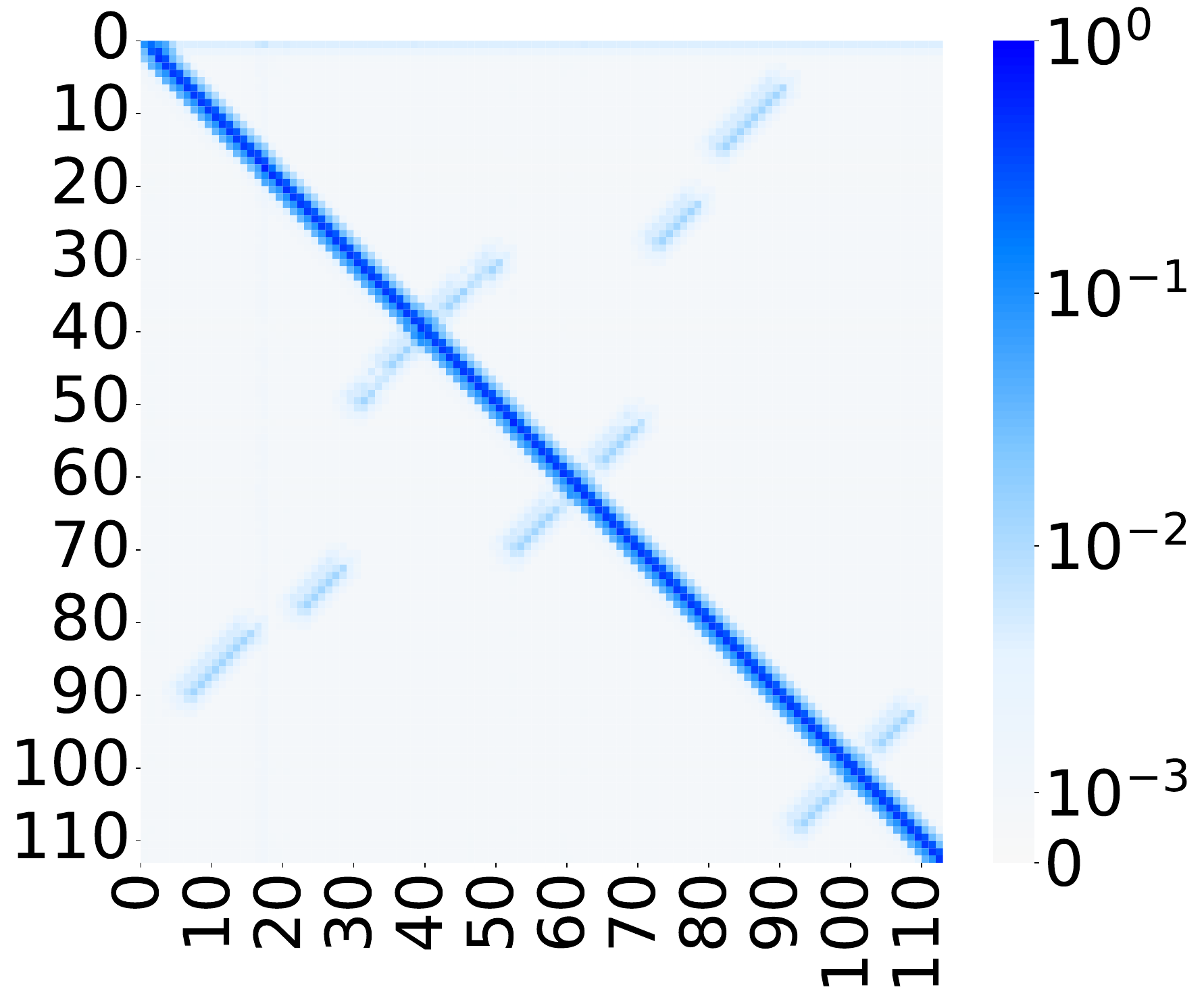}
        \caption{Transformer1D2D attention matrix.}
        \label{fig:attention_main_tsfm1d2d}
    \end{subfigure} \\ 
    \begin{subfigure}[b]{0.4\textwidth}
        \centering
        \includegraphics[width=\textwidth]{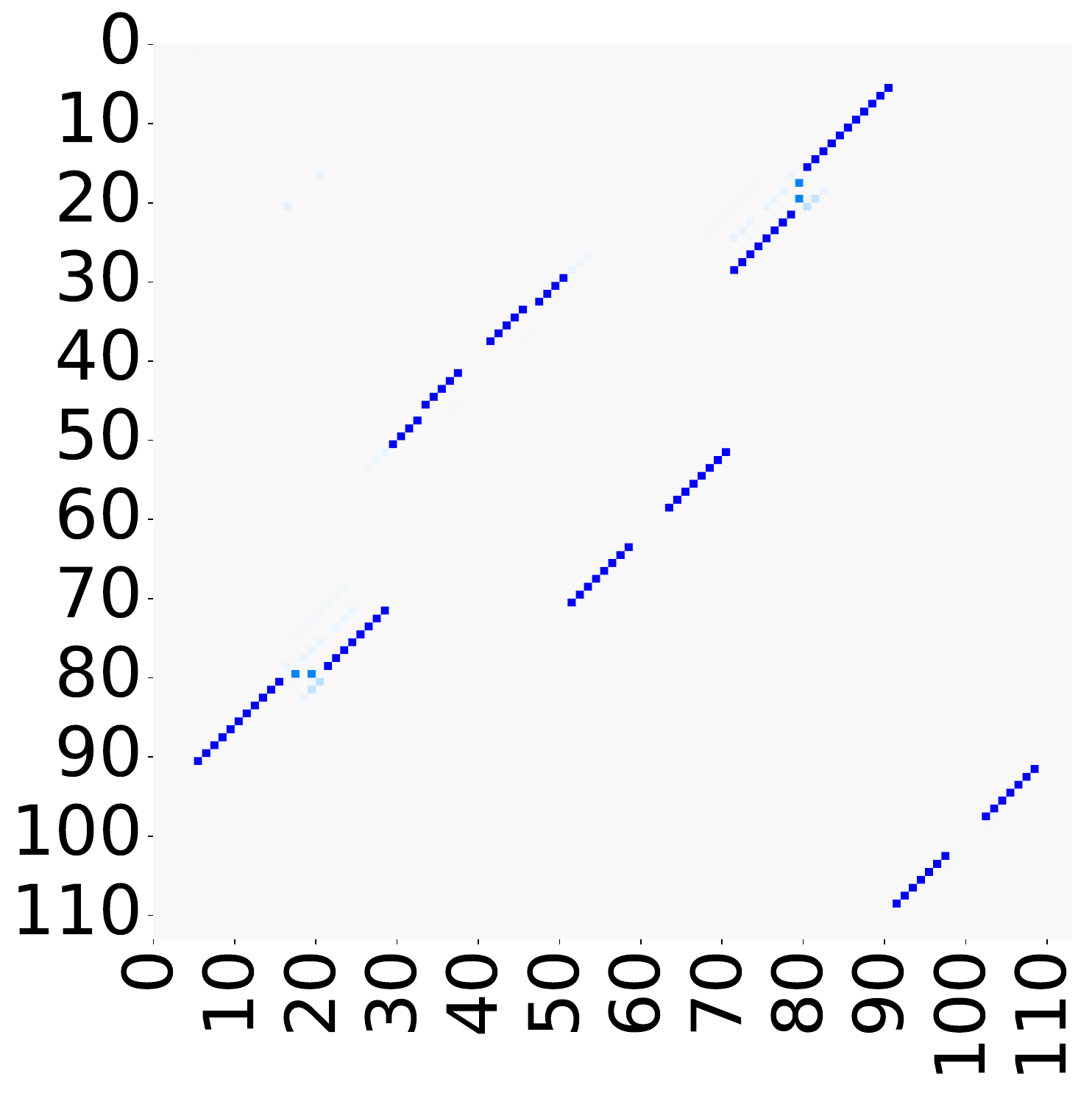}
        \caption{Base pair probability matrix.}
        \label{fig:attention_main_bpp}
    \end{subfigure} & 
    \begin{subfigure}[b]{0.4\textwidth}
        \centering
        \includegraphics[width=\textwidth]{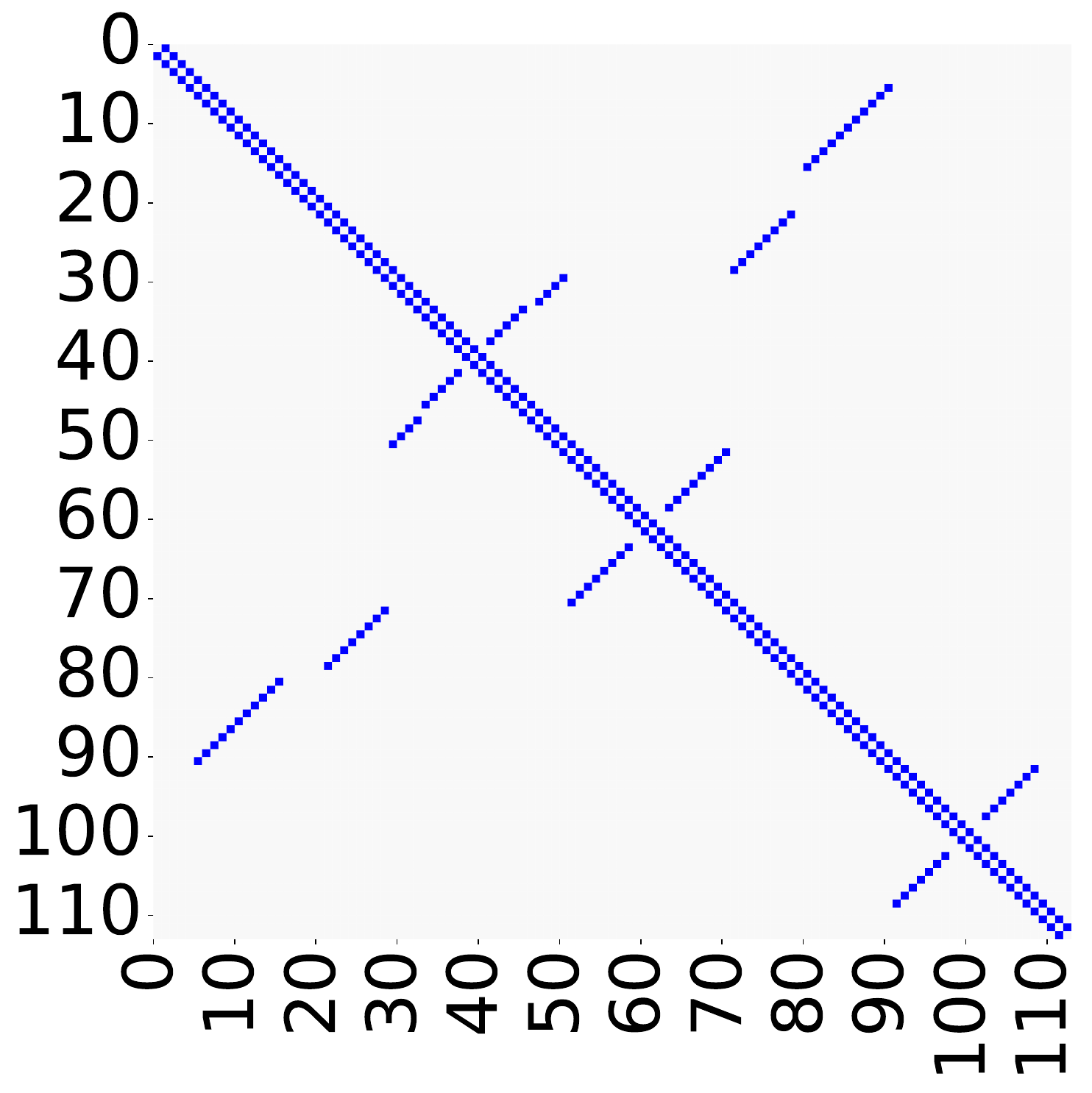}
        \caption{Graph adjacency matrix.}
        \label{fig:attention_main_adj}
    \end{subfigure}
\end{tabular}
\caption{\textbf{The heatmaps of matrices in Transformer1D and Transformer1D2D.} Attention maps from Transformer1D2D exhibit a striking resemblance to both the adjacency matrix and BPP matrix, highlighting the model's ability to learn structural features. In contrast, the standard Transformer struggles with this task, as shown by lower cosine similarity scores, reinforcing the conclusion that explicitly encoding structural information is crucial for enhanced model performance.}
\label{fig:attention_main}
\end{figure}

To further validate that incorporating structural information contributes to the final results, we analyze the attention maps generated by Transformer1D and Transformer1D2D. Fig.~\ref{fig:attention_main} illustrates the average attention maps across all heads before the final output layer for both the models for a randomly selected RNA sequence. The attention maps of Transformer1D2D exhibit a striking similarity to both the adjacency matrix and the BPP matrix, whereas the attention maps from the standard Transformer model seem to suggest that the model does not learn to attend to the structural features. Moreover, we quantify this observation by computing the cosine similarity between the attention maps of the models and the true adjacency matrix and BPP for all sequences in the COVID dataset. The results, reported in Table ~\ref{tab:cosine_similarity_values} show that Transformer1D2D achieves much higher similarity scores compared to the 1D Transformer alone. This reinforces the conclusion that explicit encoding of structural information is essential for improved model performance.

\begin{table}[ht]
\caption{\textbf{Cosine similarity values for different models.} Cosine similarity scores between the attention maps and the true adjacency and BPP matrices for all sequences in the COVID dataset demonstrate that Transformer1D2D significantly outperforms the standard Transformer. These results underscore the importance of explicitly encoding structural information for superior model performance.}
\label{tab:cosine_similarity_values}
\centering
\begin{tabular}{lcc}
\toprule
\textbf{Model} & \textbf{Cosine similarity adjacency} & \textbf{Cosine similarity BPP} \\
\midrule
Transformer1D     & 0.107 & 0.090 \\
Transformer1D2D   & 0.448 & 0.672 \\
\bottomrule
\end{tabular}
\end{table}

\section{Analysis of Noise in 3D Structures}
\label{appen:3d_structure_noise}
As mentioned in the main context, predicted 3D structures consistently exhibit noise. In this section, we analyze this issue from two perspectives: sensitivity to sequence length and variability across different prediction tools.

\subsection{Impact of Sequence Length on 3D Structure Prediction Noise}
To investigate the hypothesis that longer sequences result in greater noise in 3D structure predictions, we randomly selected a COVID and Tc-Riboswitches dataset sequence and generated structures using four state-of-the-art 3D structure prediction tools: RhoFold~\citep{shen2022e2efold}, RNAComposer~\citep{xu2014vfold}, trRosetta~\citep{baek2024accurate}, and SimRNA~\citep{boniecki2016simrna}. High variability among these predicted structures would indicate significant uncertainty in absolute atom positions. We quantified this noise by aligning the structures using the Kabsch algorithm~\citep{kabsch1976solution} and computing the pairwise RMSD values, resulting in a 4x4 matrix showing structural deviations between each pair of tools (see Table~\ref{tab:rmsd_combined}). The observed pairwise RMSD values ranged from 16 to 45 Å for the COVID dataset and from 11 to 15 Å for the Tc-Riboswitches dataset, reflecting substantial variability and suggesting considerable noise in the 3D predictions. This level of structural inaccuracy likely contributes to the poorer performance of 3D models. However, we found that 3D models outperform 1D models for shorter sequences, such as those in the Tc-Riboswitches dataset (67 to 73 nucleotides long). This improved performance is due to the lower noise in 3D predictions for shorter sequences, a phenomenon supported by previous studies~\citep{nithin2024comparative, ponce2019computational} and also exhibited by the comparatively smaller RMSD values reported in Table~\ref{tab:rmsd_combined} for Tc-Riboswitches dataset. The reduced complexity of shorter sequences allows 3D models to capture structural details more accurately, thereby enhancing performance and validating that accurate 3D structure encoding can outperform 1D models.

\begin{table}[h!]
\caption{\textbf{Pairwise RMSD values (in Å) between 3D structure prediction tools for the COVID dataset (left) and Tc-Riboswitches dataset (right).} The results indicate larger noise in predictions for longer COVID sequences.}
\label{tab:rmsd_combined}
\captionsetup{type=table} 
\centering
\begin{minipage}[b]{0.9\textwidth}
\centering
\begin{tabular}{ccccc}
\toprule
\textbf{COVID }        & \textbf{RhoFold} & \textbf{trRosetta} & \textbf{SimRNA} & \textbf{Composer} \\ 
\midrule
\textbf{RhoFold}   & 0          & 39.05192       & 44.76146      & 45.45994     \\
\textbf{trRosetta} & 39.05192   & 0              & 22.54974      & 18.17359     \\ 
\textbf{SimRNA}    & 44.76146   & 22.54974       & 0             & 16.73399     \\ 
\textbf{Composer}  & 45.45994   & 18.17359       & 16.73399      & 0            \\ 
\bottomrule
\end{tabular}
\end{minipage}
\vspace{1em}
\begin{minipage}[b]{0.9\textwidth}
\centering
\begin{tabular}{ccccc}
\toprule
\textbf{Tc-Ribo} & \textbf{RhoFold} & \textbf{trRosetta} & \textbf{SimRNA} & \textbf{Composer} \\
\midrule
\textbf{RhoFold} & 0 & 14.338 & 11.996 & 15.056 \\
\textbf{trRosetta} & 14.338 & 0 & 12.932 & 14.916 \\
\textbf{SimRNA} & 11.996 & 12.932 & 0 & 14.243 \\
\textbf{Composer} & 15.056 & 14.916 & 14.243 & 0 \\
\bottomrule
\end{tabular}
\end{minipage}
\end{table}

\subsection{Impact of different 3D prediction tools}
\begin{figure}[ht]
    \centering
    \begin{subfigure}[b]{0.32\linewidth}
        \centering
        \includegraphics[width=\linewidth]{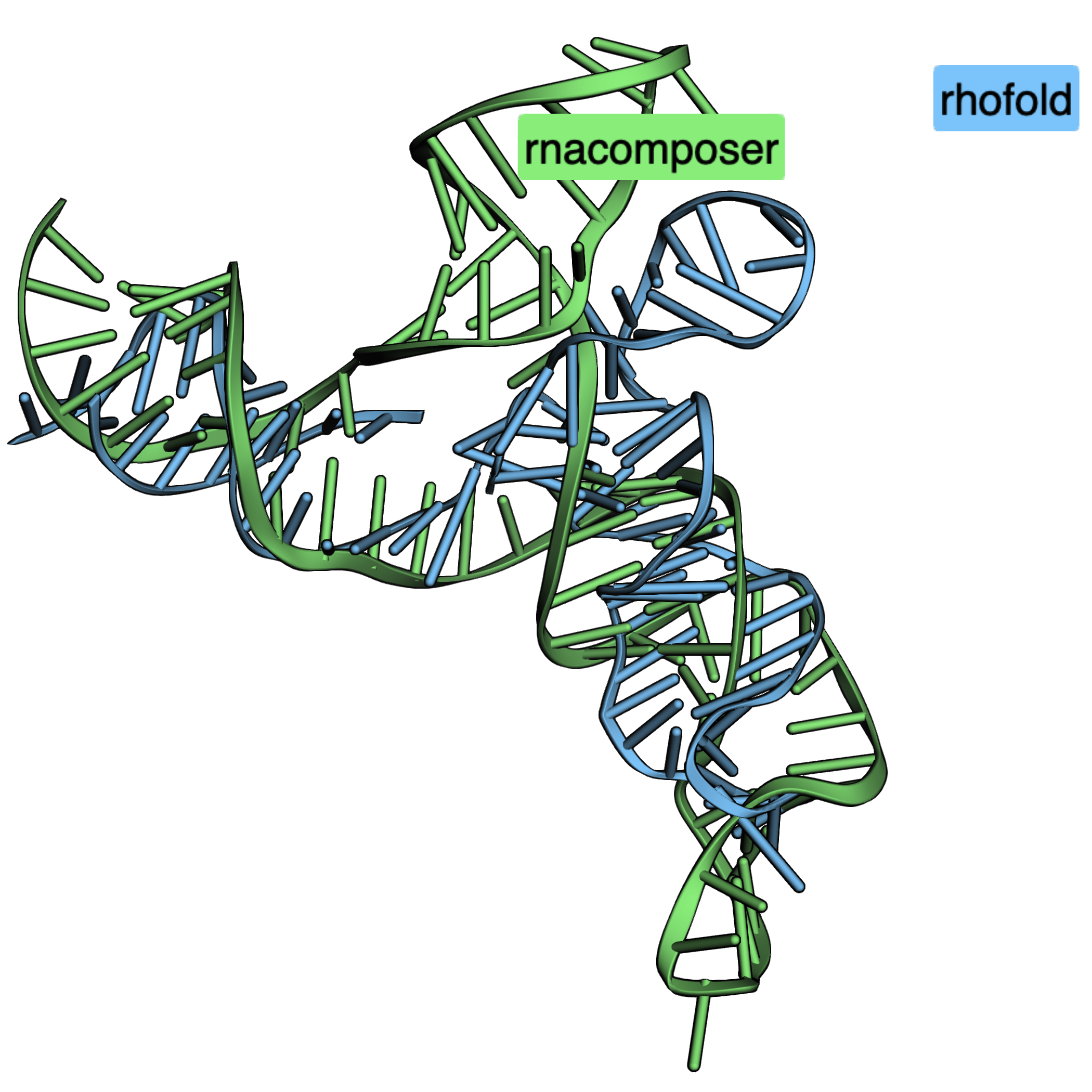}
        \caption{RhoFold vs. RNAComposer}
        \label{fig:rhofold-composer}
    \end{subfigure}
    \hfill
    \begin{subfigure}[b]{0.32\linewidth}
        \centering
        \includegraphics[width=\linewidth]{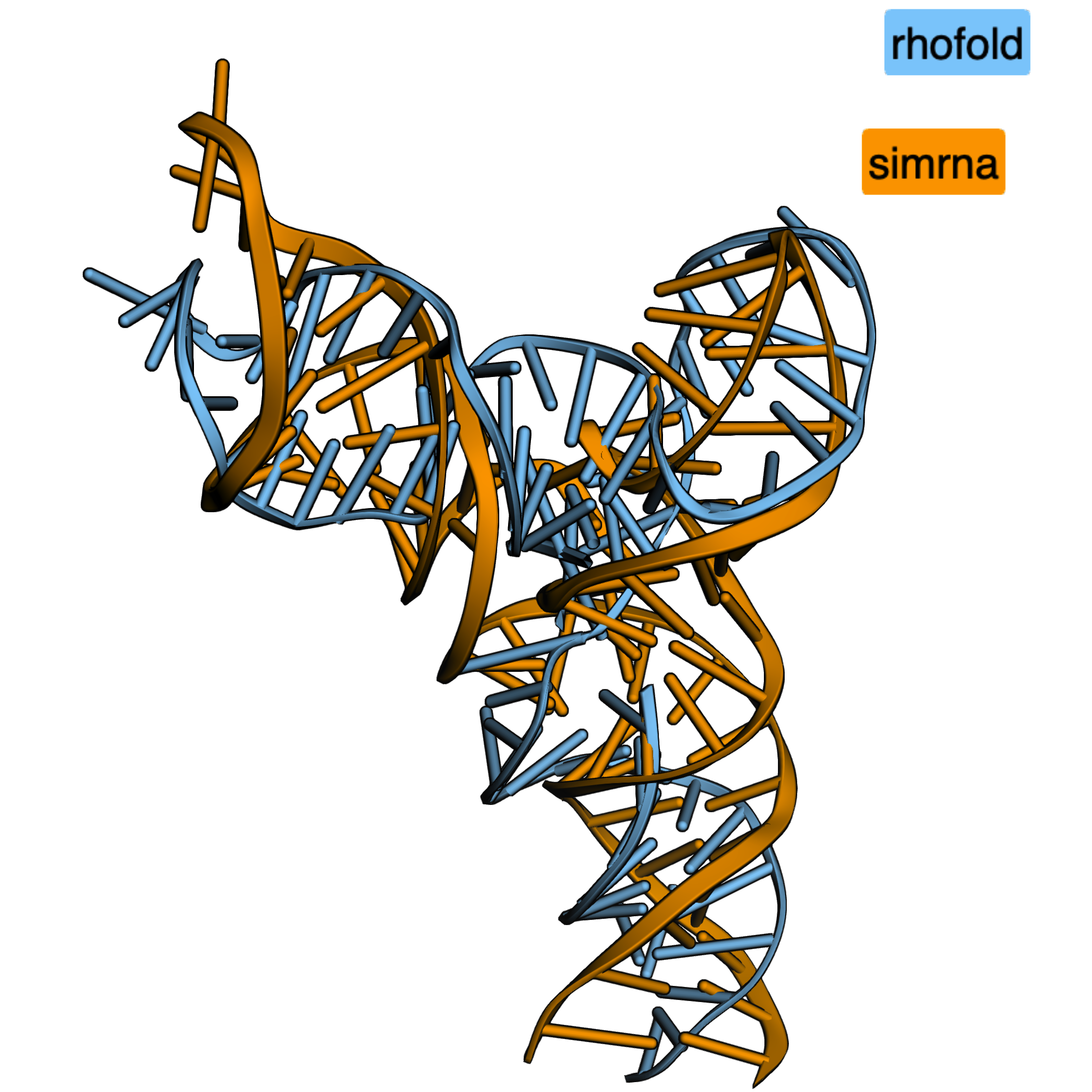}
        \caption{RhoFold vs. SimRNA}
        \label{fig:rhofold-simrna}
    \end{subfigure}
    \hfill
    \begin{subfigure}[b]{0.32\linewidth}
        \centering
        \includegraphics[width=\linewidth]{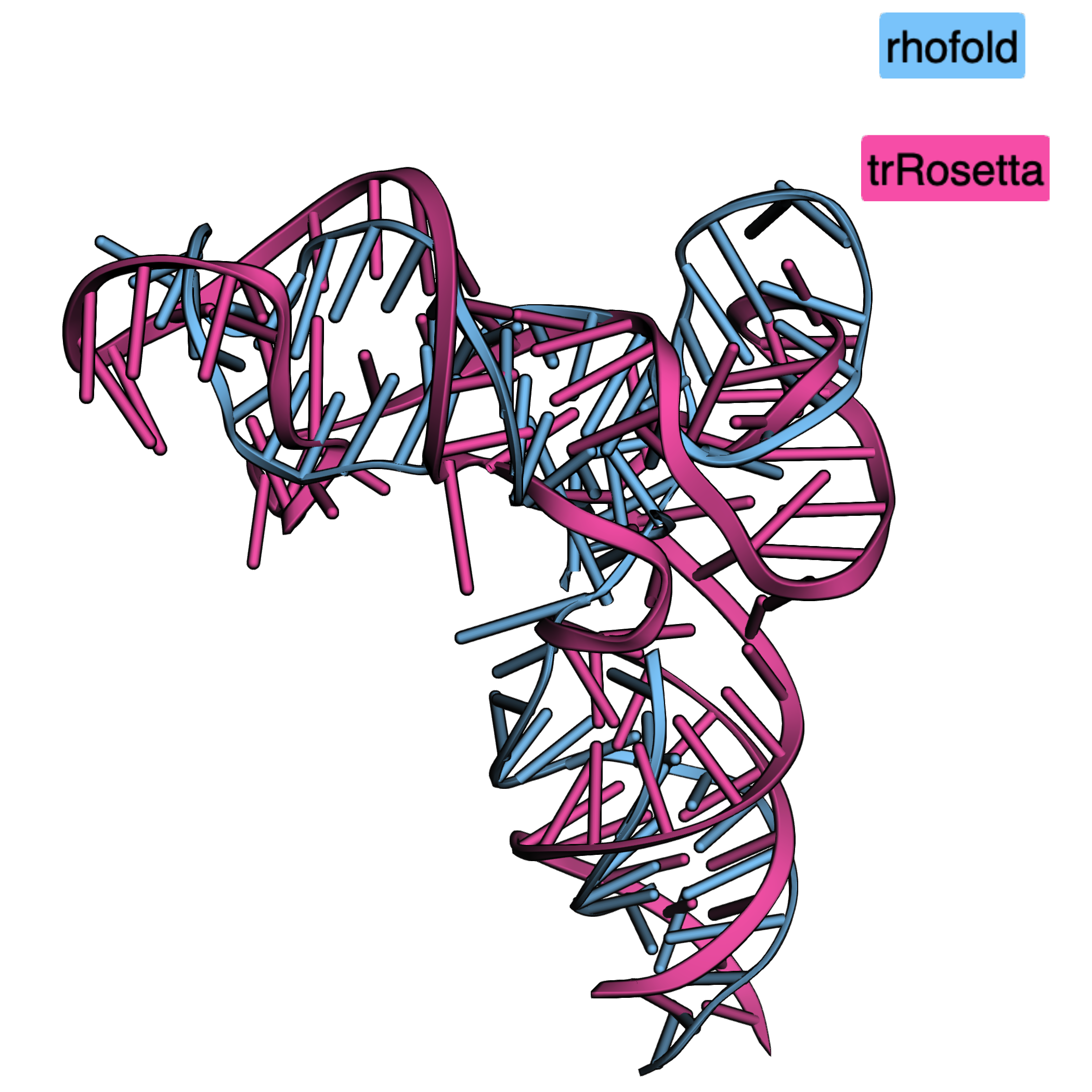}
        \caption{RhoFold vs. trRosetta}
        \label{fig:rhofold-trrosseta}
    \end{subfigure}
    \caption{Comparison of RhoFold against other 3D structure prediction tools on an example sequence from Tc-Riboswitches dataset.}
    \label{fig:rhofold-comparisons}
\end{figure}

In this section, we demonstrate the significant differences in 3D structures predicted by various tools. We compare the 3D structure obtained from RhoFold~\citep{shen2022e2efold}, which serves as our default method, with those predicted by RNAComposer~\citep{xu2014vfold}, trRosetta~\citep{baek2024accurate}, and SimRNA~\citep{boniecki2016simrna}. Each structure is visualized side by side with RhoFold in Figure~\ref{fig:rhofold-comparisons} to facilitate a more intuitive comparison. As observed, these structures predicted by each tool vary considerably.

\section{Reproduction}
\label{appen:reproduction}
This section outlines the necessary details to reproduce all experiments discussed in this paper. The code will be made publicly available upon acceptance.

\subsection{Training details}
All experiments were conducted on a single NVIDIA A100 GPU. For each baseline, hyperparameters were optimized using Optuna~\citep{akiba2019optuna}, restricting the search to models with fewer than 10 million parameters that fit within the memory constraints of an 80GB NVIDIA A100 GPU. Most baseline implementations were sourced from PyTorch Geometric~\citep{Fey/Lenssen/2019}. The Transformer1D model was adapted to Transformer1D2D as detailed in the paper. For EGNN, we utilized the authors' implementation~\citep{satorras2021n}, and for SchNet, the implementation from~\citep{joshi2023expressive} was used.

\subsection{Hyperparameters}
This section provides a comprehensive overview of the hyperparameters used in each baseline model, facilitating reproducibility and understanding of the model configurations.

Common hyperparameters across these models include \texttt{in\_channels}, which specifies the number of input features, \texttt{hidden}, which determines the number of hidden units in each hidden layer, and \texttt{out\_channels}, which defines the number of output features. The \texttt{L} parameter controls the number of layers in the network, and the \texttt{dropout} parameter sets the dropout rate for regularization. The \texttt{lr} parameter specifies the learning rate, and \texttt{weight\_decay} sets the weight decay for regularization of the optimizer. For graph-level tasks, the \texttt{pool} parameter specifies the pooling method, which can be \texttt{mean}, \texttt{max}, or \texttt{add}.

\setlength{\parskip}{-0.5pt}
\paragraph{Transformer1D} is a standard Transformer architecture for RNA sequence processing. It includes an embedding layer to convert input tokens into dense vectors, positional encoding (PE) to retain sequence order and a multi-layer Transformer encoder to capture complex dependencies within the sequence. There are some hyperparameters from the original transformer~\citep{vaswani2017attention}. \texttt{nhead}, which defines the number of attention heads in each Transformer layer; \texttt{num\_encoder\_layers}, which controls the number of encoder layers in the Transformer; \texttt{d\_model}, which determines the dimensionality of the embeddings and the model; \texttt{dim\_feedforward}, which sets the dimensionality of the feedforward network model. To shrink the search space, we set \texttt{d\_model} and \texttt{dim\_feedforward} as the same with a new hyperparameter \texttt{hidden}.

\paragraph{Transformer1D2D} is an adaptation of Transformer1D that integrates both sequence and 2D graph structure information. In addition to encoding each nucleotide, the model incorporates base pair probabilities (BPP) features for each nucleotide. It combines a standard Transformer with positional encoding and a convolutional layer applied to the graph adjacency matrix. This convolutional output is added to the Transformer’s attention matrix, allowing the model to incorporate graph structure into its attention mechanism. This design captures both the sequential and structural dependencies in RNA data, improving predictive performance. The unique hyperparameter for this model is \texttt{kernel\_size}, which specifies the size of the convolutional kernel.

\paragraph{GAT} includes the unique hyperparameters \texttt{gat\_heads}, which specify the number of attention heads in each GAT layer.

\paragraph{ChebNet} model has the unique hyperparameter \texttt{power}, which specifies the polynomial order for the Chebyshev convolution.

\paragraph{GraphGPS} and \textbf{Graph Transformer} includes \texttt{heads}, which specifies the number of attention heads in each layer, and \texttt{pe\_dim}, which defines the dimensionality of positional encoding.

\paragraph{EGNN} and \textbf{SchNet} are 3D models that operate at two granularities within the network: atom layers and nucleotide layers. The two types of layers are connected through nucleotide pooling. Atom layers use atoms as nodes, while nucleotide layers use nucleotides as nodes. Both the atom layer and nucleotide layer employ a point cloud setting and calculate edges based on the distance between two nodes. An edge is considered to exist if the distance is smaller than a certain threshold. Therefore, EGNN and SchNet share the following hyperparameters: \texttt{L\_atom}, which denotes the number of atom layers; \texttt{L\_nt}, which specifies the number of nucleotide layers; \texttt{threshold\_atom}, which is the threshold for edges in atom layers; and \texttt{threshold\_nt}, which is the threshold for edges in nucleotide layers.

For SchNet, the unique hyperparameters include \texttt{num\_filters}, which refers to the number of filters used in convolutional layers, and \texttt{num\_gaussians}, which indicates the number of Gaussian functions used for radial filters. For a more detailed explanation of these hyperparameters, please refer to \citep{schutt2017schnet}.

To ensure a fair comparison, the best hyperparameter configuration for each method was selected based on validation set performance. We report the mean performance and standard deviation across 5 random splits on the test set. For the COVID and Ribonanza datasets, we performed hyperparameter searching only on the COVID dataset and applied the same configuration to Ribonanza, as the two datasets share similar properties. The optimal hyperparameters are shown in Table~\ref{tab:optimal_hyperparameter}.

\begin{table}[]
\centering
\resizebox{0.85\columnwidth}{!}{%
\begin{tabular}{cc|ccc}
\toprule
                          & \textbf{Hyperparameter}       & \textbf{COVID \& Ribonanza} & \textbf{Tc-riboswitches} & \textbf{Fungal} \\ \midrule
\multirow{6}{*}{Transformer1D}                                               & lr & 0.001  & 0.0005 & 0.001                \\
                          & weight\_decay        & 0     & 0.0005  & 0.0005 \\
                          & hidden               & 128   & 64      & 32     \\
                          & nhead                & 8     & 8       & 8      \\
                          & num\_encoder\_layers & 8     & 6       & 6      \\
                          & pool                 & /     & mean    & mean   \\ \midrule
\multirow{7}{*}{Transformer1D2D}                                             & lr & 0.001  & 0.005  & \multirow{7}{*}{OOM} \\
                          & weight\_decay        & 0     & 0       &        \\
                          & hidden               & 256   & 32      &        \\
                          & nhead                & 16    & 4       &        \\
                          & num\_encoder\_layers & 8     & 4       &        \\
                          & pool                 & mean  & mean    &        \\
                          & kernel\_size         & 5     & 5       &        \\ \midrule
\multirow{6}{*}{GCN}      & lr                   & 0.001 & 0.0001  & 0.0001 \\
                          & weight\_decay        & 0     & 0       & 0      \\
                          & hidden               & 1024  & 512     & 512    \\
                          & L                    & 7     & 5       & 3      \\
                          & dropout              & 0.3   & 0.1     & 0.7    \\
                          & pool                 & /     & max     & add    \\ \midrule
\multirow{7}{*}{ChebNet}  & lr                   & 0.001 & 0.005   & 0.0001 \\
                          & weight\_decay        & 0     & 0.0005  & 0      \\
                          & hidden               & 512   & 256     & 256    \\
                          & L                    & 5     & 7       & 5      \\
                          & dropout              & 0.3   & 0.3     & 0.3    \\
                          & power                & 6     & 2       & 2      \\
                          & pool                 & /     & max     & mean   \\ \midrule
\multirow{7}{*}{GAT}      & lr                   & 0.001 & 0.005   & 0.0005 \\
                          & weight\_decay        & 0     & 0       & 0.0005 \\
                          & hidden               & 256   & 1024    & 256    \\
                          & L                    & 7     & 3       & 7      \\
                          & dropout              & 0.1   & 0.1     & 0.3    \\
                          & heads                & 4     & 2       & 1      \\
                          & pool                 & /     & add     & add    \\ \midrule
\multirow{6}{*}{\begin{tabular}[c]{@{}c@{}}Graph\\ Transformer\end{tabular}} & lr & 0.001  & 0.005  & 0.005                \\
                          & weight\_decay        & 0     & 0       & 0.0005 \\
                          & hidden               & 128   & 64      & 256    \\
                          & L                    & 7     & 5       & 7      \\
                          & heads                & 4     & 1       & 2      \\
                          & pool                 & /     & add     & mean   \\ \midrule
\multirow{6}{*}{GraphGPS} & lr                   & 0.001 & 1e-5    & 0.0005 \\
                          & weight\_decay        & 0     & 0.0005  & 0.0005 \\
                          & hidden               & 256   & 256     & 512    \\
                          & L                    & 5     & 7       & 5      \\
                          & heads                & 2     & 1       & 2      \\
                          & pool                 & /     & add     & max    \\ \midrule
\multirow{7}{*}{EGNN}                                                        & lr & 0.0005 & 0.001  & \multirow{7}{*}{OOM} \\
                          & weight\_decay        & 0     & 0       &        \\
                          & hidden               & 256   & 256     &        \\
                          & L\_atom              & 3     & 3       &        \\
                          & L\_nt                & 2     & 1       &        \\
                          & threshold\_atom      & 1.6   & 1.6     &        \\
                          & threshold\_nt        & 22    & 22      &        \\ \midrule
\multirow{9}{*}{SchNet}                                                      & lr & 0.0005 & 0.001  & \multirow{9}{*}{OOM} \\
                          & weight\_decay        & 0     & 0       &        \\
                          & hidden               & 128   & 128     &        \\
                          & L\_atom              & 1     & 1       &        \\
                          & L\_nt                & 2     & 4       &        \\
                          & threshold\_atom      & 1.6   & 1.8     &        \\
                          & threshold\_nt        & 44    & 88      &        \\
                          & num\_filters         & 128   & 256     &        \\
                          & num\_gaussians       & 50    & 50      &        \\
\bottomrule
\end{tabular}%
}
\caption{Best hyperparameters for each model and dataset. Hyperparameters are searched by Optuna. COVID and Ribonanza share the same hyperparameters.}
\label{tab:optimal_hyperparameter}
\end{table}

\section{Additional Results}
\label{appen:detailed_data}

In this section, we present the additional results supporting Figures~\ref{fig:train_fraction}, \ref{fig:label-ratio-efficiency}, \ref{fig:robustness}, and \ref{fig:generalization} in main text.

\subsection{Impact of data availability}
\label{app_sec:data_availability}
The detailed results of partial training data from Figure~\ref{fig:train_fraction} are shown in Tables~\ref{tab:covid_partial_training}, \ref{tab:ribonanza_partial_training}, and \ref{tab:fungal_partial_training}.

\begin{table}[h]
\caption{Performance (MCRMSE) of different models across various training data (mean $\pm$ standard deviation) fractions on COVID dataset.}
\label{tab:covid_partial_training}
\centering
\begin{tabular}{lcccc}
\toprule
\textbf{COVID}& \textbf{0.25} & \textbf{0.50} & \textbf{0.75} & \textbf{1.00} \\
\midrule
Transformer1D & 0.429 $\pm$ 0.018 & 0.410 $\pm$ 0.016 & 0.375 $\pm$ 0.018 & 0.361 $\pm$ 0.017 \\
Transformer1D2D & 0.345 $\pm$ 0.010 & 0.330 $\pm$ 0.011 & 0.306 $\pm$ 0.014 & 0.305 $\pm$ 0.012 \\
GCN & 0.389 $\pm$ 0.008 & 0.377 $\pm$ 0.009 & 0.358 $\pm$ 0.014 & 0.359 $\pm$ 0.009 \\
GAT & 0.352 $\pm$ 0.011 & 0.342 $\pm$ 0.009 & 0.320 $\pm$ 0.014 & 0.315 $\pm$ 0.006 \\
ChebNet & 0.320 $\pm$ 0.011 & 0.309 $\pm$ 0.009 & 0.286 $\pm$ 0.018 & 0.279 $\pm$ 0.007 \\
Graph Transformer & 0.356 $\pm$ 0.008 & 0.344 $\pm$ 0.007 & 0.324 $\pm$ 0.016 & 0.318 $\pm$ 0.008 \\
GraphGPS & 0.367 $\pm$ 0.018 & 0.362 $\pm$ 0.005 & 0.344 $\pm$ 0.010 & 0.332 $\pm$ 0.013 \\
EGNN & 0.398 $\pm$ 0.001 & 0.391 $\pm$ 0.013 & 0.368 $\pm$ 0.013 & 0.364 $\pm$ 0.003 \\
SchNet & 0.419 $\pm$ 0.003 & 0.414 $\pm$ 0.011 & 0.392 $\pm$ 0.011 & 0.390 $\pm$ 0.006 \\
FastEGNN & 0.460 $\pm$ 0.018 & 0.452 $\pm$ 0.015 & 0.443 $\pm$ 0.016 & 0.444 $\pm$ 0.003  \\

\bottomrule
\end{tabular}
\end{table}

\begin{table}[h]
\caption{Performance (MCRMSE) of different models across various fractions of training data (mean $\pm$ standard deviation) on Ribonanza dataset.}
\label{tab:ribonanza_partial_training}
\centering
\begin{tabular}{lcccc}
\toprule
\textbf{Ribonanza} & \textbf{0.25} & \textbf{0.50} & \textbf{0.75} & \textbf{1.00} \\
\midrule
Transformer1D & 0.777 $\pm$ 0.014 & 0.740 $\pm$ 0.005 & 0.739 $\pm$ 0.001 & 0.705 $\pm$ 0.015 \\
Transformer1D2D & 0.630 $\pm$ 0.016 & 0.553 $\pm$ 0.015 & 0.541 $\pm$ 0.018 & 0.514 $\pm$ 0.004 \\
GCN & 0.668 $\pm$ 0.018 & 0.618 $\pm$ 0.017 & 0.612 $\pm$ 0.013 & 0.595 $\pm$ 0.006 \\
GAT & 0.600 $\pm$ 0.018 & 0.553 $\pm$ 0.026 & 0.544 $\pm$ 0.012 & 0.534 $\pm$ 0.006 \\
ChebNet & 0.537 $\pm$ 0.019 & 0.494 $\pm$ 0.022 & 0.499 $\pm$ 0.007 & 0.468 $\pm$ 0.002 \\
Graph Transformer & 0.567 $\pm$ 0.019 & 0.529 $\pm$ 0.013 & 0.529 $\pm$ 0.010 & 0.515 $\pm$ 0.001 \\
GraphGPS & 0.581 $\pm$ 0.021 & 0.529 $\pm$ 0.015 & 0.540 $\pm$ 0.004 & 0.523 $\pm$ 0.003 \\
EGNN & 0.694 $\pm$ 0.010 & 0.650 $\pm$ 0.010 & 0.632 $\pm$ 0.015 & 0.619 $\pm$ 0.007 \\
SchNet & 0.768 $\pm$ 0.008 & 0.724 $\pm$ 0.013 & 0.715 $\pm$ 0.015 & 0.685 $\pm$ 0.006 \\
FastEGNN & 0.795 $\pm$ 0.007 & 0.788 $\pm$ 0.002 & 0.774 $\pm$ 0.016 & 0.753 $\pm$ 0.015 \\
\bottomrule
\end{tabular}
\end{table}


\begin{table}[h]
\caption{Performance (MCRMSE) of different models across various fractions of training data (mean $\pm$ standard deviation) on Fungal dataset.}
\label{tab:fungal_partial_training}
\centering
\begin{tabular}{lcccc}
\toprule
\textbf{Fungal} & \textbf{0.25} & \textbf{0.50} & \textbf{0.75} & \textbf{1.00} \\
\midrule
Transformer1D & 1.510 $\pm$ 0.006 & 1.446 $\pm$ 0.014 & 1.475 $\pm$ 0.035 & 1.417 $\pm$ 0.005 \\
GCN & 1.243 $\pm$ 0.064 & 1.244 $\pm$ 0.128 & 1.151 $\pm$ 0.077 & 1.192 $\pm$ 0.077 \\
GAT & 1.211 $\pm$ 0.125 & 1.168 $\pm$ 0.033 & 1.146 $\pm$ 0.109 & 1.112 $\pm$ 0.035 \\
ChebNet & 1.125 $\pm$ 0.097 & 1.011 $\pm$ 0.010 & 1.008 $\pm$ 0.009 & 0.973 $\pm$ 0.003 \\
Graph Transformer & 1.415 $\pm$ 0.014 & 1.331 $\pm$ 0.163 & 1.306 $\pm$ 0.127 & 1.317 $\pm$ 0.002 \\
GraphGPS & 1.289 $\pm$ 0.071 & 1.377 $\pm$ 0.127 & 1.357 $\pm$ 0.106 & 1.025 $\pm$ 0.081 \\
\bottomrule
\end{tabular}
\end{table}

\subsection{Impact of partial labeling}
\label{app_sec:partial_labeling}
The detailed results of the partial labeling sequence from Figure~\ref{fig:label-ratio-efficiency} are shown in Tables~\ref{tab:covid_partial_label} and \ref{tab:ribonanza_partial_label}.

\begin{table}[h]
\caption{Performance (MCRMSE) of different models across various fractions of sequence labeling (mean $\pm$ standard deviation) on COVID dataset.}
\label{tab:covid_partial_label}
\centering
\resizebox{\textwidth}{!}{%
\begin{tabular}{lccccc}
\toprule
\textbf{COVID} & \textbf{0.2} & \textbf{0.4} & \textbf{0.6} & \textbf{0.8} & \textbf{1.0} \\
\midrule
Transformer1D & 0.654 $\pm$ 0.040 & 0.559 $\pm$ 0.011 & 0.480 $\pm$ 0.004 & 0.429 $\pm$ 0.034 & 0.361 $\pm$ 0.017 \\
Transformer1D2D & 0.502 $\pm$ 0.002 & 0.470 $\pm$ 0.052 & 0.374 $\pm$ 0.007 & 0.325 $\pm$ 0.006 & 0.305 $\pm$ 0.012 \\
GCN & 0.450 $\pm$ 0.012 & 0.416 $\pm$ 0.012 & 0.397 $\pm$ 0.012 & 0.378 $\pm$ 0.011 & 0.359 $\pm$ 0.009 \\
GAT & 0.411 $\pm$ 0.010 & 0.376 $\pm$ 0.012 & 0.360 $\pm$ 0.012 & 0.336 $\pm$ 0.009 & 0.315 $\pm$ 0.006 \\
ChebNet & 0.380 $\pm$ 0.007 & 0.344 $\pm$ 0.008 & 0.325 $\pm$ 0.009 & 0.299 $\pm$ 0.007 & 0.279 $\pm$ 0.007 \\
Graph Transformer & 0.415 $\pm$ 0.012 & 0.379 $\pm$ 0.011 & 0.362 $\pm$ 0.011 & 0.338 $\pm$ 0.004 & 0.318 $\pm$ 0.008 \\
GraphGPS & 0.428 $\pm$ 0.015 & 0.400 $\pm$ 0.017 & 0.376 $\pm$ 0.013 & 0.351 $\pm$ 0.007 & 0.332 $\pm$ 0.013 \\
EGNN & 0.436 $\pm$ 0.014 & 0.421 $\pm$ 0.010 & 0.407 $\pm$ 0.004 & 0.385 $\pm$ 0.006 & 0.364 $\pm$ 0.003 \\
SchNet & 0.442 $\pm$ 0.004 & 0.429 $\pm$ 0.005 & 0.413 $\pm$ 0.001 & 0.407 $\pm$ 0.005 & 0.390 $\pm$ 0.006 \\
FastEGNN & 0.497 $\pm$ 0.004 & 0.490 $\pm$ 0.007 & 0.466 $\pm$ 0.007 & 0.469 $\pm$ 0.009 & 0.444 $\pm$ 0.003 \\ 
\bottomrule
\end{tabular}}
\end{table}

\begin{table}[h]
\caption{Performance (MCRMSE) of different models across various fractions of sequence labeling (mean $\pm$ standard deviation) on Ribonanza dataset.}
\label{tab:ribonanza_partial_label}
\centering
\resizebox{\textwidth}{!}{%
\begin{tabular}{lccccc}
\toprule
\textbf{Ribonanza} & \textbf{0.2} & \textbf{0.4} & \textbf{0.6} & \textbf{0.8} & \textbf{1.0} \\
\midrule
Transformer1D & 1.137 $\pm$ 0.163 & 0.929 $\pm$ 0.023 & 0.823 $\pm$ 0.018 & 0.742 $\pm$ 0.013 & 0.705 $\pm$ 0.015 \\
Transformer1D2D & 0.859 $\pm$ 0.025 & 0.638 $\pm$ 0.013 & 0.632 $\pm$ 0.028 & 0.568 $\pm$ 0.013 & 0.514 $\pm$ 0.004 \\
GCN & 1.191 $\pm$ 0.031 & 1.026 $\pm$ 0.079 & 1.111 $\pm$ 0.206 & 1.070 $\pm$ 0.137 & 0.595 $\pm$ 0.006 \\
GAT & 0.703 $\pm$ 0.015 & 0.632 $\pm$ 0.025 & 0.612 $\pm$ 0.030 & 0.560 $\pm$ 0.010 & 0.534 $\pm$ 0.006 \\
ChebNet & 0.614 $\pm$ 0.013 & 0.546 $\pm$ 0.008 & 0.540 $\pm$ 0.008 & 0.514 $\pm$ 0.006 & 0.468 $\pm$ 0.002 \\
Graph Transformer & 0.719 $\pm$ 0.043 & 0.607 $\pm$ 0.020 & 0.584 $\pm$ 0.015 & 0.552 $\pm$ 0.013 & 0.515 $\pm$ 0.001 \\
GraphGPS & 0.743 $\pm$ 0.058 & 0.663 $\pm$ 0.026 & 0.627 $\pm$ 0.024 & 0.651 $\pm$ 0.026 & 0.523 $\pm$ 0.003 \\
EGNN & 0.882 $\pm$ 0.010 & 0.722 $\pm$ 0.021 & 0.687 $\pm$ 0.008 & 0.665 $\pm$ 0.013 & 0.619 $\pm$ 0.007 \\
SchNet & 0.810 $\pm$ 0.002 & 0.781 $\pm$ 0.009 & 0.750 $\pm$ 0.009 & 0.725 $\pm$ 0.004 & 0.685 $\pm$ 0.006 \\
FastEGNN & 1.223 $\pm$ 0.008 & 0.929 $\pm$ 0.008 & 0.860 $\pm$ 0.004 & 0.837 $\pm$ 0.004 & 0.753 $\pm$0.015 \\
\bottomrule
\end{tabular}}
\end{table}

\subsection{Robustness to sequencing noise}
\label{app_sec:robustness}
The results of the robustness experiment from Figure~\ref{fig:robustness} are shown in Tables~\ref{tab:robustness_covid}, \ref{tab:robustness_ribonanza}, \ref{tab:robustness_fungal}, and \ref{tab:robustness_tcribo}.

\begin{table}[h]
\caption{Performance (MCRMSE) of various models in \textbf{robustness} experiments on the \textbf{COVID} dataset (mean ± standard deviation).}
\label{tab:robustness_covid}
\centering
\resizebox{\columnwidth}{!}{%
\begin{tabular}{lccccccc}
\toprule
\textbf{Model} & \textbf{0.00} & \textbf{0.05} & \textbf{0.10} & \textbf{0.15} & \textbf{0.20} & \textbf{0.25} & \textbf{0.30} \\
\midrule
Transformer1D & 0.361 $\pm$ 0.017 & 0.386 $\pm$ 0.015 & 0.400 $\pm$ 0.010 & 0.409 $\pm$ 0.006 & 0.428 $\pm$ 0.005 & 0.435 $\pm$ 0.003 & 0.449 $\pm$ 0.011 \\
Transformer1D2D & 0.305 $\pm$ 0.012 & 0.373 $\pm$ 0.007 & 0.403 $\pm$ 0.007 & 0.428 $\pm$ 0.011 & 0.444 $\pm$ 0.015 & 0.457 $\pm$ 0.009 & 0.463 $\pm$ 0.009 \\
GCN & 0.359 $\pm$ 0.009 & 0.436 $\pm$ 0.009 & 0.464 $\pm$ 0.011 & 0.481 $\pm$ 0.010 & 0.491 $\pm$ 0.012 & 0.497 $\pm$ 0.009 & 0.501 $\pm$ 0.009 \\
GAT & 0.315 $\pm$ 0.006 & 0.409 $\pm$ 0.009 & 0.448 $\pm$ 0.011 & 0.471 $\pm$ 0.010 & 0.484 $\pm$ 0.012 & 0.494 $\pm$ 0.011 & 0.500 $\pm$ 0.010 \\
ChebNet & 0.279 $\pm$ 0.007 & 0.368 $\pm$ 0.003 & 0.423 $\pm$ 0.009 & 0.456 $\pm$ 0.007 & 0.471 $\pm$ 0.009 & 0.481 $\pm$ 0.010 & 0.487 $\pm$ 0.008 \\
Graph Transformer & 0.318 $\pm$ 0.008 & 0.403 $\pm$ 0.008 & 0.441 $\pm$ 0.012 & 0.467 $\pm$ 0.011 & 0.480 $\pm$ 0.012 & 0.487 $\pm$ 0.010 & 0.494 $\pm$ 0.011 \\
GraphGPS & 0.332 $\pm$ 0.013 & 0.408 $\pm$ 0.012 & 0.441 $\pm$ 0.010 & 0.464 $\pm$ 0.014 & 0.475 $\pm$ 0.012 & 0.484 $\pm$ 0.012 & 0.487 $\pm$ 0.008 \\
EGNN & 0.364 $\pm$ 0.003 & 0.432 $\pm$ 0.012 & 0.467 $\pm$ 0.009 & 0.486 $\pm$ 0.009 & 0.499 $\pm$ 0.011 & 0.505 $\pm$ 0.012 & 0.511 $\pm$ 0.011 \\
SchNet & 0.390 $\pm$ 0.006 & 0.447 $\pm$ 0.012 & 0.477 $\pm$ 0.011 & 0.496 $\pm$ 0.009 & 0.507 $\pm$ 0.014 & 0.513 $\pm$ 0.012 & 0.517 $\pm$ 0.010 \\
FastEGNN & 0.444 $\pm$ 0.003 & 0.49283 $\pm$ 0.008 & 0.516 $\pm$ 0.005 & 0.516 $\pm$ 0.004 & 0.522 $\pm$ 0.002 & 0.527 $\pm$ 0.003 & 0.540 $\pm$ 0.004 \\
\bottomrule
\end{tabular}}
\end{table}

\begin{table}[h]
\caption{Performance (MCRMSE) of various models in \textbf{robustness} experiments on the \textbf{Ribonanza} dataset (mean ± standard deviation).}
\label{tab:robustness_ribonanza}
\centering
\resizebox{\columnwidth}{!}{%
\begin{tabular}{lccccccc}
\toprule
\textbf{Model} & \textbf{0.00} & \textbf{0.05} & \textbf{0.10} & \textbf{0.15} & \textbf{0.20} & \textbf{0.25} & \textbf{0.30} \\
\midrule
Transformer1D & 0.705 $\pm$ 0.015 & 0.733 $\pm$ 0.010 & 0.769 $\pm$ 0.014 & 0.782 $\pm$ 0.005 & 0.794 $\pm$ 0.010 & 0.805 $\pm$ 0.017 & 0.823 $\pm$ 0.005 \\
Transformer1D2D & 0.514 $\pm$ 0.004 & 0.635 $\pm$ 0.004 & 0.714 $\pm$ 0.014 & 0.763 $\pm$ 0.008 & 0.790 $\pm$ 0.009 & 0.811 $\pm$ 0.014 & 0.830 $\pm$ 0.008 \\
GCN & 0.595 $\pm$ 0.006 & 0.750 $\pm$ 0.014 & 0.846 $\pm$ 0.008 & 0.893 $\pm$ 0.003 & 0.912 $\pm$ 0.005 & 0.924 $\pm$ 0.005 & 0.929 $\pm$ 0.005 \\
GAT & 0.534 $\pm$ 0.006 & 0.691 $\pm$ 0.015 & 0.785 $\pm$ 0.006 & 0.850 $\pm$ 0.003 & 0.877 $\pm$ 0.001 & 0.904 $\pm$ 0.007 & 0.915 $\pm$ 0.007 \\
ChebNet & 0.468 $\pm$ 0.002 & 0.611 $\pm$ 0.012 & 0.720 $\pm$ 0.006 & 0.802 $\pm$ 0.011 & 0.841 $\pm$ 0.003 & 0.876 $\pm$ 0.007 & 0.897 $\pm$ 0.008 \\
Graph Transformer & 0.515 $\pm$ 0.001 & 0.670 $\pm$ 0.011 & 0.768 $\pm$ 0.011 & 0.833 $\pm$ 0.008 & 0.870 $\pm$ 0.006 & 0.893 $\pm$ 0.010 & 0.908 $\pm$ 0.007 \\
GraphGPS & 0.523 $\pm$ 0.003 & 0.677 $\pm$ 0.017 & 0.772 $\pm$ 0.006 & 0.832 $\pm$ 0.006 & 0.872 $\pm$ 0.004 & 0.896 $\pm$ 0.011 & 0.912 $\pm$ 0.006 \\
EGNN & 0.619 $\pm$ 0.007 & 0.764 $\pm$ 0.003 & 0.847 $\pm$ 0.003 & 0.889 $\pm$ 0.005 & 0.904 $\pm$ 0.003 & 0.917 $\pm$ 0.000 & 0.922 $\pm$ 0.002 \\
SchNet & 0.685 $\pm$ 0.006 & 0.814 $\pm$ 0.006 & 0.873 $\pm$ 0.004 & 0.897 $\pm$ 0.004 & 0.908 $\pm$ 0.004 & 0.918 $\pm$ 0.005 & 0.922 $\pm$ 0.005 \\
FastEGNN & 0.753 $\pm$ 0.015 & 0.857 $\pm$ 0.001 & 0.884 $\pm$ 0.005 & 0.908 $\pm$ 0.001 & 0.914 $\pm$ 0.004 & 0.920 $\pm$ 0.003 & 0.922 $\pm$ 0.003 \\
\bottomrule
\end{tabular}
}
\end{table}

\begin{table}[h]
\caption{Performance (MCRMSE) of various models in \textbf{robustness} experiments on the \textbf{Fungal} dataset (mean ± standard deviation).}
\label{tab:robustness_fungal}
\centering
\resizebox{\columnwidth}{!}{%
\begin{tabular}{lccccccc}
\toprule
\textbf{Model} & \textbf{0.00} & \textbf{0.05} & \textbf{0.10} & \textbf{0.15} & \textbf{0.20} & \textbf{0.25} & \textbf{0.30} \\
\midrule
Transformer1D & 1.417 $\pm$ 0.005 & 1.545 $\pm$ 0.045 & 1.546 $\pm$ 0.044 & 1.546 $\pm$ 0.046 & 1.543 $\pm$ 0.048 & 1.543 $\pm$ 0.051 & 1.550 $\pm$ 0.041 \\
GCN & 1.192 $\pm$ 0.077 & 1.222 $\pm$ 0.044 & 1.255 $\pm$ 0.002 & 1.277 $\pm$ 0.014 & 1.269 $\pm$ 0.011 & 1.328 $\pm$ 0.026 & 1.294 $\pm$ 0.025 \\
GAT & 1.112 $\pm$ 0.035 & 1.244 $\pm$ 0.074 & 1.391 $\pm$ 0.155 & 1.334 $\pm$ 0.099 & 1.468 $\pm$ 0.056 & 1.444 $\pm$ 0.094 & 1.446 $\pm$ 0.092 \\
ChebNet & 0.978 $\pm$ 0.000 & 1.031 $\pm$ 0.003 & 1.091 $\pm$ 0.007 & 1.108 $\pm$ 0.009 & 1.243 $\pm$ 0.005 & 1.210 $\pm$ 0.014 & 1.269 $\pm$ 0.007 \\
Graph Transformer & 1.342 $\pm$ 0.087 & 1.267 $\pm$ 0.116 & 1.409 $\pm$ 0.046 & 1.426 $\pm$ 0.051 & 1.442 $\pm$ 0.038 & 1.413 $\pm$ 0.020 & 1.450 $\pm$ 0.018 \\
GraphGPS & 1.083 $\pm$ 0.131 & 1.048 $\pm$ 0.095 & 1.133 $\pm$ 0.057 & 1.109 $\pm$ 0.040 & 1.173 $\pm$ 0.071 & 1.256 $\pm$ 0.016 & 1.328 $\pm$ 0.041 \\
\bottomrule
\end{tabular}
}
\end{table}

\begin{table}[h]
\caption{Performance (MCRMSE) of various models in \textbf{robustness} experiments on the \textbf{Tc-riboswitches} dataset (mean ± standard deviation).}
\label{tab:robustness_tcribo}
\centering
\resizebox{\columnwidth}{!}{%
\begin{tabular}{lccccccc}
\toprule
\textbf{Model} & \textbf{0.00} & \textbf{0.05} & \textbf{0.10} & \textbf{0.15} & \textbf{0.20} & \textbf{0.25} & \textbf{0.30} \\
\midrule
Transformer1D & 0.705 $\pm$ 0.079 & 0.698 $\pm$ 0.071 & 0.736 $\pm$ 0.004 & 0.672 $\pm$ 0.003 & 0.739 $\pm$ 0.008 & 0.694 $\pm$ 0.011 & 0.675 $\pm$ 0.047 \\
Transformer1D2D & 0.633 $\pm$ 0.001 & 0.697 $\pm$ 0.031 & 0.742 $\pm$ 0.003 & 0.708 $\pm$ 0.008 & 0.681 $\pm$ 0.001 & 0.762 $\pm$ 0.022 & 0.738 $\pm$ 0.016 \\
GCN & 0.701 $\pm$ 0.004 & 0.758 $\pm$ 0.003 & 0.747 $\pm$ 0.005 & 0.744 $\pm$ 0.004 & 0.733 $\pm$ 0.013 & 0.740 $\pm$ 0.011 & 0.765 $\pm$ 0.009 \\
GAT & 0.685 $\pm$ 0.024 & 0.749 $\pm$ 0.017 & 0.770 $\pm$ 0.047 & 0.734 $\pm$ 0.021 & 0.737 $\pm$ 0.009 & 0.753 $\pm$ 0.001 & 0.747 $\pm$ 0.011 \\
ChebNet & 0.621 $\pm$ 0.022 & 0.766 $\pm$ 0.014 & 0.754 $\pm$ 0.021 & 0.738 $\pm$ 0.014 & 0.763 $\pm$ 0.039 & 0.778 $\pm$ 0.048 & 0.739 $\pm$ 0.004 \\
Graph Transformer & 0.703 $\pm$ 0.054 & 0.754 $\pm$ 0.005 & 0.754 $\pm$ 0.006 & 0.773 $\pm$ 0.008 & 0.810 $\pm$ 0.087 & 0.742 $\pm$ 0.004 & 0.754 $\pm$ 0.005 \\
GraphGPS & 0.702 $\pm$ 0.028 & 0.785 $\pm$ 0.053 & 0.805 $\pm$ 0.092 & 0.750 $\pm$ 0.006 & 0.755 $\pm$ 0.060 & 0.769 $\pm$ 0.031 & 1.078 $\pm$ 0.469 \\
EGNN & 0.663 $\pm$ 0.010 & 0.750 $\pm$ 0.001 & 0.739 $\pm$ 0.002 & 0.749 $\pm$ 0.005 & 0.749 $\pm$ 0.001 & 0.749 $\pm$ 0.001 & 0.756 $\pm$ 0.013 \\
SchNet & 0.655 $\pm$ 0.038 & 0.762 $\pm$ 0.005 & 0.742 $\pm$ 0.002 & 0.771 $\pm$ 0.037 & 0.746 $\pm$ 0.005 & 0.791 $\pm$ 0.016 & 0.730 $\pm$ 0.016 \\
FastEGNN & 0.710 $\pm$ 0.010 & 0.733 $\pm$ 0.007 & 0.749 $\pm$ 0.006 & 0.748 $\pm$ 0.006 & 0.752 $\pm$ 0.008 & 0.758 $\pm$ 0.017 & 0.761 $\pm$ 0.010 \\ 

\bottomrule
\end{tabular}
}
\end{table}

\subsection{Generalization to OOD data}
\label{app_sec:generalization_ood}
The results of the generalization experiment from Figure~\ref{fig:generalization} are shown in Tables~\ref{tab:generalization_covid}, \ref{tab:generalization_ribonanza}, \ref{tab:generalization_fungal}, and \ref{tab:generalization_tcribo}.

\begin{table}[h]
\caption{Performance (MCRMSE) of various models in \textbf{generalization} experiments on the \textbf{COVID} dataset (mean ± standard deviation).}
\label{tab:generalization_covid}
\centering
\resizebox{\columnwidth}{!}{%
\begin{tabular}{lccccccc}
\toprule
\textbf{Model} & \textbf{0.00} & \textbf{0.05} & \textbf{0.10} & \textbf{0.15} & \textbf{0.20} & \textbf{0.25} & \textbf{0.30} \\
\midrule
Transformer1D & 0.361 $\pm$ 0.017 & 0.382 $\pm$ 0.022 & 0.402 $\pm$ 0.018 & 0.436 $\pm$ 0.015 & 0.461 $\pm$ 0.021 & 0.478 $\pm$ 0.015 & 0.494 $\pm$ 0.016 \\
Transformer1D2D & 0.305 $\pm$ 0.012 & 0.406 $\pm$ 0.016 & 0.466 $\pm$ 0.017 & 0.513 $\pm$ 0.016 & 0.545 $\pm$ 0.027 & 0.581 $\pm$ 0.025 & 0.596 $\pm$ 0.018 \\
GCN & 0.359 $\pm$ 0.009 & 0.459 $\pm$ 0.011 & 0.508 $\pm$ 0.011 & 0.550 $\pm$ 0.014 & 0.572 $\pm$ 0.016 & 0.601 $\pm$ 0.014 & 0.612 $\pm$ 0.008 \\
GAT & 0.315 $\pm$ 0.006 & 0.437 $\pm$ 0.013 & 0.490 $\pm$ 0.013 & 0.528 $\pm$ 0.008 & 0.555 $\pm$ 0.013 & 0.580 $\pm$ 0.015 & 0.592 $\pm$ 0.012 \\
ChebNet & 0.279 $\pm$ 0.007 & 0.415 $\pm$ 0.017 & 0.483 $\pm$ 0.023 & 0.538 $\pm$ 0.025 & 0.571 $\pm$ 0.029 & 0.604 $\pm$ 0.030 & 0.621 $\pm$ 0.028 \\
Graph Transformer & 0.318 $\pm$ 0.008 & 0.449 $\pm$ 0.015 & 0.501 $\pm$ 0.018 & 0.543 $\pm$ 0.015 & 0.571 $\pm$ 0.019 & 0.596 $\pm$ 0.014 & 0.609 $\pm$ 0.014 \\
GraphGPS & 0.332 $\pm$ 0.013 & 0.443 $\pm$ 0.011 & 0.496 $\pm$ 0.006 & 0.536 $\pm$ 0.005 & 0.559 $\pm$ 0.010 & 0.586 $\pm$ 0.007 & 0.593 $\pm$ 0.005 \\
EGNN & 0.365 $\pm$ 0.011 & 0.458 $\pm$ 0.014 & 0.504 $\pm$ 0.018 & 0.530 $\pm$ 0.020 & 0.549 $\pm$ 0.021 & 0.565 $\pm$ 0.022 & 0.572 $\pm$ 0.022 \\
SchNet & 0.390 $\pm$ 0.006 & 0.457 $\pm$ 0.011 & 0.491 $\pm$ 0.008 & 0.515 $\pm$ 0.007 & 0.531 $\pm$ 0.010 & 0.543 $\pm$ 0.009 & 0.556 $\pm$ 0.002 \\
FastEGNN & 0.444 $\pm$ 0.003 & 0.491 $\pm$ 0.020 & 0.511 $\pm$ 0.014 & 0.524 $\pm$ 0.009 & 0.533 $\pm$ 0.006 & 0.541 $\pm$ 0.003 & 0.543 $\pm$ 0.001 \\
\bottomrule
\end{tabular}%
}
\end{table}

\begin{table}[h]
\caption{Performance (MCRMSE) of various models in \textbf{generalization} experiments on the \textbf{Ribonanza} dataset (mean ± standard deviation).}
\label{tab:generalization_ribonanza}
\centering
\resizebox{\columnwidth}{!}{%
\begin{tabular}{lccccccc}
\toprule
\textbf{Model} & \textbf{0.00} & \textbf{0.05} & \textbf{0.10} & \textbf{0.15} & \textbf{0.20} & \textbf{0.25} & \textbf{0.30} \\
\midrule
Transformer1D & 0.705 $\pm$ 0.015 & 0.747 $\pm$ 0.005 & 0.796 $\pm$ 0.006 & 0.828 $\pm$ 0.008 & 0.860 $\pm$ 0.013 & 0.886 $\pm$ 0.013 & 0.899 $\pm$ 0.003 \\
Transformer1D2D & 0.514 $\pm$ 0.004 & 0.685 $\pm$ 0.014 & 0.857 $\pm$ 0.008 & 0.986 $\pm$ 0.015 & 1.055 $\pm$ 0.007 & 1.142 $\pm$ 0.020 & 1.192 $\pm$ 0.034 \\
GCN & 0.595 $\pm$ 0.006 & 0.857 $\pm$ 0.018 & 0.993 $\pm$ 0.012 & 1.054 $\pm$ 0.034 & 1.094 $\pm$ 0.043 & 1.129 $\pm$ 0.061 & 1.139 $\pm$ 0.075 \\
GAT & 0.534 $\pm$ 0.006 & 0.778 $\pm$ 0.021 & 0.919 $\pm$ 0.030 & 1.003 $\pm$ 0.056 & 1.047 $\pm$ 0.073 & 1.076 $\pm$ 0.082 & 1.093 $\pm$ 0.091 \\
ChebNet & 0.468 $\pm$ 0.002 & 0.699 $\pm$ 0.005 & 0.881 $\pm$ 0.038 & 1.025 $\pm$ 0.095 & 1.083 $\pm$ 0.111 & 1.165 $\pm$ 0.185 & 1.200 $\pm$ 0.220 \\
Graph Transformer & 0.515 $\pm$ 0.001 & 0.752 $\pm$ 0.005 & 0.930 $\pm$ 0.013 & 1.067 $\pm$ 0.036 & 1.124 $\pm$ 0.033 & 1.194 $\pm$ 0.080 & 1.224 $\pm$ 0.104 \\
GraphGPS & 0.523 $\pm$ 0.003 & 0.771 $\pm$ 0.026 & 0.958 $\pm$ 0.068 & 1.087 $\pm$ 0.142 & 1.116 $\pm$ 0.195 & 1.154 $\pm$ 0.202 & 1.165 $\pm$ 0.196 \\
EGNN & 0.691 $\pm$ 0.006 & 0.815 $\pm$ 0.004 & 0.975 $\pm$ 0.026 & 1.138 $\pm$ 0.078 & 1.228 $\pm$ 0.079 & 1.350 $\pm$ 0.173 & 1.395 $\pm$ 0.187 \\
SchNet & 0.685 $\pm$ 0.006 & 0.844 $\pm$ 0.006 & 0.949 $\pm$ 0.022 & 1.068 $\pm$ 0.035 & 1.157 $\pm$ 0.069 & 1.270 $\pm$ 0.049 & 1.342 $\pm$ 0.117 \\
FastEGNN & 0.753 $\pm$ 0.015 & 0.857 $\pm$ 0.001 & 0.912 $\pm$ 0.007 & 0.939 $\pm$ 0.011 & 0.940 $\pm$ 0.010 & 0.952 $\pm$ 0.008 & 0.957 $\pm$ 0.001 \\
\bottomrule
\end{tabular}
}
\end{table}

\begin{table}[h]
\caption{Performance (MCRMSE) of various models in \textbf{generalization} experiments on the \textbf{Fungal} dataset (mean ± standard deviation).}
\label{tab:generalization_fungal}
\centering
\resizebox{\columnwidth}{!}{%
\begin{tabular}{lccccccc}
\toprule
\textbf{Model} & \textbf{0.00} & \textbf{0.05} & \textbf{0.10} & \textbf{0.15} & \textbf{0.20} & \textbf{0.25} & \textbf{0.30} \\
\midrule
Transformer1D & 1.417 $\pm$ 0.005 & 1.575 $\pm$ 0.002 & 1.575 $\pm$ 0.002 & 1.575 $\pm$ 0.002 & 1.575 $\pm$ 0.002 & 1.575 $\pm$ 0.002 & 1.575 $\pm$ 0.002 \\
GCN & 1.192 $\pm$ 0.077 & 1.230 $\pm$ 0.061 & 1.256 $\pm$ 0.050 & 1.280 $\pm$ 0.052 & 1.290 $\pm$ 0.044 & 1.328 $\pm$ 0.032 & 1.333 $\pm$ 0.046 \\
GAT & 1.112 $\pm$ 0.035 & 1.262 $\pm$ 0.115 & 1.284 $\pm$ 0.102 & 1.312 $\pm$ 0.092 & 1.334 $\pm$ 0.087 & 1.364 $\pm$ 0.080 & 1.373 $\pm$ 0.072 \\
ChebNet & 0.973 $\pm$ 0.003 & 1.118 $\pm$ 0.008 & 1.257 $\pm$ 0.018 & 1.382 $\pm$ 0.022 & 1.525 $\pm$ 0.020 & 1.686 $\pm$ 0.030 & 1.777 $\pm$ 0.033 \\
Graph Transformer & 1.317 $\pm$ 0.002 & 1.407 $\pm$ 0.053 & 1.418 $\pm$ 0.046 & 1.427 $\pm$ 0.040 & 1.439 $\pm$ 0.033 & 1.447 $\pm$ 0.029 & 1.456 $\pm$ 0.028 \\
GraphGPS & 1.025 $\pm$ 0.081 & 1.083 $\pm$ 0.011 & 1.160 $\pm$ 0.006 & 1.217 $\pm$ 0.012 & 1.316 $\pm$ 0.026 & 1.405 $\pm$ 0.040 & 1.463 $\pm$ 0.052 \\
\bottomrule
\end{tabular}
}
\end{table}

\begin{table}[ht]
\caption{Performance (MCRMSE) of various models in \textbf{generalization} experiments on the \textbf{Tc-riboswitches} dataset (mean ± standard deviation).}
\label{tab:generalization_tcribo}
\centering
\resizebox{\columnwidth}{!}{%
\begin{tabular}{lccccccc}
\toprule
\textbf{Model} & \textbf{0.00} & \textbf{0.05} & \textbf{0.10} & \textbf{0.15} & \textbf{0.20} & \textbf{0.25} & \textbf{0.30} \\
\midrule
Transformer1D & 0.705 $\pm$ 0.079 & 0.711 $\pm$ 0.038 & 0.732 $\pm$ 0.007 & 0.753 $\pm$ 0.019 & 0.815 $\pm$ 0.091 & 0.803 $\pm$ 0.062 & 0.796 $\pm$ 0.079 \\
Transformer1D2D & 0.633 $\pm$ 0.001 & 0.705 $\pm$ 0.007 & 0.745 $\pm$ 0.008 & 0.749 $\pm$ 0.017 & 0.800 $\pm$ 0.034 & 0.766 $\pm$ 0.017 & 0.754 $\pm$ 0.014 \\
GCN & 0.701 $\pm$ 0.004 & 0.774 $\pm$ 0.026 & 0.781 $\pm$ 0.051 & 0.782 $\pm$ 0.062 & 0.825 $\pm$ 0.093 & 0.845 $\pm$ 0.072 & 0.858 $\pm$ 0.126 \\
GAT & 0.685 $\pm$ 0.024 & 0.829 $\pm$ 0.074 & 0.958 $\pm$ 0.131 & 0.926 $\pm$ 0.152 & 1.037 $\pm$ 0.297 & 0.998 $\pm$ 0.213 & 1.036 $\pm$ 0.392 \\
ChebNet & 0.621 $\pm$ 0.022 & 0.824 $\pm$ 0.185 & 0.862 $\pm$ 0.251 & 0.888 $\pm$ 0.302 & 0.997 $\pm$ 0.439 & 0.983 $\pm$ 0.412 & 1.045 $\pm$ 0.514 \\
Graph Transformer & 0.710 $\pm$ 0.041 & 0.759 $\pm$ 0.035 & 0.770 $\pm$ 0.047 & 0.776 $\pm$ 0.054 & 0.815 $\pm$ 0.101 & 0.795 $\pm$ 0.085 & 0.822 $\pm$ 0.115 \\
GraphGPS & 0.715 $\pm$ 0.012 & 0.751 $\pm$ 0.016 & 0.801 $\pm$ 0.023 & 0.795 $\pm$ 0.014 & 0.833 $\pm$ 0.025 & 0.814 $\pm$ 0.022 & 0.830 $\pm$ 0.018 \\
EGNN & 0.663 $\pm$ 0.010 & 0.760 $\pm$ 0.043 & 0.898 $\pm$ 0.134 & 0.901 $\pm$ 0.123 & 0.849 $\pm$ 0.053 & 1.058 $\pm$ 0.239 & 1.192 $\pm$ 0.453 \\
SchNet & 0.655 $\pm$ 0.038 & 1.368 $\pm$ 0.449 & 1.099 $\pm$ 0.320 & 1.419 $\pm$ 0.491 & 0.900 $\pm$ 0.119 & 1.957 $\pm$ 0.858 & 2.025 $\pm$ 0.779 \\
FastEGNN & 0.710 $\pm$ 0.011 & 0.854 $\pm$ 0.079 & 0.806 $\pm$ 0.032 & 0.863 $\pm$ 0.020 & 0.846 $\pm$ 0.110 & 0.849 $\pm$ 0.079 & 0.988 $\pm$ 0.131 \\
\bottomrule
\end{tabular}
}
\end{table}

\end{document}